\documentclass[12pt]{article}
\pdfoutput=1
\usepackage{amssymb,amsmath,dsfont,verbatim}
\usepackage{pdfsync}
\usepackage{colonequals}
\usepackage[normalem]{ulem}
\usepackage{slashed}
\usepackage{epsfig}
\usepackage{epstopdf}
\usepackage{latexsym}
\usepackage{graphicx}
\usepackage{booktabs}
\usepackage{bbm}
\usepackage{xspace}
\usepackage{cancel}
\usepackage{enumitem}
\usepackage[numbers,compress]{natbib}
\usepackage[T1]{fontenc}
\usepackage[margin=20pt,small]{caption}
\usepackage{subfig}
\usepackage[toc]{appendix}
\usepackage{color}
\usepackage{float}
\usepackage{datetime}
\usepackage[
colorlinks=false,
linkcolor=darkblue,  
urlcolor=blue,    
filecolor=blue,
citecolor=red,
linktocpage=true,
pdfstartview=FitV,
bookmarksopen=true,
hidelinks
]{hyperref}
\numberwithin{equation}{section}

\newcommand*{\boxedcolor}{red}
\makeatletter
\renewcommand{\boxed}[1]{\textcolor{\boxedcolor}{%
		\fbox{\normalcolor\m@th$\displaystyle#1$}}}
\makeatother

\newcommand{\vev}[1]{\ensuremath{\langle #1 \rangle}\xspace}

\newcommand{\LL}{\mathcal L}

\newcommand{\op}{\mathcal O}

\newcommand{\del}{\partial} 

\newcommand{\du}[2]{_{ #1 }^{\phantom{ #1 } #2 }}

\newcommand{\D}{\Delta}
\newcommand{\hD}{\hat\Delta}

\newcommand{\Disp}{\text{D}}
\newcommand{\uhp}{{\mathbb{H}^+}}
\newcommand{\lhp}{{\mathbb{H}^-}}

\newcommand{\Cd}{\hC_{\Disp}}
\newcommand{\Cdd}{\hC_{\Disp^2}}
\newcommand{\hC}{\hat{C}}
\newcommand{\AdStwo}{\text{AdS}_2}

\newcommand{\ish}[1]{|#1\rangle\!\rangle}
\newcommand{\bc}[1]{{\bf #1}}
 \def\im{{\rm Im}}
\newcommand{\id}{{\mathds 1}}

\newcommand{\indmetr}{\hat{g}}
\newcommand{\phiR}{\phi^R}
\newcommand{\psiR}{\psi^R}
\newcommand{\psihR}{\psih^R}
\newcommand{\hid}{{\hat \id}}
\newcommand{\hop}{{\hat\op}}

\newcommand{\deltahD}{\delta{\hD}}
\newcommand{\deltaD}{\delta \D}
\newcommand{\psih}{{\hat\psi}}
\newcommand{\tieta}{\tilde{\eta}}
\newcommand{\TTb}{T\bar{T}}
\newcommand{\ddxsqrtg}[3]{\ensuremath{d^{#1} #2\,\sqrt{#3}\,}}

\definecolor{cardinal}{rgb}{0.6,0,0}
\definecolor{darkgreen}{rgb}{0,0.5,0}
\definecolor{golden}{rgb}{0.92, 0.7, 0}
\definecolor{midnight}{rgb}{0, 0, 0.5}
\definecolor{darkblue}{rgb}{0.2, 0, 0.8}

\begin{document}  
	
	\medskip
	\begin{center} 
		{\Large \bf Perturbative RG flows in AdS: an \'etude}

		\bigskip
		\bigskip
		\bigskip
		
		{\bf  Edoardo Lauria$^{1,2}$, Michael N. Milam$^{2,3}$ and Balt C. van Rees$^2$\\ }
		\bigskip
		\bigskip
		${}^{1}$
		LPENS, Département de physique, École Normale Supérieure - PSL\\
		Centre Automatique et Systèmes (CAS), Mines Paris - PSL\\
		Université PSL, Sorbonne Université, CNRS, Inria, 75005 Paris\\
		\vskip 5mm
		${}^{2}$
		CPHT, CNRS, École polytechnique, Institut Polytechnique de Paris, 91120 Palaiseau, France\\
		\vskip 5mm
		${}^{3}$
		Institut de Physique Théorique, Université Paris-Saclay, CNRS, CEA, F-9119, Gif-sur-Yvette, France\\
		\vskip 5mm				
		\texttt{edoardo.lauria@minesparis.psl.eu,~michael.milam@ipht.fr\\~balt.van-rees@polytechnique.edu} \\
	\end{center}
	
	\bigskip
	\bigskip
	
	\begin{abstract}
		\noindent 
		We discuss general properties of perturbative RG flows in AdS with a focus on the treatment of boundary conditions and infrared divergences. In contrast with flat-space boundary QFT, general covariance in AdS implies the absence of independent boundary flows. We illustrate how boundary correlation functions remain conformally covariant even if the bulk QFT has a scale. We apply our general discussion to the RG flow between consecutive unitary diagonal minimal models which is triggered by the $\phi_{(1,3)}$ operator. For these theories we conjecture a flow diagram whose form is significantly simpler than that in flat-space boundary QFT. In several stand-alone appendices we discuss two-dimensional BCFTs in general and the minimal model BCFTs in particular. These include both an extensive review as well as the computation of several new BCFT correlation functions.
	\end{abstract}
	
\newpage	

\setcounter{tocdepth}{3}	
\tableofcontents

\section{Introduction}
This work concerns perturbative and covariant renormalization group flows for QFTs in AdS. Such flows are interesting because the boundary correlation functions remain conformally invariant along the entire flow which means they can be studied with conformal bootstrap techniques \cite{Antunes:2021abs,Antunes:2024hrt}. Some early works on QFT in AdS are \cite{Callan:1989em,Aharony:2012jf}. More recently interesting results at large $N$ were obtained in \cite{Carmi:2018qzm,Ankur:2023lum,Kakkar:2022hub,Kakkar:2023gzu}, and the structure of certain correlation functions involving the stress tensor was analyzed in \cite{meineri2023renormalization}. Of course, in the flat-space limit there are many connections to the S-matrix bootstrap \cite{Kruczenski:2022lot}, both for gapped \cite{QFTinAdS} and gapless \cite{Caron-Huot:2021enk} theories, but in this paper we will mostly discuss theories with non-trivial IR fixed points.

The first part of this paper discusses several general aspects of these flows. Compared to flat space, one novelty of RG flows in AdS is the non-zero background curvature. The intricacies this introduces in perturbation theory are however well-known and discussed in many places; see for example \cite{Birrell:1982ix} for a classic reference. We therefore focus on the second novelty of RG flows in AdS: the treatment of boundary conditions and infrared divergences.

Let us start with the UV. A foolproof way to set up a QFT in AdS, at the level of correlation functions at least, is to start with a CFT on a half-space and to apply the usual Weyl rescaling rules to the local operators to obtain their AdS counterparts. This procedure appears to work for any BCFT setup, so any conformal boundary condition immediately leads to a consistent AdS boundary condition.

Next we can consider deformations that break conformal invariance. As we discuss below, one can expect infrared divergences if the boundary spectrum involves marginal or relevant operators. Infrared divergences in flat-space scattering amplitudes signal a problem with the observable itself (more specifically, individual amplitudes are simply zero), but in AdS the infrared divergences can be mitigated with suitable counterterms which make both bulk and boundary correlation functions finite.

In this work we will analyze the structure of these divergences at one loop in some generality. We note in particular that the boundary action is often fully fixed by requiring boundary conformal invariance. This observation, which we can summarize by saying that ``the boundary follows the bulk'', is in sharp contrast to the BQFT setup where the boundary can flow independently from the bulk theory. Below we also explicitly discuss the renormalization of boundary two-point functions. We show that they remain conformally invariant but that extra logarithmic divergences appear which give rise to anomalous dimensions of the boundary operators.

In the second part of this paper we analyze the RG flow between the $m$'th and $(m-1)$'th diagonal Virasoro minimal model in AdS. We will do so perturbatively at large $m$ in order to maintain control. In an upcoming paper we will analyze the same flow for finite $m$ numerically \cite{Antunes:2024hrt}.

As we discuss in more detail in appendix \ref{app:minmodelsbcft}, the consistent elementary boundary conditions for the minimal models on the half-space are well-known to be the Cardy boundary conditions. They are labeled by two integers:
\begin{equation}
	\textbf{a} \equiv (a_1,a_2)_m, \quad 1 \leq a_1 \leq m-1, \quad1 \leq a_2 \leq m, \quad (a_1,a_2) \simeq (m - a_1, m+1 - a_2)~.
\end{equation}
Each of these allowed elementary boundary conditions is also a consistent starting point of the RG flow in AdS.

The flow itself is triggered by deforming the bulk theory by the $\phi_{(1,3)}$ operator. In AdS the integrability of these flows \cite{Zamolodchikov:1987jf} does not appear to survive, but at large $m$ the flow is under perturbative control \cite{Zamolodchikov:1987ti} (see also \cite{Mussardo:2020rxh} for a review). By all accounts the bulk flow ends in the $(m-1)$'th minimal model, but the fate of the boundary conditions is less clear. A central question in this work is how precisely the flow in the space of boundary conditions on AdS differs from that on the flat half-space.

On the flat half-space the RG flow was analyzed perturbatively in \cite{Fredenhagen:2009tn}. For example, for $a_1, a_2 \ll m$ and $a_2 > 1$ the authors of \cite{Fredenhagen:2009tn} found evidence for a natural flow from $(a_1,a_2)_m$ to $(a_2-1,a_1)_{m-1}$. Together with earlier discovered boundary flows within the $m$'th minimal model \cite{Recknagel:2000ri,Graham:2001pp}, there is an interesting web of flows as shown diagrammatically in figure 2 of \cite{Fredenhagen:2009tn}. In particular, the diagram shows that reaching $(a_2,a_1)_{m-1}$ from $(a_1,a_2)_{m}$ can only be done through a sequence of flows that passes through two intermediate fixed points:
\begin{equation}
	(a_1,a_2)_m \to \bigoplus_{l = 1}^{\min(a_1,a_2)} (a_1 + a_2 + 1 - 2l, 1)_m \to \bigoplus_{l = 1}^{\min(a_1,a_2)} (1, a_1 + a_2 + 1 - 2l)_{m-1} \to (a_2,a_1)_{m-1}~,
\end{equation}
where the first and last arrow correspond to the boundary flow of \cite{Recknagel:2000ri,Graham:2001pp} (with opposite signs) and the second arrow to a bulk flow.

The ``boundary follows the bulk'' principle means that, in contrast to \cite{Fredenhagen:2009tn}, we should not have to contend with independent couplings and beta functions for the boundary operators in AdS. This considerably simplifies our analysis. We provide a bit of evidence for a much simpler picture and, based on the anomalous dimensions of boundary operators, speculate that
\begin{equation}
	(a_1,a_2)_{m} \to (a_2,a_1)_{m-1}~,
\end{equation}
is the covariant flow in AdS.

Below we will analyze the perturbed one-point functions of bulk operators and two-point functions of boundary operators. One peculiar feature is the appearance of divergences linear in $m$ in certain integrals which are cancelled by $1/m$ behavior in the (OPE) coefficients that multiply the integrals. The divergences themselves are entirely analogous to the $1/\epsilon$ poles in dimensional regularization (and one may call them `$1/(1/m)$ poles'). Normally one would add suitable counterterms to renormalize these divergences, but there is no need to do so if they multiply suppressed coefficients. This phenomenon however invalidates any estimate of the magnitude of certain contributions (in the $1/m$ expansion) that is based purely on OPE coefficients.

\subsubsection*{Material in the appendices}
Our perturbative QFT in AdS results are obtained by Weyl rescaling and integrating correlation functions of two-dimensional BCFTs. These BCFT correlators warrant a separate discussion, which we present in appendices \ref{app:2dbcfts} (for general 2d BCFTs) and \ref{app:minmodelsbcft} (for the minimal models specifically). The material presented there is a substantial part of the work done for this paper and contains several new results.\footnote{In particular, to the best of our knowledge the equations \eqref{phi13twoptblocks}, \eqref{allblocks}, \eqref{phi31twoptblocks}, \eqref{allblocks31} are all new. We would also like to highlight equations \eqref{bulkbdgenm}, \eqref{largemalpha13}, \eqref{largemalpha31}, \eqref{bulkbd4mmm}, \eqref{bulkbdgenm31}. These can in principle be derived using F-matrices, but the latter are usually difficult to implement for continuous values of the parameters.} As such it could have been included in the main text, but its inclusion as an appendix hopefully clarifies that it is stand-alone material which might be useful beyond the QFT in AdS discussion. We advise the reader to at least briefly consult these appendices before starting with subsection \ref{anomalousdimAdS} and section \ref{sec:minmoddeformations}.

\subsubsection*{Ancillary material}
With the submission of this work to the arXiv e-print archive we have included a \texttt{Mathematica} notebook \texttt{ancillary.nb} which contains all the substantial computations done in this work.

\section{General principles}
\label{sec:generalities}
In this and the next section we will discuss some general principles of QFT in AdS as well as one-loop perturbation theory. Many of these principles have appeared (sometimes implicitly) in the literature to date, but we think that it is useful to present a systematic viewpoint.

\subsection{From BCFT to AdS}
In Poincar\'e coordinates the AdS metric reads
\begin{equation}
	ds^2 = g_{\mu \nu} dx^\mu dx^\nu = \frac{R^2}{z^2} (dz^2 + d \vec x^2)~,
\end{equation}
and it is immediately seen that AdS$_{d+1}$ is Weyl equivalent to the half-space defined as the subspace of flat $\mathbb R^{d+1}$ with $z > 0$. Therefore any consistent CFT defined on a half-space, so any BCFT, can immediately be put in AdS by the usual Weyl rescaling rules. To illustrate this consider for example the general form of the BCFT one- and two-point functions of scalar conformal primaries:
\begin{equation}\label{BCFTdefs}
\begin{split}
	\vev{\op(\vec x,z)}_\text{BCFT} &= \frac{B_\op}{(2z)^\Delta}~,\\
	\vev{\op_1(\vec x_1,z_1) \op_2(\vec x_2,z_2)}_\text{BCFT} &= \frac{f(\xi)}{(2z_1)^{\Delta_1} (2z_2)^{\Delta_2}}, \qquad \xi = \frac{(\vec x_1 - \vec x_2)^2 + (z_1 - z_2)^2}{4 z_1 z_2}~,
\end{split}
\end{equation}
with $B_\op$ a BCFT-dependent coefficient and $f(\xi)$ a BCFT-dependent function. For these operators the Weyl rescaling rule is simply
\begin{equation}
	\vev{\op(\vec x, z)\ldots}_\text{AdS} = (z/R)^\Delta \vev{\op(\vec x, z)\ldots}_\text{BCFT}~,
\end{equation}
and therefore
\begin{equation}
\label{cftinadscorrelators}
\begin{split}
	(2R)^\Delta  \vev{\op(\vec x,z)}_\text{AdS} &= B_\op~,\\
	(2R)^{\Delta_1 + \Delta_2} \vev{\op_1(\vec x_1,z_1) \op_2(\vec x_2,z_2)&}_\text{AdS} = f(\xi)~.
\end{split}
\end{equation}
In words, we see that one-point functions are constant and two-point functions depend only on the geodesic distance $d_{1,2}$ since
\begin{equation}
	2 \xi + 1=  \cosh(d_{1,2}/R)~.
\end{equation}

Let us now draw some more general lessons. The form \eqref{cftinadscorrelators} is also the one required by AdS general covariance, fundamentally because the BCFT conformal algebra $\mathfrak{so}(d+1,1)$ is precisely the same as the AdS$_{d+1}$ isometry algebra. There are in fact no \emph{further} Ward identities (at fixed $R$) that follow from conformal invariance of the theory: the form \eqref{cftinadscorrelators} is the most general one for an arbitrary non-conformal QFT in AdS.\footnote{This phenomenon extends to spinning operators. For example, the two-point function of (not necessarily traceless) spin 2 operators in AdS as computed recently in \cite{meineri2023renormalization} essentially agrees with the corresponding BCFT correlator \cite{Liendo:2012hy}.} Therefore one needs to work a little harder to see that the bulk theory is conformal. One method is to inspect the undetermined functions like $f(\xi)$ in order to find out that they are compatible with a bulk conformal OPE. Another is to compare the theories at different values of $R$. For example, in a general QFT with a mass scale $\mu$ the one-point function coefficients $B_\op$ can be complicated functions of $\mu R$, whereas in a CFT the $R$-dependence must be trivial.

The above analysis immediately extends to higher-point functions. Covariant correlation functions must depend only on the geodesic distances between points and, for non-conformal theories, on the dimensionless combination $\mu R$. These dependences are not otherwise constrained by first principles.

In the above we used the dimensionless combinations $R^\Delta \op(\vec x,z)$. In a general non-conformal theory the bulk operator may of course obtain a non-trivial scaling, which can however similarly be offset with powers of $R$. It is convenient to always work with such a dimensionless version of the bulk operators.

\subsection{Towards the boundary}
Consider now sending the insertion points towards the conformal boundary of AdS by taking $z \to 0$. For BCFT correlation functions this limit is defined by the boundary operator expansion. For scalar primaries we can for example write: 
\begin{equation}
\vev{\op(\vec x, z)\ldots}_\text{BCFT} = \sum_{\hop} z^{{\hD}_{\hop} - \Delta_\op} B_\op^{\phantom{\op}\hop} \vev{\hop(\vec x)\ldots}_\text{BCFT}~,
\end{equation}
with $\hop(\vec x)$ a boundary operator. The AdS version then reads:
\begin{equation}
\label{bulkboundaryinads}
(2R)^{\Delta_\op} \vev{\op(\vec x, z)\ldots}_\text{AdS} = \sum_{\hop} z^{{\hD}_{\hop}} B_\op^{\phantom{\op}\hop} \vev{\hop(\vec x)\ldots}_\text{AdS}~,
\end{equation}
so we Weyl rescale the bulk operator but not the boundary operator. The boundary operator correlation functions are then unchanged, for example:
\begin{equation}
	\vev{\hop(\vec x) \hop(\vec y)}_\text{BCFT} = \frac{1}{(\vec x - \vec y)^{2 \hD_{\hop}}} = \vev{\hop(\vec x) \hop(\vec y)}_\text{AdS}~,
\end{equation}
and similarly we find the usual conformal structures for any $n$-point function of boundary operators.

How does this work for a non-conformal QFT in AdS? The general idea, formulated for example in \cite{QFTinAdS}, is that there is still a convergent state-operator correspondence, or more precisely a correspondence between bulk states and boundary operators. If the AdS radial Hamiltonian can be diagonalized and has discrete spectrum then it defines a set of boundary operators that transform covariantly under conformal transformations. The expansion as in \eqref{bulkboundaryinads} is then fixed by symmetry; one quick way to conclude this is to notice the functional form of the $\vev{\op(\vec x, z) \hop(\vec y)}$ two-point function is completely fixed. For a non-conformal field theory the coefficients $b_\op^{\phantom{\op}\hop}$ and $\hD$ can depend non-trivially on the dimensionless ratio $\mu R$.

\subsubsection{Distance versus curvature}
Consider a non-conformal field theory in AdS and hold fixed its mass scale $\mu$. Correlation functions of local bulk operators can have an arbitrary dependence on invariant distances, say $d(\cdot,\cdot)/R$, and on the combination $\mu R$. Physically, however, it is not immediately clear what happens when we consider the near-boundary or large-distance limit of these correlation functions. Do we always observe the infrared behavior of the theory, as would be the case for theories in flat space? Or does the curvature intervene and cut off the RG flow at the scale $\mu R$? It appears to us that the latter option must be realized. (This viewpoint was also taken in \cite{Aharony:2012jf}. As argued in that paper, see also \cite{Carmi:2018qzm}, observables do not need to be smooth as a function of $\mu R$.) To see this we appeal to the boundary operator expansion which ensures that every correlation function has a convergent boundary block decomposition. The dimensionless parameters $\hD_\op$ and $b_\op^{\phantom{\op}\hop}$ appearing in this decomposition are insensitive to any distances, and therefore can only depend on $\mu R$. So we conclude that, by varying the curvature radius, we should be able to see the entire RG flow even if we restrict ourselves to boundary correlation functions (or, more generally, the longest distances in AdS).

\section{Hyperbolic life at one loop}
Suppose we switch on a marginal or relevant deformation in AdS$_{D}$ by a local operator $\op(x)$ with scaling dimension $\Delta_\op$. The correlation functions in the deformed theory can be computed perturbatively by expanding
\begin{equation}
  \vev{\ldots \exp\left( - g  R^{\Delta_\op - D} \int \ddxsqrtg{D}{x}{g}\op(x)+ S_\text{c.t.} \right)}~,
\end{equation}
in the dimensionless coupling $g$. (We do not expect denoting the metric by $g_{\mu \nu}$ and the coupling by $g$ to cause any confusion; in particular, the latter will always appear outside integrals.) Rather than introducing a new dimensionful parameter, we have introduced the naturally available AdS curvature radius to make the perturbation dimensionless.

The counterterm action in the above expression is necessary to cancel both bulk and boundary divergences. The bulk divergences are of the UV type. They are essentially unchanged compared to those in flat space, although new counterterms involving the AdS curvature may be needed. Their renormalization is well understood, see for example \cite{Birrell:1982ix}, so we will only briefly consider them in the next subsection. Afterwards we will focus on the \emph{infrared} divergences which arise close to the AdS boundary.

\subsection{Ultraviolet behavior}
\label{subsec:uvbehaviorgeneral}
In this subsection we will consider a nearly marginal deformation, so we set
\begin{equation}
	\Delta_\op = D - \epsilon~,
\end{equation}
and consider the limit of small $\epsilon$.

Let us first investigate the one-loop running in AdS and in BCFT. To this end we consider
\begin{equation}
\frac{1}{2} g^2 \mu^{2 \epsilon} \int \ddxsqrtg{D}{x_1}{g}\op(x_1) \int \ddxsqrtg{D}{x_2}{g}\op(x_2)~,
\end{equation}
in either setup. We can assume the self-OPE of $\op(x)$ to take the form
\begin{equation}
\label{selfopenearlymarginal}
	\op(x) \op(y) = \frac{1}{|x-y|^{2\Delta_\op}} + \frac{C_{\op\op\op} \op(y)}{|x-y|^{\Delta_\op}} + \ldots~,
\end{equation}
so the term leading to renormalization of the coupling constant reads
\begin{equation}
\frac{1}{2} g^2 \mu^{2 \epsilon} C_{\op\op\op} \int \ddxsqrtg{D}{x_2}{g}\op(x_2) \times
\begin{cases}
\int d^d \vec x \int_0 dz_1 |x_1 - x_2|^{-\Delta_\op} & \qquad \text{BCFT}\\
\int d^d \vec x \int_0 dz_1 R^{D - \Delta_\op} z_1^{\Delta_\op -D} |x_1 - x_2|^{-\Delta_\op} & \qquad \text{AdS}
\end{cases}
\end{equation}
with the extra factors in the AdS integral originating from the metric determinant and the Weyl rescaling, with $D=d+1$. Doing the parallel integral yields
\begin{equation}
\frac{1}{2} g^2 \mu^{2 \epsilon} C_{\op\op\op} \frac{\pi^{d/2} \Gamma((\D - d)/2)}{\Gamma(\Delta/2)} \int \ddxsqrtg{D}{x_2}{g} \op(x_2) \times
\begin{cases}
\int_0 dz_1 |z_1 - z_2|^{\epsilon - 1} & \quad \text{BCFT}\\
\int_0 dz_1 R^{\epsilon} z_1^{-\epsilon} |z_1 - z_2|^{\epsilon - 1}	& \quad \text{AdS}\\
\end{cases}
\end{equation}
The integral over the orthogonal direction diverges, but since we are only interested in the short-distance singularities we can just cut it off at a distance $L$ from $z_2$. In a more physical setup such an $L$ would more naturally be provided by another operator insertion, and so one should think of $L$ as the typical scale of the observable under consideration. With such a cutoff we find:
\begin{equation}
\frac{1}{2 \epsilon} g^2 \mu^{2 \epsilon} C_{\op\op\op} \frac{\pi^{d/2} \Gamma((\D - d)/2)}{\Gamma(\Delta/2)} \int \ddxsqrtg{D}{x_2}{g} \op(x_2) \times
\begin{cases}
L^\epsilon & \qquad \text{BCFT}\\
\begin{cases}
L^\epsilon \qquad L \ll R\\
R^\epsilon \qquad R \ll L
\end{cases}
& \qquad \text{AdS}
\end{cases}
\end{equation}
where we ignored non-pole terms in $\epsilon$ --- in AdS these are a non-trivial function of $R/L$ for which we show the dominant term. From this computation we can first of all conclude that the coupling self-renormalization is the same in both theories. Secondly, to the extent that we can trust this one-loop result, we see that in AdS the natural dimensionless combination is $\mu R$ for sufficiently large distances. On the other hand, in BCFT the natural dimensionless ratio is $\mu L$ (as long as we can ignore boundary effects). This is again in agreement with the general discussion presented above, where we argued that in AdS the coupling constants naturally run with the curvature radius $R$ and not with the size $L$ of the observable. This also provides a concrete illustration of the picture sketched in the introduction of \cite{Carmi:2018qzm}.

Another interesting aspect of renormalization in curved space is the existence, for every local operator $\op(x)$ of dimension $\Delta$, of sequences of additional local operators of the schematic form
\begin{equation}
	(\text{Riemann})^n \op(x)~,
\end{equation}
where $(\text{Riemann})^n$ is a degree $n$ homogeneous polynomial in the Riemann tensor with index contractions chosen at will. These operators will have dimensions $\Delta + 2n$ and it is essential to incorporate them in order to understand operator mixing in perturbation theory.

This phenomenon arises for example when we consider the integral of the first term in equation \eqref{selfopenearlymarginal}. Let us use point-splitting to regulate the corresponding short-distance divergence in the integral. In flat space we then find a simple power-law divergence $\Lambda^{2\Delta - D}$, which can be absorbed by including a (divergent) operator mixing with the identity operator $\id$. In AdS we will of course find the same divergence to leading order but there can also be subleading terms of the form $\Lambda^{2\Delta - D - 2n}$. If some of these terms are divergent then they need to be renormalized, and for scalar operators the correct way to do this is by including operator mixing with the additional operator $(\text{Ric})^n \id$. Notice finally that, if $\Delta = D - \epsilon$ and $D$ is even then there might in principle appear $1/\epsilon$ poles as well, but for AdS an explicit computation shows that this is not the case.

Consider now a theory where there exist two operators $\op_1(x)$ and $\op_2(x)$ with scaling dimensions such that $\Delta_2 = \Delta_1 + 2$ (plus perhaps perturbatively small corrections) and otherwise the same quantum numbers. In that case there is an ambiguity: for any real $\alpha$ the operator
\begin{equation}
	\op_2'(x) = \op_2(x) + \alpha (\text{Ric})(x) \op_1(x)~,
\end{equation}
will be just as good as $\op_2(x)$. On a general curved manifold there is no scheme-independent way to fix such ambiguities, although exceptions exist in special cases. For example, in \cite{Gerchkovitz:2016gxx} the coefficients were fixed on the round $S^4$ by demanding that two-point functions are diagonal and one-point functions are vanishing. The extra symmetries of AdS do not help much in this regard, although at a fixed point one can fix $\alpha$ by demanding that the one-point functions match those of the corresponding BCFT.

\subsection{Infrared divergences}
\label{subsec:infrareddivergencesgeneral}
Infrared divergences in conformal perturbation theory are not uncommon. For example, in flat space conformal correlation functions fall off like $x^{-2 \Delta}$, so infrared divergences arise whenever $\Delta \leq d/2$. An example of such a strongly relevant deformation is the mass deformation for the CFT of a free scalar field $\phi$. A divergence can already be found at first order in perturbation theory:
\begin{equation}
  \delta \vev{\phi(0)\phi(y)} = -\mu^2 \int d^D x\,\vev{\phi(0)\phi(y) \phi^2(x)} \approx - \mu^2 S_D \int \frac{dr}{r^{D-3}}= - \mu^2 S_D \frac{R^{4-D}}{4-D}~,
\end{equation}
where we had to cut off the integral at $|x| < R$ to make it finite. One might now decide to stop the computation here, because the infrared divergence indicates that the theory becomes strongly coupled at large distances. This is however too pessimistic. After all, the flat-space propagator of a massive free scalar field has a good expansion in $\mu^2 x^2$ which we should be able to compute perturbatively. And indeed the divergence can easily be cancelled if we add a boundary counterterm at $|x| = R$. Since
\begin{equation}
\frac{\mu^2 R}{4 - D}\int_{|x| = R} d^{D-1} x\,\vev{\phi^2(0) \phi^2(x)}  =  \mu^2 S_D \frac{R^{4-D}}{4-D}~,
\end{equation}
a deformation of the form:
\begin{equation}
  \vev{\ldots \exp\left( - \mu^2 \int_{|x| \leq R} d^D x\,\phi^2 (x) + \frac{\mu^2R}{4-D} \int_{|x| = R} d^{D-1} x\,\phi^2(x)~ \right)}~,
\end{equation}
is infrared finite to first order in perturbation theory.

In AdS the situation is richer. For a bulk operator $\op$ (not necessarily free), substituting the boundary operator expansion, we find for small $z$ that
\begin{equation}
  \int \ddxsqrtg{D}{x}{g} \op(x) \approx \int d^d \vec x \int_0 dz \, B_{\op \hop} z^{\hD_{\hop} - d - 1} \hop(\vec x)~,
\end{equation}
which is divergent for any marginal or relevant boundary operator with $\hD_{\hop} \leq d$. We can remove the leading divergence by cutting the $z$ integral off at a small value $a$ and adding a counterterm
\begin{equation}
  \frac{R}{\hD_{\hop} - d} \int \ddxsqrtg{d}{\vec{x}}{h} \op(\vec x, a)~,
\end{equation}
with $h$ the induced metric at the cutoff surface and an extra $R$ inserted for dimensional reasons. Notice that the counterterm involves the \emph{bulk} operator $\op(\vec x, a)$. Subleading divergences can be removed by adding other operators or by a counterterm involving normal derivatives, for example
\begin{equation}
 \int \ddxsqrtg{d}{\vec{x}}{h} n^\mu \del_\mu \op(\vec x, a)~,
\end{equation}
and appropriately tuning the coefficients.

When $\hD_{\hop} = d$ the divergence is logarithmic and we need a counterterm of the form
\begin{equation}
  \log(a/R)\int \ddxsqrtg{d}{\vec{x}}{h} \op(\vec x, a)~,
\end{equation}
where we inserted a factor $R$ on dimensional grounds. This leads to a breaking of conformal invariance, which we may capture in terms of a beta function for the boundary coupling. We will investigate this in more detail in section \ref{ss:marginacase}.

\subsubsection{Covariance and comparison to BCFT}
The near-boundary divergences in conformal perturbation theory in AdS are similar to those found in BCFT. Indeed, at any loop order the integrands in an AdS background differ from those in a flat BCFT background only by the insertion of Weyl factors and the metric determinant. There is however an important qualitative difference: the AdS divergences are of an \emph{infrared} nature and the BCFT divergences are of an \emph{ultraviolet} nature. This completely changes the way in which the divergences are cancelled by counterterms.

First, in AdS the counterterms are naturally defined on the cutoff surface in order to retain covariance. To see this, consider a Killing vector field $\xi^\mu$ in AdS, and consider the invariance of the first-order correction to any correlation function $\vev{\ldots}$ along $\xi$. Concretely this means that we should impose that transporting every operator in the perturbed correlation function along $\xi$ by means of the Lie derivative $\LL_\xi$ gives zero. Since the bare correlation function is by assumption invariant, we can move the Lie derivative to the perturbing operator $\op(x)$, leading to the following demand:
\begin{equation}
	0 = \int_{a} \ddxsqrtg{D}{x}{g}\xi^\mu\nabla_\mu \vev{\op(x) \ldots} + \LL_\xi S_\text{c.t.}~,
\end{equation}
where we assumed that $\op(x)$ is a scalar. After integration by parts we find a boundary term at the cutoff surface:
\begin{equation}
	0 =\left. \int \ddxsqrtg{d}{\vec{x}}{h} \xi^\mu n_\mu \vev{\op(x) \ldots} \right|_a + \LL_\xi S_\text{c.t.}~.
\end{equation}
This is a simple differential equation for the counterterm action. It is most naturally solved by a counterterm \emph{on} the near-boundary cutoff surface, which is why such counterterms are the correct covariant ones.

In BCFT no such constraint arises because all the remaining Killing vector fields leave the boundary invariant and therefore obey $n_\mu \xi^\mu = 0$. It is then customary, and indeed much simpler, to define the counterterm action on the actual boundary rather than at the cutoff surface. Such a procedure is of course entirely in keeping with the usual way that UV divergences are renormalized. For example, consider the point-splitting regularization procedure to renormalize a bare local operator $\op_i(x)$. The renormalization is always done by replacing the bare operator by a renormalized operator $Z\du{i}{j}\op_j(x)$, which in particular is inserted at exactly the same point as the original operator rather than, say, integrated along the cutoff surface.

As we have seen, demanding general covariance also means that the boundary counterterms are \emph{completely determined} by the bulk RG flow. In BQFT it is well known that there can be independent bulk and boundary RG flows; in particular, the boundary can undergo an RG flow while the bulk theory remains critical. In AdS/CFT a similar phenomenon occurs for example in the form of the double-trace flow \cite{Witten:2001ua}, where a double-trace deformation of the CFT is encoded holographically by a changing of the boundary conditions of the bulk fields. However such flows are unnatural in the precise sense that they do not retain AdS covariance. If we insist that the deformed correlation functions remain covariant (and if there is no exactly marginal boundary coupling) then there is no freedom left to apply independent boundary deformations. In this way we can say that ``the boundary follows the bulk'' for QFT in AdS.

\subsubsection{Marginal boundary deformations and covariance}\label{ss:marginacase}
It is worth discussing how the above discussion is modified in the presence of an exactly marginal boundary operator. As is nicely discussed in section 6 \cite{Hogervorst:2021spa} and more recently also in \cite{toappearLorenzo}, the general idea is that in such cases a boundary RG flow (dis)appears through a so-called saddle-node bifurcation, just as happens in ordinary field theories (see for example \cite{Gorbenko:2018ncu}). Here we will confirm this picture by establishing that boundary covariance is indeed violated without fine-tuning a boundary coupling, and consequently that unitarity covariant boundary conditions generically disappear at this point.\footnote{This subsection arose from discussions with L.~Di Pietro, based on results obtained in \cite{toappearLorenzo}.}

Suppose then that along the bulk RG flow we have arrived at a point with an exactly marginal boundary operator $\hop(\vec x)$. If it appears in the boundary operator expansion then we may write:
\begin{equation}
    (2R)^{\D_{\op}}\op(\vec x, z) =  z^{\hD} B_\op^{\phantom{\op}\hop} \hop(\vec x) + \ldots~,
\end{equation}
with $\hD \equiv \hD_{\hop} = d$ and a non-zero coefficient $B_\op^{\phantom{\op}\hop}$. As we mentioned above, the first-order correction then features a logarithmic divergence:
\begin{equation}
    \delta\vev{\ldots} = - \lambda B_\op^{\phantom{\op}\hop} \int_a \frac{dz}{z} \int d^d \vec x\, \vev{\hop(\vec x)\ldots}  = - \lambda B_\op^{\phantom{\op}\hop} \log(a / R) \int d^d \vec x\, \vev{\hop(\vec x)\ldots}~,
\end{equation}
and to cancel it we add a counterterm at the cutoff surface:
\begin{equation}
    \label{firstordercounterterm}
    \lambda \log(a/R) \int \ddxsqrtg{d}{\vec{x}}{h} \op(\vec{x},a)    + \hat g \int \ddxsqrtg{d}{\vec{x}}{h}  \op(\vec{x},a)~.
\end{equation}
Here the first logarithmic term cancels the divergence and we also added a finite term with an arbitrary coefficient $\hat g$ because why not.

We will now show that this counterterm breaks AdS covariance, by repeating the argument above. Recall that the AdS Killing vector fields $\xi^\mu \partial_\mu$ in Poincar\'e coordinates $(\vec x, z)$ are all of the form
\begin{equation}\label{killvects}
    \xi^i(\vec x, z) = \beta^i(\vec x) - \frac{z^2}{2 d} \delta^{ij} \partial_j \partial_k \beta^k(\vec x)~, \qquad \xi^z = \frac{z}{d} \partial^k \beta_k(z)~,
\end{equation}
where $\beta^i(\vec x)\partial_i$ is a conformal Killing vector field on the boundary, so
\begin{equation}
    \partial_i \beta_j + \partial_j \beta_i \propto \delta_{ij}~.
\end{equation}
(For example, if $\beta^i(\vec x)$ is an ordinary Killing vector field then the AdS uplift is simply $\xi^\mu \partial_\mu = \beta^i (\vec x) \partial_i$, whereas for the scaling vector field we find $\xi^\mu \partial_\mu = z \partial_z + \vec x \cdot \partial_{\vec x}$.)
As in the previous subsubsection, we compute (with $\hat g = 0$ for brevity)
\begin{equation}
\label{firstorderbreaking}
\begin{split}
- \delta \vev{\LL_\xi \ldots} &= \lambda \int_a \ddxsqrtg{D}{x}{g}\vev{\ldots \LL_\xi\op(x)} - \lambda \log(a/R) \int \ddxsqrtg{d}{\vec{x}}{h} \vev{ \ldots \LL_\xi\op(\vec{x},a) }\\
    &= \lambda \int \ddxsqrtg{d}{\vec{x}}{h} n^\mu \xi_\mu  \vev{\ldots  \op(\vec{x},a)} - \lambda \log(a/R) \int \ddxsqrtg{d}{\vec{x}}{h} \vev{ \ldots \LL_\xi\op(\vec{x},a) }\\
    &= \frac{\lambda B_\op^{\phantom{\op}\hop} }{d} \int d^d \vec x\, (\partial_k \beta^k)\, \vev{\ldots \hop(\vec x)}~.
\end{split}
\end{equation}
Here, on the first line we again used covariance of the full unperturbed correlation function to move the Lie derivative to the perturbing operator $\op(x)$. Then we used integration by parts, with $n^\mu \partial_\mu = z \partial_z$ in Poincar\'e coordinates. The second term on the second line, which multiplies the logarithm, actually vanishes as $a \to 0$ because in that limit it becomes just a conformal transformation of an integrated operator of dimension $d$. For the same reason the term multiplying $\hat g$ in \eqref{firstordercounterterm} would also not contribute to the violation. Finally we used the bulk-boundary expansion and \eqref{killvects}.

From equation \eqref{firstorderbreaking} we see that the AdS isometries that correspond to conformal (but not ordinary) Killing vector fields on the boundary are \emph{broken} by the bulk perturbation. One can try to fix this issue by adding a finite counterterm as in equation \eqref{firstordercounterterm} with $\hat g$ proportional to $\sqrt{\lambda}$. In that case we have another term at $O(\lambda)$ of the form:
\begin{equation}
     \frac{\hat g^2}{2} \int \ddxsqrtg{d}{\vec{x}}{h} \op(\vec{x},a) \int \ddxsqrtg{d}{\vec{y}}{h} \op(\vec{y},a)~.
\end{equation}
Using first the bulk-boundary and then the boundary-boundary OPEs, we find another divergence in this integral as $\vec x$ approaches $\vec y$.\footnote{It is not a priori clear that this divergence is accurately captured by this sequence of operator expansions, since part of the integral is at the boundary of its domain of convergence. Our analysis can probably be made rigorous using the local blocks in \cite{Levine:2023ywq}. We however expect the conclusions to be the same.} To regulate this divergence we will introduce a further regulator $|\vec x - \vec y| > a$. Then a small computation shows that we need the counterterm
\begin{equation}
\begin{split}
     - \frac{\hat g^2}{2}  B_\op^{\phantom{\op}\hop}\,  S_d\, \hC_{\hop\hop\hop}\, \ln(a/R) \int \ddxsqrtg{d}{\vec{x}}{h} \op(\vec{x},a)~,
\end{split}
\end{equation}
where $\hC_{\hop\hop\hop}$ is the boundary OPE coefficient and $S_d$ the volume of a unit sphere in $d$ dimensions. The logarithmic divergence again leads to a violation of AdS covariance, this time of the form:
\begin{equation}
    \delta_{\hat g} \vev{\LL_\xi \ldots} = - \frac{\hat g^2}{2d}  B_\op^{\phantom{\op}\hop}B_{\op\hop}\,  S_d\, \hC_{\hop\hop\hop}\, \int d^d{\vec x}\, (\partial_k \beta^k)\, \vev{\ldots \hop(\vec x)}~,
\end{equation}
so we can actually cancel the non-covariance by setting: 
\begin{equation}
    \lambda - {\hat g}^2 B_{\op\hop}\, S_d\, \hC_{\hop\hop\hop}   = 0~.
\end{equation}
For non-zero $B_{\op\hop}$ and  $\hC_{\hop\hop\hop}$ we see that only one sign of $\lambda$ gives a real solution for $\hat g$. On the other side we must break either unitarity or AdS covariance.

The simplest example of this phenomenon is the Breitenlohner-Freedman bound. For a free massive bulk scalar with $m^2 R^2 = \Delta (\Delta - d)$ the bound is saturated at $\Delta =  d/2$ where the Dirichlet and Neumann boundary conditions coincide. If we now perform a mass deformation then we find a logarithmic infrared divergence at the first order, which can be canceled by the exactly marginal double twist boundary operator. If we want to retain covariance we need to tune the coefficient of its coupling, which can only be chosen real for $m^2 R^2 > - d^2/4$. We refer to \cite{Hogervorst:2021spa,Ankur:2023lum,toappearLorenzo} for other interesting examples of this phenomenon.

\subsubsection{Comparison to holographic renormalization}
The importance of the covariance of the counterterms is familiar from previous works on holographic renormalization in the gauge/gravity dualities; see \cite{Skenderis:2002wp} for an introduction. Most of that work considers classical gravity with matter on AAdS backgrounds, although recently loop corrections were also considered more systematically in \cite{Fichet:2021xfn,Banados:2022nhj}. In a gravity setup there are no local bulk operators and it is more natural to consider the partition function as a functional of the boundary conditions $\phi_{(0)}$ for the bulk fields. In the modern understanding the one-point function of the boundary operator is then given by the (renormalized) conjugate radial momentum of the bulk field. The renormalization of the divergences in the on-shell action leads to an infamous factor $2\Delta - d$ \cite{Freedman:1998tz} compared to a more naive prescription.

We do not consider theories of gravity. Therefore we do have local bulk operators and we can simply consider their limit as we send them to the boundary. One immediately verifies \cite{Klebanov:1999tb} that this leads to the same normalization for two-point functions, including the extra factor $2\Delta -d$, because of the difference between the bulk-bulk and bulk-boundary propagator:
\begin{equation}
	G_{\text{bu-bu}}(x,x') = z^\Delta \left( \frac{1}{2\D -d} G_\text{bu-bou}(\vec x, x') + O(z) \right)~,
\end{equation}
holding $x'$ fixed in the bulk. Although this prescription suffices for our purposes, it does not give all the same information as the more familiar holographic renormalization prescription. In particular, since the boundary correlation functions are never integrated (against sources) we are insensitive to their potentially ill-defined distributional nature. Correspondingly it is not straightforward to see contact terms in Ward identities of boundary currents (including a potential stress tensor) or their anomalies \cite{Henningson:1998gx}. When such questions are important for QFT in AdS we expect that a more refined analysis will be needed. It is then a matter of taste whether to take the more abstract viewpoint of conformal perturbation theory, as we do in this work, or Lagrangian perturbation theory, as was for example done in \cite{Banados:2022nhj}.\footnote{There is one more sense in which the renormalization of the on-shell action, or partition function at loop level, is richer than that of the correlation functions. According to \cite{Papadimitriou:2005ii} a specific boundary action is needed in order to have a well-defined variational principle for the bulk fields. It is unclear to us whether this condition can be understood from the conformal perturbation theory viewpoint that we take here.}

\subsubsection{Is AdS a box?}
A massive particle cannot reach the conformal boundary of AdS by moving along a geodesic, even though part of that boundary is certainly in causal contact with it. In this sense AdS confines, and can be seen as a box. On the other hand, for massless particles AdS is much less box-like since they can reach the boundary in finite global time.

Even for gapped theories, however, it is not at all immediate that (conformal) perturbation theory in an AdS background is \emph{automatically} infrared finite. The earliest claim that it is appears to date from \cite{Callan:1989em}. In that paper the authors claim that ``perturbation theory diagrams are infrared well-behaved'' based on the exponential falloff of bulk-bulk propagators with distance. However in actuality this exponential falloff can be offset by the exponentially growing area of spheres in hyperbolic space; a fact that is actually mentioned a little earlier in the same paper. As we have just seen, whether a perturbative computation is infrared finite or not depends on the spectrum of boundary operators, and there is certainly the possibility of infrared divergences. We are led to conclude that AdS is perhaps not that great of a box.

One major advantage nevertheless appears to remain: after renormalization the boundary correlation functions are likely unambiguously defined for massless and massive theories alike, even when the corresponding massless S-matrix elements are infrared divergent.

\subsection{Anomalous dimensions of boundary operators}\label{anomalousdimAdS}
It is interesting to consider the general first-order correction to two-point functions of boundary operators under a bulk deformation. Our goal will be to demonstrate that these correlation functions remain conformal at first order, and that for equal operators there generically appears a logarithmic divergence that gives rise to an anomalous dimension.

To first order the two-point function is given by:
\begin{equation}
\label{correctedtwoptgeneral}
	 \vev{\psih_1(\vec{x}_1) \psih_2(\vec{x}_2)} = \frac{\delta_{12}}{{x}_{12}^{2 \hD_1}} -  g  R^{\Delta- D}\int \ddxsqrtg{D}{x}{g} \vev{\phi(x) \psih_1(\vec{x}_1) \psih_2(\vec{x}_2)} + (\text{counterterms}) + O( g^2)~,
\end{equation}
for a generic deformation by a bulk operator denoted $\phi(x)$ with dimension $\Delta$. Notice that $\vec{x}_1$ and $\vec{x}_2$ are boundary points, $x_{12}\equiv |\vec{x}_{1} - \vec x_2|$, and $x$ is a bulk point. We will work in Poincar\'e coordinates where $z$ is the transverse coordinate. The dimensionless bulk coupling is denoted $g$.

The boundary counterterms will consist of two parts. First there will be the generic boundary counterterms discussed in the previous section, which will cancel any divergences arising as $\phi({x})$ approaches the boundary away from the insertion points of the boundary operators. Then there will be wave function renormalization factors that will cancel additional divergences that appear as $\phi(x)$ approaches $\psih(\vec{x}_1)$ or $\psih(\vec{x}_2)$.

The three-point function in equation \eqref{correctedtwoptgeneral} can be decomposed into conformal blocks:
\begin{align}
	x_{12}^{\hD_1+\hD_2} (2R)^{\D} \langle \psih_1({\vec{x}}_1)\psih_2({\vec{x}}_2)\phi(\vec{x},z)\rangle =\left( \frac{z^2+(\vec{x}_2-\vec{x})^2}{z^2+(\vec{x}_1-\vec{x})^2}\right)^{{\hD_{12}/2}} 
	\sum_{k} B\du{\phi}{k} \hC_{k \psih \psih}  f(\hD_{12}, \hD_k, v)~,
\end{align}
where $\hD_{12}\equiv \hD_1-\hD_2$ and the Weyl rescaling from BCFT to AdS means that we replace a factor $(2z)^{-\Delta}$ with $(2R)^{-\Delta}$. The cross-ratio $v$ is defined as
\begin{align}\label{vdef}
	v =\frac{z^2 x_{12}^2}{ (z^2 + ({\vec{x}}_1-{\vec{x}})^2)(z^2 + ({\vec{x}}_2-{\vec{x}})^2)}~.
\end{align}
The conformal blocks were computed in \cite{Lauria:2020emq,Nishioka:2022qmj} and in our conventions read:
\begin{align}
	f(\hD_{12}, \hD_k, v)=	(4v)^{\hD_k/2} \, _2F_1\left(\frac{\hD_k+\hD_{12}}{2},\frac{\hD_k-\hD_{12}}{2};1+\hD_k-\frac{d}{2};v\right)~.
\end{align}
This conformal block decomposition converges everywhere in AdS except on the geodesic that connects $\vec{x}_1$ and $\vec{x}_2$. Our approach will be to swap the conformal block decomposition with the integral and analyze the resulting expressions.

\subsubsection{Identical operators}
We will first focus on identical boundary operators with dimension $\hD_1$. This requires us to integrate:
\begin{align}\label{G1AdS2k}
	G_{k}({x}_{12}) \equiv \int d^d \vec{x} \int_{a}^\infty \frac{dz}{z^{d+1}}  f(0,\hD_k, v)~,
\end{align}
where we introduced a near-boundary cutoff $a$. To do this integral we power-expand $f(0,\hD_k, v)$ around $v=0$ and perform the integration term-wise using equations~\eqref{Ismalla} and \eqref{Ipsmalla} in appendix~\ref{app:integrals}. The divergences that we find as we send $a\to 0$ are now of two types.

First, for boundary operators with $\hD_k < d$ we find a power-law divergence that has exactly the form predicted by the near-boundary operator expansion discussed in the previous section. If such operators are present then we must add the corresponding counterterms to the `counterterm' expression in \eqref{correctedtwoptgeneral}. It is however easy to verify that these counterterms do not give any finite contributions and therefore we can (and will) ignore these divergences in what follows.

Secondly we find a universal \emph{logarithmic} divergence which is present for all values of $\D$, $\hD_1$ and $\hD_k \neq 0$. More precisely the expression we find is:
\begin{equation}
	G_k(\vec{x}_{12}) = X_k \log({x}_{12}^2/a^2) + Y_k~,
\end{equation}
with
\begin{align}
X_k &= - 2^{\hD_k-\D} \pi ^{d/2}	\frac{\hD_k  \Gamma \left(1-\frac{d}{2}+\hD_k\right)}{(d-\hD_k)\Gamma \left(1+\frac{\hD_k}{2}\right)^2 }~,\\
Y_k &= - 2^{\hD_k-\D}\sum_{m=0}^\infty \frac{\pi ^{d/2} \left(\frac{\hD_k}{2}\right)_{m}^2 \Gamma \left(m+\frac{\hD_k}{2}-\frac{d}{2}\right)}{m! \Gamma \left(m+\frac{\hD_k}{2}\right) \left(1-\frac{d}{2}+\hD_k\right)_{m}} \left(\psi\left(m+\frac{\hD_k}{2}\right)-\psi\left(m+\frac{\hD_k}{2}-\frac{d}{2}\right)\right) ~,\nonumber\\
\end{align}
with $\psi(n)$ the digamma function.
The logarithmic divergence was not present in our previous near-boundary analysis and must therefore be due to the presence of the boundary operators. It can be cancelled with a wave function renormalization factor. If we sum over all the blocks:
\begin{align}
\label{gammadelta}
\begin{split}
\delta \hD &\equiv\sum_{k\neq \id} B\du{\phi}{k} \hC_{k \psih \psih} X_k~,\\
\delta \hat C &\equiv\sum_{k\neq \id} B\du{\phi}{k} \hC_{k \psih \psih} Y_k~,
\end{split}
\end{align}
then the addition of the wave function renormalization factor
\begin{align}\label{WC}
	Z_{\psih} = 1 - \frac{1}{2} g\, \delta \hD \log \left(a^2/R^2\right) +O( g^2)~,
\end{align}
produces the finite one-loop result:
\begin{align}\label{G1finitemDS}
	({{x}}_{12})^{2\hD_1}\langle \psihR_1({\vec{x}}_1) \psihR_1({\vec{x}}_2)\rangle =1 -  g \left( \delta \hD \log({x}_{12}^2/R^2) + \delta \hat C \right) + O( g^2)~.
\end{align}
As expected, this expression is compatible with boundary conformal covariance, and we read off that $\delta \hD$ is exactly the one-loop anomalous dimension of the boundary operator.

One remaining question concerns the convergence of the sums in \eqref{gammadelta}. There is no a priori reason that they are convergent, since our integration domain includes a line where the sum over conformal blocks diverges. In the next section we will show how the sum is in fact divergent in certain examples. Fortunately the right answer can be recovered easily with a simple regularization procedure. 

Recently the authors of \cite{meineri2023renormalization} have also derived a non-perturbative expression for the scaling dimension of $\psih$. Their derivation uses special properties of the bulk stress tensor. It is thereby different from ours, but their `naive' expression agrees with the one presented here when expanded to first order in the coupling. (To see this one uses that the trace of the stress tensor is proportional to the deforming operator.) In contrast with the more simplistic regularizations that we use below, the paper \cite{meineri2023renormalization} also provides a first-principles derivation of an absolutely convergent modification of this sum rule. The essential ingredients of this derivation are the recently derived local blocks of \cite{Levine:2023ywq} for the decomposition of the three-point functions. 

\subsubsection{Example: anomalous dimension of \texorpdfstring{$\Disp$}{D} in \texorpdfstring{$\AdStwo$}{AdS2}}

As an example, we will take a CFT in $\AdStwo$ and compute the first-order anomalous dimension of the displacement operator $\Disp$ under a deformation by an operator $\phi$ which we can take to be a generic scalar bulk Virasoro primary. We recall that $\Disp$ is a universal (global) boundary primary operator with scaling dimension $\D_\Disp = 2$. Below we will make essential use of several results presented in appendices \ref{subapp:localconformalinvariance} and \ref{subapp:universalcorrelators}, which the reader should probably consult at this stage.

As discussed in appendix \ref{subsubapp:ddphi}, the boundary modes exchanged in the connected part of $\langle \Disp \Disp \phi\rangle$ have $\hD_k = k = 2\mathbb{N}^{>0}$ and
\begin{align}
B\du{\phi}{k} \hC_{k \Disp \Disp}=B_{\phi}  \frac{\sqrt{\pi } \D   e^{\frac{i \pi  k}{2}} (\D  (k-2) (k+1)+4) \Gamma (k)}{2^{2 k-1}\Gamma \left(k-\frac{1}{2}\right)}~.
\end{align}
According to the previous discussion, the anomalous dimension of $\Disp$ should read (taking into account the non-unit normalization of $\Disp$)
\begin{align}
\delta\D_\Disp =\frac{ 2\pi \D}{c} \frac{B_\phi}{2^\D}	\sum_{k=2,4,\dots} S_k~,
\end{align}
where
\begin{align}
	S_k =2^{-k} e^{\frac{i \pi  k}{2}} \frac{k (2 k-1) (\D  (k-2) (k+1)+4) \Gamma (k-1)}{\Gamma \left(\frac{k}{2}+1\right)^2}~.
\end{align}
Since $S_k \sim k^{1/3}$ as $k\rightarrow \infty$, this sum is divergent because of operators with high scaling dimensions. A simple way of regularizing it is to define 
\begin{align}
S(L)=\sum_{k=2,4,\dots} S_k e^{- k L}~.
\end{align}
After performing the sum we can take the $L\rightarrow 0$ limit of the expansion of $S(L)$ to find
\begin{align}
\delta\D_\Disp =\frac{2\sqrt{\pi} \D}{c}\frac{B_\phi}{2^\D}\lim_{L\rightarrow 0}  S(L)	= \frac{4\pi}{c} \frac{B_\phi}{2^\D}   (\D -2) \D~.
\end{align}
This is the correct result: one would obtain the same answer after integrating the correlator of eq.~\eqref{DDphiphi} over $\AdStwo$.

We can likewise compute the anomalous dimension of $\Disp^2$, i.e.~the first (global) boundary primary after $\Disp$ in the $\Disp \times \Disp$ OPE. The relevant correlation function is discussed in appendix \ref{subsubapp:dsquaredcorrs}, and in particular equation \eqref{D2D2phiphi_dec_bd_3pt} informs us that $\hD_k = k = 2\mathbb{N}^{>0}$ and that
\begin{align}
	B\du{\phi}{k} \hC_{k \Disp^2 \Disp^2} =B_{\phi} \left(c_k^{(1)} \D+c_k^{(2)}\D^2+c_k^{(3)}\D^3+c_k^{(4)}\D^4\right)~,
\end{align}
where the $c_k$'s are defined in eq.~\eqref{cndefs}. The anomalous dimension of $\Disp^2$ then reads (again we divided by the normalization of $\Disp^2$)
\begin{align}
\delta\D_{\Disp^2}= \frac{B_\phi}{2^\D}\frac{10 \sqrt{\pi }  }{c (5 c+22)}\sum_{k=2,4,...}S_k~,
\end{align}
where
\begin{align}
	S_k = \frac{2^{k} k \Gamma \left(k+\frac{1}{2}\right)}{(k-1) \Gamma \left(\frac{k}{2}+1\right)^2}\left(c_k^{(1)} \D+c_k^{(2)}\D^2+c_k^{(3)}\D^3+c_k^{(4)}\D^4\right)~.
\end{align}
The sum is again divergent due to boundary operators with high scaling dimensions. As before, we can regularize it by defining
\begin{align}
	S(L)=\sum_{k=2,4,\dots} S_k e^{- k L}~,
\end{align}
and $\delta\D_{\Disp^2}$ can be recovered by taking the $L\rightarrow 0$ of $S(L)$, to find
\begin{align}
\delta\D_{\Disp^2}=	\frac{B_\phi}{2^\D}  \frac{2\pi \D (\D -2) (20 c+25 (\D -2) \D +64)}{c (5 c+22)}~.
\end{align}
This is once more the correct result, as we have checked by integrating the correlator of eq.~\eqref{D2D2phiall} over $\AdStwo$.

\subsubsection{Boundary operators mixing at one-loop}
Let us also consider a boundary two-point function of different boundary primary operators. These must vanish if the bulk RG preserves the AdS isometries:
\begin{align}\label{conditionpsi12}
	\langle \psihR_1(\vec{x}_1) \psihR_2 (\vec{x}_2)\rangle = 0~.
\end{align}
We will now check that the one-loop correction indeed vanishes. The integral to perform is now:
\begin{align}\label{G1AdS2koffdiag}
	\mathcal{I}_{1,k} &=\int d^d {\vec{x}}\int_{a}^\infty \frac{ d z} {z^{d+1}}~\frac{f(\hD_{12},\hD_k, v)}{(z^2+(\vec{x}_1-\vec{x})^2)^{{\hD_{12}/2}}(z^2+(\vec{x}_2-\vec{x})^2)^{{\hD_{21}/2}}}~.
\end{align}
As before, we power-expand each block around $v=0$ and, performing the integration term by term using the contour integral representation for the hypergeometric functions, we arrive at ($u\equiv a/|\vec{x}_{12}|$)
\begin{align}
	\mathcal{I}_{1,k} &= \frac{\pi ^{d/2} 2^{\hD_{k}-\D }\Gamma\left(1-\frac{d}{2}+\hD_{k}\right)}{\Gamma \left(\frac{\hD_{k}-\hD_{12}}{2}\right) \Gamma \left(\frac{\hD_{12}+\hD_{k}}{2}\right) }|\vec{x}_{12}|^{-\hD_1-\hD_2} \int_{-i \infty}^{i \infty}\frac{ds}{2\pi i} \sum_{m=0}^\infty\,u^{2s-2m-{\hD_k}{}}  \nonumber\\
	\times& \frac{ \Gamma (s)\Gamma \left(2 m-s+\hD_{k}-\frac{d}{2}\right) \Gamma \left(m-s+\frac{\hD_{k}}{2}-{\hD_{12}/2}\right) \Gamma \left(m-s+\frac{\hD_{k}}{2}+{\hD_{12}/2}\right) }{m!\,\Gamma \left(m+\hD_{k}-\frac{d}{2}+1\right) \Gamma (2 m-2 s+\hD_{k}+1)}~.\nonumber\\
\end{align}
We want to investigate the small $u$ behavior of the expressions above. If we close the contour to the right there are three series of poles that contribute to the integral:
\begin{itemize}
	\item[I.] for  $-\frac{\hD_{12}}{2}+\frac{\hD_{k}}{2}+m-s= -i$, $i \in \mathbb{N}$, corresponding to $u^{2i -\hD_{12}}$ powers;
	\item[II.] for  $\hD_{k}+2 m-\frac{d}{2}-s=-i$, $i \in \mathbb{N}$, corresponding to $u^{2m+2i+\hD_{k}-d}$ powers;
	\item[III.] for $\frac{\hD_{12}}{2}+\frac{\hD_{k}}{2}+m-s= -i$, $i \in \mathbb{N}$, corresponding to $u^{2i +\hD_{12}}$ powers.
\end{itemize}
We recognize that the second series gives rise to the familiar infrared divergences that arise only for $\hD_k < d$. Their renormalization was described above and they do not give rise to finite terms. Depending on the sign of $\hD_{12}$ we however also find that either the type I or the type III poles have at least one divergent term.\footnote{When  $\hD_{12}=0$ both type I and type III poles contribute and generate the $\log(u^2)$ divergence associated to the anomalous dimension of the external operators.}
Let us consider for concreteness $0<\hD_{12}<2$. In this case we let
\begin{align}
	\begin{pmatrix}
		\psihR_1 \\
		\psihR_2
	\end{pmatrix}=\begin{pmatrix}
		Z_1 & Z_{12}\\
		Z_{21} & Z_{2}
	\end{pmatrix}	\begin{pmatrix}
		\psih_1 \\
		\psih_2
	\end{pmatrix}~,
\end{align}
with $Z_{1,2}$ the wave-function renormalizations (eq.~\eqref{WC}). In order to remove the leading $a^{-\hD_{12}}$ we need to set:
\begin{align}
	Z_{12} =  g\,a^{-\hD_{12}} \sum_{k}B_{\phi}{}^k {\hC}_{12 k} \delta Z_{12,k}+O( g^2)~,\quad Z_{21}= 0.
\end{align}
The quantity $ \delta Z_{12,k}$ is simply determined by the most divergent power in $\mathcal{I}_{1,k}$, that is the residue of the first pole in the I series. Once again, the removal of this power-law divergence does not produce any finite terms and we find that the renormalized two-point function of unequal operators also vanishes at one loop:
\begin{equation}
	\langle \psihR_1(\vec{x}_1) \psihR_2 (\vec{x}_2)\rangle = O( g^2)~,
\end{equation}
as expected from conformal covariance. Notice that the additional divergences that arise for highly unequal operators can be cancelled by introducing a larger mixing matrix that also includes descendants.

\section{Flows between minimal models}
\label{sec:minmoddeformations}
Our starting point is the $m$'th unitary and diagonal minimal model $\mathcal{M}_{m+1,m}$, with conformal boundary condition ${\bf a}= (a_1,a_2)_m$, on an $\AdStwo$ background. As per our general discussion, its correlation functions are simply the Weyl rescaled correlation functions of the BCFT setup. In the following we suppose the reader is familiar with the material in appendix \ref{app:minmodelsbcft} where we discuss these minimal models and their conformal boundary conditions.

Consider now deforming this theory by the operator $\phi\equiv \phi_{(1,3)}$. Perturbed correlation functions take the form:
\begin{align}\label{epsilonprimepertgenmin}
\vev{\ldots}_g = \vev{\ldots \exp\left( - g R^{\D-2}\int \ddxsqrtg{2}{x}{g} \phi(x)+S_\text{c.t.} \right)}_0~,
\end{align}
where the counterterm action is to be defined and the coupling $g$ is seen to be dimensionless if we recall the UV scaling dimension:
\begin{align}\label{phi13dim}
	\D = 2h_{1,3} = 2\frac{m-1}{m+1} = 2- 4/m+ O(m^{-2})~.
\end{align}
The one-loop beta function of $g$ reads:
\begin{align}\label{betafctlit}
	\dot g  = (2-\D)g - \pi\,C_{\phi\phi\phi}g^2+O(g^3)~.
\end{align}
Using that $C_{\phi\phi\phi}={4}/{\sqrt{3}}$, we find a fixed point at
\begin{align}\label{IRfixed}
	g_* = \frac{\sqrt{3}}{\pi m}+O(m^{-2})~.
\end{align}
This perturbative calculation is reliable at large $m$, so when $c$ is close to $1$. Notice that the presence of the AdS background does not complicate the above analysis. In particular, as we already discussed in subsection \ref{subsec:uvbehaviorgeneral}, at one loop there is no need to include mixing with the identity operator times the Ricci scalar (even though it has approximately the same dimension as $\phi_{(1,3)}$).

It is well-known that the fixed point we have found corresponds to the $(m-1)$'th minimal model, and this is simply the perturbatively controllable limit of the famous `staircase' RG flow between consecutive minimal models~\cite{Zamolodchikov:1987ti}. It is perhaps less well-known that the phase portrait for this flow on the upper half-plane is significantly richer, because the boundary can also undergo flows that are independent of the bulk. The perturbative picture was comprehensively analyzed in \cite{Fredenhagen:2009tn} and discussed in the introduction. We will now investigate how it gets modified in an AdS background.

\subsection{Infrared divergences}
The near-boundary analysis follows the logic of subsection \ref{subsec:infrareddivergencesgeneral}. To analyze the infrared divergences at one loop we are led to consider
\begin{equation}
	 - g R^{\Delta - 2} \int_a^b dy \int dx\,  \sqrt{g}\, \vev{ \phi(x) \ldots}^{\text{connected}}_{\text{AdS}_2}~,
\end{equation}
as the infrared cutoff $a$ goes to zero. Notice that we also introduced $b$, which we choose such that $z > b$ for all the other insertions. This allows us to focus on the infrared divergences alone. 

The boundary operator expansion of $\phi$ in the BCFT was given in equation \eqref{bOPEphi13} and, after subtracting the disconnected piece, contains one (marginally) relevant boundary operator. Adding the Weyl rescaling factors for AdS we find:
\begin{equation}
	 \phi(x,y) - \vev{\phi(x,y)} = B_{(1,3)}^{{\bf a}\,(1,3)} (2R)^{-\D}  (2y)^{h_{1,3}}\psi_{(1,3)}(x) + \ldots~,
\end{equation}
and therefore our target expression equals
\begin{equation}
	- g\,2^{-h_{1,3}} B_{(1,3)}^{{\bf a}\,(1,3)}    \left( \frac{b^{h_{1,3}-1}}{h_{1,3}-1}-\frac{a^{h_{1,3}-1}}{h_{1,3}-1} \right) \int dx\,  \vev{\psi_{(1,3)}(x) \ldots}_{\text{AdS}_2}~.
\end{equation}
Since $h_{1,3} < 1$, we find a power-law infrared divergence as $a \to 0$. In accordance with our general discussion, we introduce a counterterm of the form:
\begin{equation}
\label{mmctaction}
	- \frac{g R^{\D - 1}}{h_{1,3} - 1} \left. \int dx\, \sqrt{h}\, \phi(x) \right|_{y = a}~,
\end{equation}
which indeed, using $\sqrt{h} = R/y$, cancels the divergence when added to the bulk integral.

\subsubsection{Large \texorpdfstring{$m$}{m} limit}
So far our analysis of the infrared divergences was merely a repetition of our general discussion. However as $m \to \infty$ we find that $h_{1,3} \to 1$ and the nearly marginal boundary operator seemingly produces an extra divergence, namely a pole multiplying the residual finite piece proportional to $b^{h_{1,3} - 1}$. However for the minimal models this divergence is cancelled by the behavior of the bulk-boundary OPE coefficient: from equation \eqref{B13B15sqrt} we find that
\begin{equation}
	B_{(1,3)}^{{\bf a}\,(1,3)} \propto \frac{1}{m}~,
\end{equation}
as $m \to \infty$. So the whole expression actually remains finite\footnote{Notice that we are assuming that the correlator $\vev{\psi(x) \ldots}$ remains finite in the large $m$ limit. If this is not the case our analysis will need to be modified.} and there is no new divergence!

We would like to stress that we introduce here the following general principle: an integral that diverges as $m \to \infty$ does not need to be renormalized if it is multiplied by an OPE coefficient that cancels that divergence. We believe this is justified, and in any case we obtain manifestly finite answers in the end which, absent some hidden inconsistency, need to be given a physical interpretation. It would be interesting to understand this phenomenon better, and perhaps verify whether similar issues arise in the flat-space BQFT analysis.

\subsection{Anomalous dimensions of boundary operators}\label{sec:anomaldimbd}
We now turn to the one-loop computation of the two-point function of boundary operators.

\subsubsection{Spectrum of boundary Virasoro primaries}
Let us consider a generic conformal boundary condition ${\bf a} = (a_1,a_2)_m$ that supports a boundary Virasoro primary $\psi_{(r,s)}$  with scaling dimension
\begin{align}\label{hrsUVexp}
	\hD_{r,s}=h_{r,s} =\frac{\bigl( (m+1)r-ms \bigr)^2-1}{4m (m+1)}~.
\end{align}
The one-loop correction to the two-point correlation function of $\psi_{(r,s)}$ is then
\begin{align}\label{listfirstordIminimal13app}
R^{-\D} G_1(x_{12})&\equiv	\int_{-\infty}^{\infty}d{}x\int_{a}^\infty \frac{d{}y} {y^2}~\langle \psi_{(r,s)}(x_1)\psi_{(r,s)}(x_2)\phi(x+i y,x-iy)\rangle^c\nonumber\\
	&-\frac{1}{1-\D/2}\int_{-\infty}^{\infty}\frac{d{}x}{a}~\langle \psi_{(r,s)}(x_1)\psi_{(r,s)}(x_2)\phi(x+i a,x-ia)\rangle^c~.
\end{align}
The correlation function in the integrands above is\footnote{This can be obtained from eq.~\eqref{finalcorrectgen}, upon taking the connected contribution and going to $\AdStwo$ via a Weyl rescaling.}
\begin{align}\label{finalcorrectgenAdS2}
	\langle \psi_{(r,s)}(x_1)\psi_{(r,s)}(x_2)\phi(x+i y,x-iy)\rangle^c=\frac{\tilde{ \mathcal{G}}(\tilde\eta)-\alpha_{(1,1)}}{(x_{12}^2)^{\hD_{r,s}} (2R)^{\D}}~,
\end{align}
where  $\tilde{\mathcal{G}}(\tilde\eta)$ is a sum of three Virasoro blocks
\begin{align}\label{finalcorrectgenmain}
\tilde{ \mathcal{G}}(\tilde\eta)=\alpha_{(1,1)}V_{(1,1)}(\tilde\eta)+	\alpha_{(1,3)}V_{(1,3)}(\tilde\eta)+	\alpha_{(1,5)}V_{(1,5)}(\tilde\eta)~,
\end{align}
which, as shown in appendix \ref{app:bdbdbulk3pt}, read (here and below $q\equiv s-r \in \mathbb{Z}$)
\begin{align}\label{allblocksmain}
V_{(1,1)}(\tilde\eta)&=\, _3F_2\left(\frac{1-m}{m+1},\frac{1+m q-r}{m+1},\frac{1-m q+r}{m+1};\frac{2-m}{m+1},\frac{m+3}{2 m+2};\frac{\tieta}{4}\right)~,\nonumber\\
V_{(1,3)}(\tilde\eta)&=\tieta^{h_{1,3}/2} \, _3F_2\left(\frac{1-m}{2m+2},\frac{m+2 m q-2 r+1}{2m+2},\frac{m-2 m q+2 r+1}{2m+2};\frac{3-m}{2m+2},\frac{3 m+1}{2m+2};\frac{\tieta}{4}\right),\nonumber\\
V_{(1,5)}(\tilde\eta)&=\tieta^{h_{1,5}/2}\, _3F_2\left(\frac{m}{m+1},\frac{2 m+m q-r}{m+1},\frac{2 m-mq+r}{m+1};\frac{3 m}{m+1},\frac{5 m+1}{2m+2};\frac{\tieta}{4}\right)~,
\end{align}
with cross-ratio $\tieta$ defined as:
\begin{align}\label{etatildedefmain}
	\tieta=\frac{4 y^2 (x_{12})^2}{\left((x_1-x)^2+y^2\right) \left((x_2-x)^2+y^2\right)}~,\quad 0\leq \tieta\leq 4~. 
\end{align}
The coefficients in \eqref{finalcorrectgenAdS2} and \eqref{finalcorrectgenmain} are
\begin{align}
\alpha_{(1,1)}\equiv B_{(1,3)}^{{\bf a}\,(1,1)}~,\quad 	\alpha_{(1,3)}\equiv {\hC}_{(r,s)(r,s)(1,3)}^{{\bf a}}B_{(1,3)}^{{\bf a}\,(1,3)}~,\quad \alpha_{(1,5)}\equiv {\hC}_{(r,s)(r,s)(1,5)}^{{\bf a}}B_{(1,3)}^{{\bf a}\,(1,5)}~,
\end{align}
with $B_{(1,3)}^{{\bf a}\,(1,1)}$,  $B_{(1,3)}^{{\bf a}\,(1,3)}$, and $B_{(1,3)}^{{\bf a}\,(1,5)}$ the bulk-boundary coefficients discussed in detail in appendix~\ref{app:minmodelsbcft}.

\paragraph{Computation at finite $m$\\}
In order to compute the integrals in eq.~\eqref{listfirstordIminimal13app} it is convenient to discuss separately the various contributions: from $V_{(1,1)}$, $V_{(1,3)}$ and $V_{(1,5)}$. 
Let us start from the $V_{(1,1)}$ contribution, which from eq.~\eqref{listfirstordIminimal13app} reads
\begin{align}\label{finalcorrectgenAdS2id}
		\alpha_{(1,1)}\int_{-\infty}^{\infty}d{}x\int_{a}^\infty \frac{d{}y} {y^2}~\frac{V_{(1,1)}(\tilde\eta)-1}{(x_{12}^2)^{\hD_{r,s}} {2^\D}}-\frac{	\alpha_{(1,1)}}{1-\D/2} \int_{-\infty}^\infty \frac{d{}x} {a}~\frac{V_{(1,1)}(\tilde\eta)-1}{(x_{12}^2)^{\hD_{r,s}} {2^\D}}\bigg\rvert_{y=a}~.
\end{align}
To compute these integrals we write $V_{(1,1)}$ as a power-series expansion around $\tieta=0$ and perform the integration term by term using the integration rules of appendix~\ref{app:integrals}.
Upon performing the infinite sums and neglecting terms that vanish as we send $a\rightarrow 0$ we find
\begin{align}\label{resV1}
\mathcal{I}_{(1,1)}\equiv& 	\alpha_{(1,1)}\int_{-\infty}^{\infty}d{}x\int_{a}^\infty \frac{d{}y} {y^2}~\frac{V_{(1,1)}(\tilde\eta)-1}{(x_{12}^2)^{\hD_{r,s}} {2^\D}}= \deltahD_{r,s}^{(1,1)}(m) \frac{\log \left({x_{12}^2}/{a^2}\right)}{(x_{12}^2)^{\hD_{r,s}}} +\frac{f^{(1,1)}_{r,s}(m)}{(x_{12}^2)^{\hD_{r,s}}}~,\nonumber\\
\mathcal{I}^{\text{c.t.}}_{(1,1)}\equiv& -\frac{\alpha_{(1,1)}}{1-\D/2} \int_{-\infty}^\infty \frac{d{}x} {a}~\frac{V_{(1,1)}(\tilde\eta)-1}{(x_{12}^2)^{\hD_{r,s}} {2^\D}}\bigg\rvert_{y=a}= -\frac{2\deltahD_{r,s}^{(1,1)}(m)}{1-\D/2}\frac{1}{(x_{12}^2)^{\hD_{r,s}}}~.
\end{align}
The function $f^{(1,1)}_{r,s}(m)$ does not seem to admit a simple closed-form expression, but it is finite in the $m\rightarrow \infty$ limit.\footnote{One can verify this by computing the integral directly at $m=\infty$, or by noticing that the $O(a^0)$ piece of the integrand reduces to a finite sum of $q=r-s$ terms  at $m=\infty$.} The coefficient of the logarithm in the first of eq.~\eqref{resV1} reads  
\begin{align}\label{hrs11}
	\deltahD_{r,s}^{(1,1)}(m)&= 	-\alpha_{(1,1)}\frac{\pi  (m-1) (m r-m s+r-1) (m r-m s+r+1)}{2^{1-\frac{4}{m+1}}(m-2) (m+1) (m+3)}\nonumber\\
	&\times\sum_{k=0}^\infty \frac{\left(\frac{1}{2}\right)_k \left(\frac{2}{m+1}\right)_k \left(\frac{2+r+r m-s m+m}{m+1}\right)_k \left(\frac{2-r-r m+s m+m}{m+1}\right)_k}{k! (2)_k \left(\frac{3}{m+1}\right)_k \left(\frac{3}{2}+\frac{1}{m+1}\right)_k}~.
\end{align}

For the $V_{(1,3)}$  contribution we have
\begin{align}\label{finalcorrectgenAdS213}
	\alpha_{(1,3)}\int_{-\infty}^{\infty}d{}x\int_{a}^\infty \frac{d{}y} {y^2}~\frac{V_{(1,3)}(\tilde\eta)}{(x_{12}^2)^{\hD_{r,s}} {2^\D}}-\frac{\alpha_{(1,3)}}{1-\D/2}\int_{-\infty}^\infty \frac{d{}x} {a}~\frac{V_{(1,3)}(\tilde\eta)}{(x_{12}^2)^{\hD_{r,s}} {2^\D}}\bigg\rvert_{y=a}~.
\end{align}
It is not difficult to show that, again up to terms that vanish as we send $a\rightarrow 0$,
\begin{align}\label{resV13}
\mathcal{I}_{(1,3)}&\equiv	\alpha_{(1,3)}\int_{-\infty}^{\infty}d{}x\int_{a}^\infty \frac{d{}y} {y^2}~\frac{V_{(1,3)}(\tilde\eta)}{(x_{12}^2)^{\hD_{r,s}} {2^\D}}\nonumber\\
	&=\frac{{f}^{(1,3)}_{\infty}(m)}{(x_{12}^2)^{\hD_{r,s}}}\left(\frac{|x_{12}|}{a}\right)^{\frac{2}{1+m}}+ \deltahD_{r,s}^{(1,3)}(m) \frac{\log \left({x_{12}^2}/{a^2}\right)}{(x_{12}^2)^{\hD_{r,s}}} +\frac{f^{(1,3)}_{r,s}(m)}{(x_{12}^2)^{\hD_{r,s}}}~,\nonumber\\
\mathcal{I}^{\text{c.t.}}_{(1,3)}&\equiv	-\frac{\alpha_{(1,3)}}{1-\D/2}\int_{-\infty}^\infty \frac{d{}x} {a}~\frac{V_{(1,3)}(\tilde\eta)}{(x_{12}^2)^{\hD_{r,s}} {2^\D}}\bigg\rvert_{y=a}\nonumber\\
	&=-\frac{{f}^{(1,3)}_{\infty}(m)}{(x_{12}^2)^{\hD_{r,s}}}\left(\frac{|x_{12}|}{a}\right)^{\frac{2}{1+m}}-\frac{2\deltahD_{r,s}^{(1,3)}(m)}{1-\D/2}\frac{1}{(x_{12}^2)^{\hD_{r,s}}}~.
\end{align}
As is clear from the expression above, the power-law divergent terms cancel out from the sum $\mathcal{I}_{(1,3)}+\mathcal{I}_{(1,3)}^{\text{c.t.}}$.\footnote{The coefficient that multiplies the power-law divergence in eq.~\eqref{resV13} is
	\begin{align}
		{f}^{(1,3)}_{\infty}(m)=-\frac{\alpha_{(1,3)}}{\pi } \sin \left(\frac{\pi }{m+1}\right) \cos ^3\left(\frac{\pi }{m+1}\right) \Gamma \left(-\frac{4}{m+1}\right) \Gamma \left(\frac{2}{m+1}\right)^2 2^{\frac{3 m+5}{m+1}}~.
\end{align}}
The function $f^{(1,3)}_{r,s}(m)$, for which we do not have a simple closed-form expression, behaves as
\begin{align}\label{f13divpart}
	f^{(1,3)}_{r,s}(m)&= -\frac{1}{2}\alpha_{(1,3)} m^2-{\alpha_{(1,3)}}{} \left(1+\log 2\right)m+O(m^0)~.
\end{align}
The coefficient of the logarithm in the first of eq.~\eqref{resV13} reads
\begin{align}\label{hrs13}
	\deltahD_{r,s}^{(1,3)}(m) &=\alpha_{(1,3)}\frac{\sqrt{\pi}\,\Gamma \left(-\frac{1}{m+1}\right)}{2^{1-\frac{2}{m+1}} \Gamma \left(\frac{1}{2}-\frac{1}{m+1}\right)}~\nonumber\\
	&\quad\times \sum_{k=0}^\infty\frac{\left(-\frac{1}{m+1}\right)_k \left(\frac{1}{m+1}-\frac{1}{2}\right)_k \left(\frac{1}{2}+r-\frac{m s}{m+1}\right)_k \left(\frac{1}{2}-r+\frac{m s}{m+1}\right)_k}{k! \left(\frac{1}{2}-\frac{1}{m+1}\right)_k \left(\frac{3}{2}-\frac{1}{m+1}\right)_k \left(\frac{2}{m+1}-\frac{1}{2}\right)_k}~.
\end{align}
The r.h.s. of eq.~\eqref{hrs13} features a `potential' linear divergence at large $m$ since $\Gamma\left(-\frac{1}{m+1}\right)\sim \Gamma(0)$ in this limit. On general grounds, this divergence would be removed by a boundary counterterm of the type $\sqrt{g} \int \psi_{(1,3)}$ with appropriately tuned coefficient, which might induce an instability of the boundary condition, see e.g.~\cite{Hogervorst:2021spa}. In our case this divergence is however compensated by the large-$m$ falloff of $\alpha_{(1,3)}$ and, as we discussed previously, no counterterm is needed.

Finally, for the $V_{(1,5)}$  contribution we have
\begin{align}\label{finalcorrectgenAdS215}
	\alpha_{(1,5)}\int_{-\infty}^{\infty}d{}x\int_{a}^\infty \frac{d{}y} {y^2}~\frac{V_{(1,5)}(\tilde\eta)}{(x_{12}^2)^{\hD_{r,s}} {2^\D}}-\frac{\alpha_{(1,5)}}{1-\D/2}\int_{-\infty}^\infty \frac{d{}x} {a}~\frac{V_{(1,5)}(\tilde\eta)}{(x_{12}^2)^{\hD_{r,s}} {2^\D}}\bigg\rvert_{y=a}~,
\end{align}
for which we find (up to regular terms as we send $a\rightarrow 0$ while keeping $m$ finite)
\begin{align}\label{resV15}
\mathcal{I}_{(1,5)}&\equiv\alpha_{(1,5)}\int_{-\infty}^{\infty}d{}x\int_{a}^\infty \frac{d{}y} {y^2}~\frac{V_{(1,5)}(\tilde\eta)}{(x_{12}^2)^{\hD_{r,s}} {2^\D}}~= \deltahD_{r,s}^{(1,5)}(m)\frac{\log \left({x_{12}^2}/{a^2}\right)}{(x_{12}^2)^{\hD_{r,s}}} +\frac{f^{(1,5)}_{r,s}(m)}{(x_{12}^2)^{\hD_{r,s}}}~,\nonumber\\
\mathcal{I}^{\text{c.t.}}_{(1,5)}&\equiv-\frac{\alpha_{(1,5)}}{1-\D/2}\int_{-\infty}^\infty \frac{d{}x} {a}~\frac{V_{(1,5)}(\tilde\eta)}{(x_{12}^2)^{\hD_{r,s}} {2^\D}}\bigg\rvert_{y=a}=-\frac{2\deltahD_{r,s}^{(1,5)}(m)}{1-\D/2}\frac{1}{(x_{12}^2)^{\hD_{r,s}}}~.
\end{align}
The function $f^{(1,5)}_{r,s}(m)$ is constant in the $m\rightarrow \infty$ limit, and the coefficient of the logarithm in the first of eq.~\eqref{resV15} reads  
\begin{align}\label{hrs15}
	\deltahD_{r,s}^{(1,5)}(m) &=\alpha_{(1,5)} \frac{\sqrt{\pi }\, \Gamma \left(\frac{3}{2}-\frac{3}{m+1}\right)}{2^{-\frac{2m}{m+1}}\Gamma \left(2-\frac{3}{m+1}\right)} \sum_{k=0}^\infty \frac{\left(\frac{m}{m+1}\right)_k \left(\frac{3}{2}-\frac{3}{m+1}\right)_k \left(\frac{2m+r+m r-m s}{m+1}\right)_k \left(\frac{2m-r+m s-m r}{m+1}\right)_k}{k! \left(\frac{3 m}{m+1}\right)_k \left(2-\frac{3}{m+1}\right)_k \left(\frac{5}{2}-\frac{2}{m+1}\right)_k}~.
\end{align}

Putting everything together, the one-loop, unrenormalized two-point correlation function of $\psi_{(r,s)}$ at finite $m$ is
\begin{align}\label{G1pert}
	G_1(x_{12})=\mathcal{I}_{(1,1)}+\mathcal{I}^{\text{c.t.}}_{(1,1)}+\mathcal{I}_{(1,3)}+\mathcal{I}^{\text{c.t.}}_{(1,3)}+\mathcal{I}_{(1,5)}+\mathcal{I}^{\text{c.t.}}_{(1,5)}~.
\end{align}

\paragraph{Renormalization and anomalous dimensions at finite $m$\\}

Having obtained $G_1(x_{12})$, we can now renormalize it to extract the anomalous dimension of $\psi_{(r,s)}$ at $O(g)$. At finite $m$, we can remove the $\log \left({x_{12}^2}/{a^2}\right)$ divergence in eq.~\eqref{G1pert} via a wave-function renormalization; i.e. we define $\psiR_{(r,s)}\equiv Z_{r,s} \psi_{(r,s)}$ with
\begin{align}\label{wfZfinitem}
Z_{r,s} =1 - \frac{1}{2}g\,\delta Z_{r,s} \log \left({a^2/R^2}\right) +O(g^2)~,
\end{align}
and choose
\begin{align}
\delta Z_{r,s} =\deltahD_{r,s}^{(1,1)}(m)+\deltahD_{r,s}^{(1,3)}(m)+\deltahD_{r,s}^{(1,5)}(m)~.
\end{align}
The renormalized two-point correlation function at one-loop reads
\begin{align}\label{G1finitem}
	(x_{12}^2)^{\hD_{r,s}}\langle \psiR_{(r,s)}(x_1) \psiR_{(r,s)}(x_2)\rangle =1 -g\,\delta Z_{r,s} \log \left({a^2/R^2}\right)- g \,(x_{12}^2)^{\hD_{r,s}}G_{1}(x_{12}) +O(g^2)~,
\end{align}
and the one-loop anomalous dimension of $\psi_{r,s}$ is
\begin{align}\label{hrsfinitem}
\deltaD_{r,s} = \delta Z_{r,s}~.
\end{align}

\paragraph{The large-$m$ limit\\}

The result of eq.~\eqref{G1finitem} is valid as long as $m$ is kept finite. For $m$ large, there are additional divergences that should be removed from the physical correlator before we can meaningfully take the large $m$ limit. As we discuss in appendix~\ref{app:minmodelsbcft}, at large $m$ and  for $a_1,a_2\ll m$ we have that\footnote{On general grounds, $\alpha_{(1,3)}=O(m^{-1})$ follows from requiring unphysical singularities in the correlator to cancel, see e.g. appendix~\ref{app:removingunphdiv}. That $\alpha_{(1,5)}=O(m^{-2})$ follows from $B_{(1,3)}^{{\bf a}\,(1,5)}=O(m^{-2})$ -- see e.q.~\eqref{B13B15sqrt} -- and upon assuming that the boundary three-point functions ${\hC}_{(r,s)(r,s)}^{{\bf a}(1,5)}$ remain finite in the large-$m$ limit.}
\begin{align}
	\alpha_{(1,1)}=\sqrt{3}+O(m^{-1})~,\quad \alpha_{(1,3)}={\alpha_{(1,3)}^{(1)}}{m^{-1}}+O(m^{-2}), \quad \alpha_{(1,5)}=O(m^{-2})~,
\end{align}
where
\begin{align}
\alpha_{(1,3)}^{(1)}=\frac{2 \pi  (r-s) (s-\text{sgn}(r-s))}{\sqrt{3}}~.
\end{align}
From the explicit expressions given earlier in this section, it is not difficult to verify that
\begin{align}
	\deltahD_{r,s}^{(1,1)}(\infty)&= -\frac{\pi\sqrt{3}}{2}(r-s)^2 \left(\frac{1}{3}+\frac{2}{3}|s-r|^{-1}\right)+O(m^{-1})~,\nonumber\\
	\deltahD_{r,s}^{(1,3)}(\infty)&= -\frac{\alpha_{(1,3)}^{(1)}}{2}+O(m^{-1})~,\nonumber\\
	\deltahD_{r,s}^{(1,5)}(\infty)&=O(m^{-2})~.
\end{align}
With these results in hand we can easily verify that at large $m$ the correlator of eq.~\eqref{G1finitem} features a divergence linear in $m$. It originates from the $V_{(1,1)}$ contribution to the counterterm and reads
\begin{align}\label{largemectra}
(x_{12}^2)^{\deltahD_{r,s}}	(\mathcal{I}_{(1,1)}+\mathcal{I}^{\text{c.t.}}_{(1,1)})&=-m\,\delta h_{r,s}^{(1,1)}(\infty)+O(m^0)~.
\end{align}
We might expect further large-$m$ divergent contributions coming from $\mathcal{I}_{(1,3)}+\mathcal{I}^{\text{c.t.}}_{(1,3)}$; however, thanks to a special cancellation we find
\begin{align}
 f_{r,s}^{(1,3)}(m)-\frac{2 \deltahD_{r,s}^{(1,3)}(m)}{1-\D/2}=O(m^0)~.
\end{align}
We can remove the divergence of eq.~\eqref{largemectra} via an extra (multiplicative) wave-function renormalization i.e. we define 
$Z_{r,s} Z_{r,s}^{\text{extra}}\psi_{(r,s)}$ with $Z_{r,s}$ given by eq.~\eqref{wfZfinitem} and with
\begin{align}
Z_{r,s}^{\text{extra}}=1 + \frac{1}{2}g\,m\, \delta Z^{\text{extra}}_{r,s} +O(g^2)~,\quad \delta Z^{\text{extra}}_{r,s} = -\deltahD_{r,s}^{(1,1)}(\infty)~,
\end{align}
so that the renormalized one-loop two-point correlation function
\begin{align}
(x_{12}^2)^{\hD_{r,s}}\langle \psiR_{(r,s)}(x_1) \psiR_{(r,s)}(x_2)\rangle =1& +g m\,\delta Z^{\text{extra}}_{r,s}-g\,\delta Z_{r,s} \log \left({a^2/R^2}\right)\nonumber\\
&- g(x_{12}^2)^{\hD_{r,s}}G_{1}(x_{12}) +O(g^2)~,
\end{align}
is completely finite.
From the result above we can read off the boundary spectrum at the IR fixed point:
\begin{align}\label{bdspectrum}
	\hD^{\text{IR}}_{r,s}=\hD_{r,s}+g\deltahD_{r,s}^{(1,1)}(\infty) +g\deltahD_{r,s}^{(1,3)}(\infty)+O(m^{-2})=\frac{1}{4} (r-s)^2-\frac{r^2-s^2}{4 m}+O(m^{-2})~,
\end{align}
where we used that
\begin{align}\label{hrsUVexplargem}
	\hD_{r,s}= \frac{1}{4} (r-s)^2+ \frac{r^2-s^2}{4 m}+O(m^{-2})~.
\end{align} 

\paragraph{Interpretation\\}
The result of eq.~\eqref{bdspectrum} can be interpreted as follows. Assuming that the endpoint of the RG flow is a conformal boundary condition (or a superposition thereof) for the $\mathcal{M}_{m,m-1}$ minimal model, and assuming that $\psi_{(r,s)}$ flows to another Virasoro primary $\psi_{(r',s')}$ in such an IR conformal boundary condition, then there should exist two positive integers $(r',s')$ such that
\begin{align}\label{eq:condition}
	\hD_{r,s}-\hD_{r',s'}^{\text{IR}}=O(m^{-2})~.
\end{align}
By comparing equations~\eqref{bdspectrum} and~\eqref{hrsUVexplargem}, we find that if $r\neq s$ one possible solution to the condition above is that $(r',s')=(s,r)$. This would mean that
\begin{align}
	\psi_{(r,s)}\longrightarrow \psi_{(s,r)}~,
\end{align}
under the RG flow.

This spectral flow is very natural: it would be consistent with the following large-$m$ flow between conformal b.c.
\begin{align}\label{RGflowbc}
	(a_1,a_2)_{m}\longrightarrow (a_2, a_1)_{m-1}~,\quad a_1,a_2\ll m~,
\end{align}
since the selection rules imply that, if $(r,s)$ appears in $(a_1,a_2)_m$ then $(s,r)$ appears in $(a_2,a_1)_{m-1}$. However we should keep in mind that other possibilities (including flows to linear combinations of conformal b.c.) are still logically possible. In particular, it need not even be true that the infrared operator is still a Virasoro primary.

\paragraph{Mixing\\}
The reader may wonder why we do not consider operator mixing, since by equation \eqref{hrsUVexplargem} there are often many operators whose dimensions become equal at large $m$. In contrast with flat space there is however no need to consider this mixing. This is because our one-loop result for the anomalous dimension is reliable and physical (i.e. scheme-independent) for any finite $m$: after all, the boundary is always conformal so $\hD$ is always a good observable. At finite $m$ all the scaling dimensions are different, so the matrix of two-point functions remains diagonal (cf. our discussion in subsection \ref{anomalousdimAdS}) and there is no operator mixing. We have seen there are extra divergences at large $m$ (those cancelled by the $\delta Z^{\text{extra}}_{r,s}$ above) but obviously these can only appear if the correlation function is non-zero in the first place. Therefore, in contrast with a flat-space analysis where conformal invariance is completely lost along the flow, nearby operators on the boundary of AdS do not mix. 

\subsection{Bulk one-point function of \texorpdfstring{$\phi_{(1,3)}$}{phi13}}\label{bulk1ptmm}
In this section we compute the one-point correlation function of $\phi = \phi_{(1,3)}$ on the AdS disk
\begin{align}
	ds^2  = \left(\frac{2R}{1-r^2} \right)^2 (dr^2 + r^2 d\theta^2)~,\quad 0\leq r\leq 1~,-\pi\leq \theta\leq \pi~,
\end{align}
at one loop. We again consider an elementary conformal boundary condition $(a_1,a_2)_m$, with $a_1, a_2 \ll m$. At tree-level, putting for simplicity $\phi$ at the center of the disk, we have (see eq.~\eqref{B1315})
\begin{align}
(2R)^\D	\langle \phi(0,0) \rangle_{0,{\bf a}} = {B_{(1,3)}^{{\bf a}\,(1,1)}}&=\left(\sqrt{3}-\frac{2 \pi ^2 \left(2 a_2^2-1\right)}{\sqrt{3} m^2}+\frac{4 \pi ^2 \left(2a_2^2-1\right)}{\sqrt{3} m^3}+O(m^{-4})\right)~,
\end{align}
and the one-loop correction is given by
\begin{align}\label{phicorrpertmain}
    \delta \langle \phi(0,0) \rangle_{{\bf a}} &=  -g \mu^{2-\Delta} \int \ddxsqrtg{2}{x}{g} \langle \phi(0,0) \phi(r,\theta)\rangle^c_{\bf a} \nonumber\\
	&\qquad + \text{IR counterterms} + \text{UV counterterms}~,
\end{align}
where the superscript `c' means that we will consider the connected correlator only. Notice that the coupling $g$ is dimensionless and we inserted a scale $\mu$.

\paragraph{Computing the integrand\\}
As shown in appendix~\ref{sec:phi13twoptsol}, the (connected) correlator in the integrand of eq.~\eqref{phicorrpertmain} reads
\begin{align}\label{finalcorrectgenAdS2twopt}
	(2R)^{2\D}\langle \phi(0,0)\phi(r,\theta)\rangle_{\bf a}^c={\tilde{ \mathcal{G}}(\tilde\eta)-(B_{(1,3)}^{{\bf a}\,(1,1)})^2}~,
\end{align}
with
\begin{align}\label{finalcorrectgenmaintwo}
	\tilde{ \mathcal{G}}(\tilde\eta)=(B_{(1,3)}^{{\bf a}\,(1,1)})^2V_{(1,1)}(\tilde\eta)+(B_{(1,3)}^{{\bf a}\,(1,3)})^2V_{(1,3)}(\tilde\eta)+	(B_{(1,3)}^{{\bf a}\,(1,5)})^2V_{(1,5)}(\tilde\eta)~,
\end{align}
where
\begin{align}\label{allblockstwopt}
V_{(1,1)}(\tilde\eta)&=\, _3F_2\left(\frac{1-m}{m+1},\frac{2m}{m+1},-\frac{2m-2}{m+1};\frac{3+m}{2m+2},\frac{2-m}{m+1};-\frac{\tieta}{4}\right)~,\nonumber\\
V_{(1,3)}(\tilde\eta)&=\tieta^{{h_{1,3}}/{2}} \, _3F_2\left(\frac{5 m-1}{2m+2},\frac{1-m}{2m+2},-\frac{3m-3}{2m+2};\frac{3-m}{2m+2},\frac{3 m+1}{2m+2};-\frac{\tieta}{4}\right),\nonumber\\
V_{(1,5)}(\tilde\eta)&=\tieta^{{h_{1,5}}/{2}}\, _3F_2\left(\frac{1}{m+1},\frac{m}{m+1},\frac{4 m-1}{m+1};\frac{3 m}{m+1},\frac{5 m+1}{2m+2};-\frac{\tieta}{4}\right)~.
\end{align}
The cross-ratio $\tieta$ reads
\begin{align}\label{tietadefmain}
	\tieta= \frac{\left( 1-r^2\right)^2}{r^2}~.
\end{align}
We discuss separately the contributions from $V_{(1,1)}$, $V_{(1,3)}$, and $V_{(1,5)}$. 

\paragraph{The $V_{(1,1)}$ contribution\\}
The relevant integrals in the first line of eq.~\eqref{phicorrpertmain} are
\begin{align}\label{finalcorrectgenAdS2idtwo}
I_{(1,1)}\equiv \int_{\epsilon<r<\sqrt{1-a}} \sqrt{g}~(V_{(1,1)}(\tilde\eta)-1)+c_{\text{c.t.}} \int\sqrt{\indmetr}~(V_{(1,1)}(\tilde\eta)-1)\big\rvert_{r=\sqrt{1-a}}~,
\end{align}
for a properly chosen $c_{\text{c.t.}}$, as explained below.
For convenience we have set an IR cut-off at $r=\sqrt{1-a}$ and a UV cut-off at $\epsilon>0$. At finite $m$ the counterterm contribution is completely regular in the limit of $a\rightarrow 0$:
\begin{align}
	\int\sqrt{\indmetr}~(V_{(1,1)}(\tilde\eta)-1)\big\rvert_{r=\sqrt{1-a}}&=-\frac{4\pi R\Delta ^2 (\Delta +2)}{(\Delta -4) (3 \Delta -2)}a+O(a^2)~.
\end{align}
This is expected since there are no relevant boundary operators contributing to $I_{(1,1)}$.

\begin{figure}
	\centering
	\includegraphics[width=0.5\textwidth]{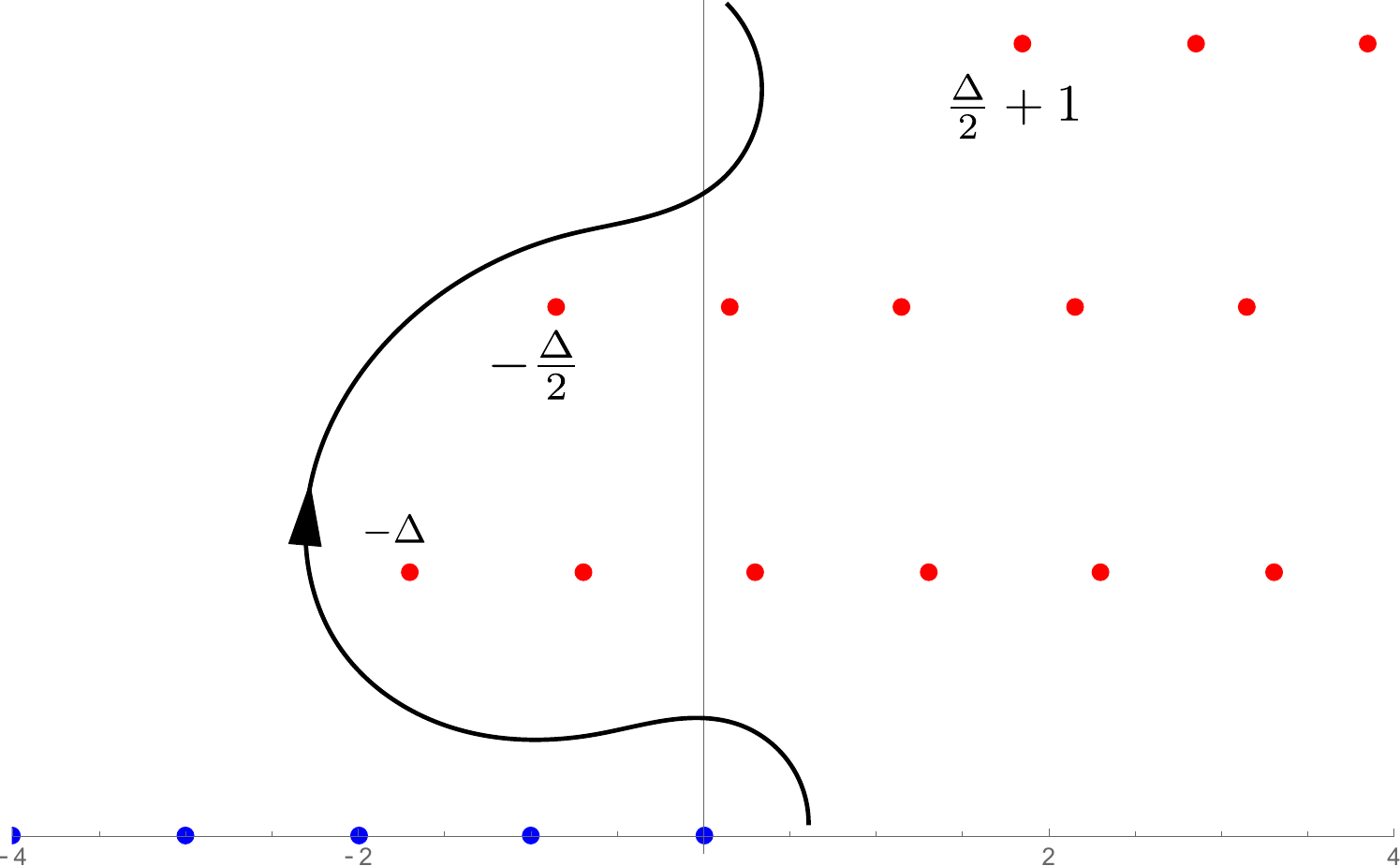}
	\caption{}
	\label{V11poles}
\end{figure}

The bulk contribution is more subtle due to UV divergences. Upon invoking the Mellin-Barnes representation of ${}_3 F_2$:
\begin{align}
\, _3F_2(a_1,a_2,a_3;b_1,b_2;x)= \frac{\Gamma (b_1) \Gamma (b_2)}{\Gamma (a_1) \Gamma (a_2) \Gamma (a_3)}\int_{C} \frac{ds}{2\pi i}  \frac{ (-x)^{-s} \Gamma (s)\Gamma (a_1-s) \Gamma (a_2-s) \Gamma (a_3-s)}{\Gamma (b_1-s) \Gamma (b_2-s)}~,
\end{align}
where the contour $C$ separates the poles $\Gamma(\dots+s)$ from those of $\Gamma(\dots-s)$, we find
\begin{align}\label{V11integral}
\int_{\epsilon<r<\sqrt{1-a}}\sqrt{g}~{V_{(1,1)}(\tilde\eta)}= \int_{C} \frac{ds}{2\pi i}\,K_{(1,1)}(\D,s)\left[B_{1-a}(s+1,-2 s-1)-B_{\epsilon ^2}(s+1,-2 s-1)\right]~,
\end{align}
where $B_z(a,b)$ is the incomplete beta function and
\begin{align}
	R^{-2}K_{(1,1)}(\D,s)\equiv \frac{  4^s \pi  \Delta\Gamma \left(\frac{1}{2}-\frac{3 \Delta }{4}\right) \Gamma \left(-\frac{\Delta }{4}\right) \Gamma (s) \Gamma (-s-\Delta ) \Gamma \left(-s-\frac{\Delta }{2}\right) \Gamma \left(1-s+\frac{\Delta }{2}\right)}{\Gamma \left(1-\frac{\Delta }{2}\right) \Gamma (-\Delta ) \Gamma \left(\frac{\Delta }{2}\right) \Gamma \left(\frac{1}{2}-s-\frac{3 \Delta }{4}\right) \Gamma \left(1-s-\frac{\Delta }{4}\right)}~.
\end{align}
The contour $C$ is the black line shown in fig.~\ref{V11poles} where the dots represent the poles of the expression in eq.~\eqref{V11integral} (slightly displaced along the imaginary axis for convenience of representation). As is clear from its definition in terms of $_2F_1$, the $B_z(a,b)$ has poles at $-a=n \in \mathbb{N}$. 
 For the $B_{1-a}(s+1,-2 s-1)$ bit in \eqref{V11integral}, since 
\begin{align}\label{betarule}
&	B_{1-a}(s+1,-2 s-1)=\nonumber\\
	&\frac{1}{s+1} \left(\frac{a^{-2 s-1} (1-a)^{s+1}\Gamma (s+2) \Gamma (2 s+1) \, _2F_1(1,-s;-2 s;a)}{\Gamma (s+1) \Gamma (2 s+2)}+\frac{\Gamma (s+2) \Gamma (-2 s-1)}{\Gamma (-s)}\right)~,
\end{align}
we shall close the contour to the left and pick up the residues for the poles at $-s=n \in \mathbb{N}$, corresponding to the blue dots of fig.~\ref{V11poles}.
The result is
\begin{align}\label{11bulk1}
\int_{C} \frac{ds}{2\pi i}\,K_{(1,1)}(\D,s)B_{1-a}(s+1,-2 s-1)=\frac{4\pi R^2}{a}+R^2f_{\text{fin}}^{(1,1)}(m)+O(a)~.
\end{align}
The finite term $f_{\text{fin}}^{(1,1)}(m)$ does not have a nice closed form, but in the large-$m$ expansion it reads
\begin{align}
f_{\text{fin}}^{(1,1)}(m)=\frac{8}{3} \pi  m -8 \pi -\frac{8 \pi  \left(15+2 \pi ^2\right) }{9 m}+O(m^{-2})~.
\end{align}
Note that the IR divergent term in the expression above drops out of the connected correlator, since
\begin{align}\label{11bulk2}
\int_{\epsilon<r<\sqrt{1-a}}\sqrt{g}=\frac{4 \pi  R^2 \left(a+\epsilon ^2-1\right)}{a \left(\epsilon ^2-1\right)} = \frac{4 \pi  R^2}{a}-4 \pi  R^2 + O(a,\epsilon)~.
\end{align}
 For the $B_{\epsilon ^2}(s+1,-2 s-1)$ bit in \eqref{V11integral}, since 
 \begin{align}
 	B_{\epsilon ^2}(s+1,-2 s-1)=\frac{\epsilon ^{2 s+2}}{1+s}\left(1+\frac{2 (s+1)^2\epsilon^2}{s+2}+O(\epsilon^4)\right)~,
 \end{align}
 we shall close the contour to the right, and picking up residues at the red poles in fig.~\ref{V11poles} we find
 \begin{align}\label{11UV}
 	\int_{C} \frac{ds}{2\pi i}\,K_{(1,1)}(\D,s)(-B_{\epsilon ^2}(s+1,-2 s-1))=\underset{\equiv R^2 f^{(1,1)}_\epsilon (m)}{\underbrace{\frac{4 \pi  R^2 \epsilon ^{2-2 \Delta }}{(\Delta -1) \left(1-2 \cos \left(\frac{\pi  \Delta }{2}\right)\right)}}}+O(\epsilon^\#)~,
 \end{align}
where $O(\epsilon^\#)$ are positive powers of $\epsilon$, which will go to zero as we send $\epsilon\rightarrow 0$ while keeping $\D$ finite. Putting it all together, and taking the $a,\epsilon \rightarrow 0$ limit we get
\begin{align}\label{resV11twopt}	
	I_{(1,1)}&= 4\pi R^2+R^2f_{\text{fin}}^{(1,1)}(m)+R^2 f^{(1,1)}_\epsilon (m)~.
\end{align}
In particular the final result features a constant piece (in $m$) that comes from taking the difference between \eqref{11bulk1} and \eqref{11bulk2}.

\begin{figure}
	\centering
	\includegraphics[width=0.5\textwidth]{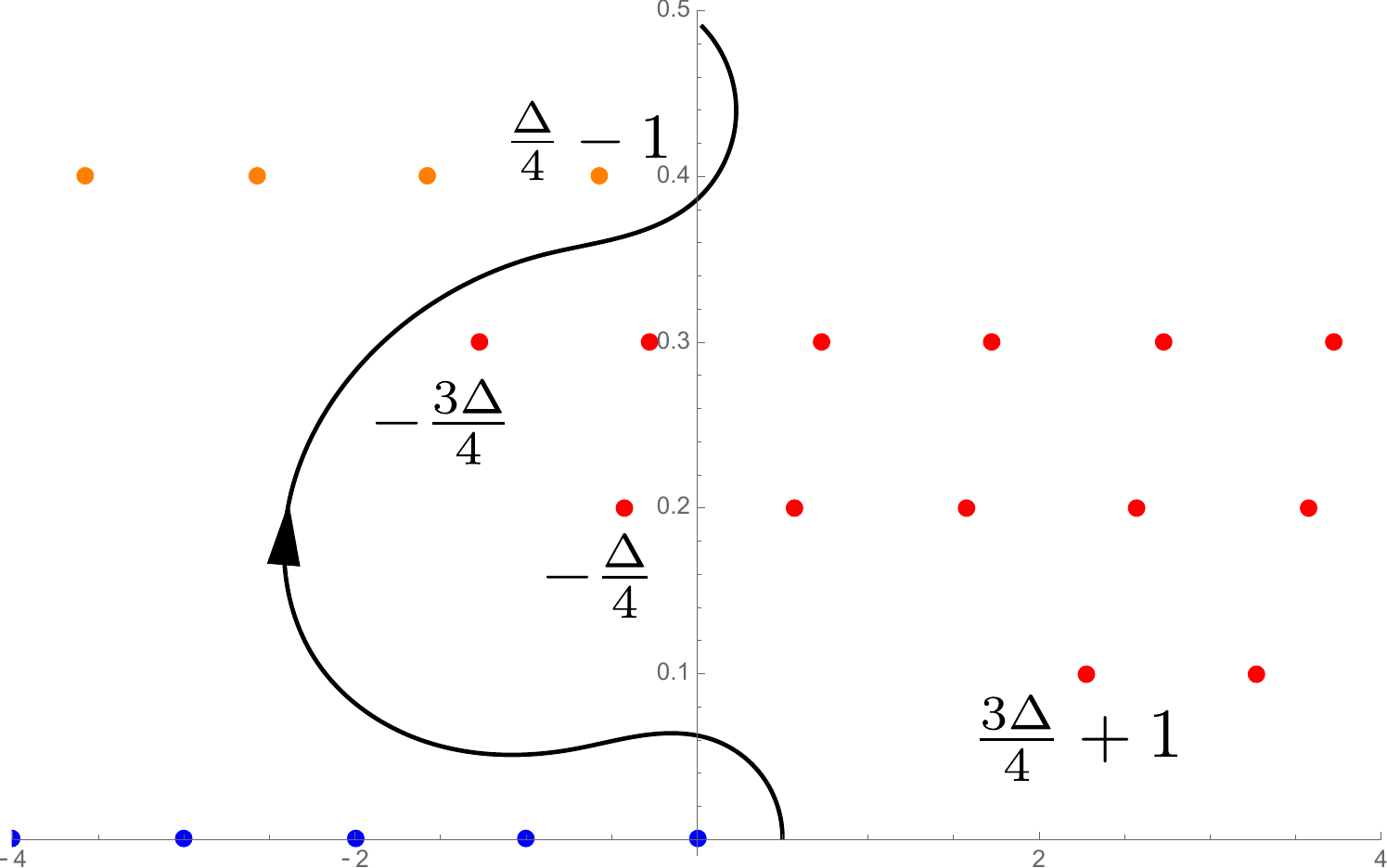}
	\caption{}
	\label{V13poles}
\end{figure}

\paragraph{The $V_{(1,3)}$ and the $V_{(1,5)}$ contributions\\}
We proceed similarly for the other two contributions, but we skip some of the details. For the $V_{(1,3)}$  contribution we have
\begin{align}
I_{(1,3)}\equiv \int_{\epsilon<r<\sqrt{1-a}}\sqrt{g}~{V_{(1,3)}(\tilde\eta)}+c_{\text{c.t.}} \int\sqrt{\indmetr}~{V_{(1,3)}(\tilde\eta)}\big\rvert_{r=\sqrt{1-a}}~.
\end{align}
The counterterm contribution this time is divergent as $a\rightarrow 0$ and equals
\begin{align}\label{V13ct}
\int\sqrt{\indmetr}~{V_{(1,3)}(\tilde\eta)}\big\rvert_{r=\sqrt{1-a}}=4 \pi  R a^{\frac{\Delta }{2}-1}(1+O(a))~.
\end{align}
From the bulk term we find
\begin{align}\label{V13integral}
	&\int_{\epsilon<r<\sqrt{1-a}}\sqrt{g}~{V_{(1,3)}(\tilde\eta)}\nonumber\\
	&= \int_{C}\frac{ds}{2\pi i}\,K_{(1,3)}(\D,s)\left(B_{1-a}\left(s-\frac{\D }{4}+1,-2 s+\frac{\D }{2}-1\right)-B_{\epsilon ^2}\left(s-\frac{\D }{4}+1,-2 s+\frac{\D }{2}-1\right)\right)~,
\end{align}
where we defined
\begin{align}
	R^{-2}K_{(1,3)}(\D,s)\equiv-\frac{4^s\pi  \Delta   \Gamma \left(\frac{1}{2}-\frac{\Delta }{2}\right) \Gamma \left(\frac{\Delta }{4}\right) \Gamma (s) \Gamma \left(-s-\frac{3 \Delta }{4}\right) \Gamma \left(-s-\frac{\Delta }{4}\right) \Gamma \left(1-s+\frac{3 \Delta }{4}\right) }{\Gamma \left(1-\frac{3 \Delta }{4}\right) \Gamma \left(-\frac{\Delta }{4}\right) \Gamma \left(\frac{3 \Delta }{4}\right) \Gamma \left(\frac{1}{2}-s-\frac{\Delta }{2}\right) \Gamma \left(1-s+\frac{\Delta}{4}\right)}~.
\end{align}
Besides the poles coming from $K_{(1,3)}(\D,s)$, in eq.~\eqref{V13integral} we have additional poles from the incomplete beta functions at $s=\D/4-1$, $\D/4-2,$ etc. The original contour $C$ has not moved, and we are free to decide where these new poles lie with respect to $C$. A possible choice is depicted in fig.~\ref{V13poles}.
With this choice, for the  $B_{1-a}\left(s-\frac{\D }{4}+1,-2 s+\frac{\D }{2}-1\right)$ bit we close to the left to pick up residues at the blue and orange poles in fig.~\ref{V13poles}, while for the $B_{\epsilon^2}\left(s-\frac{\D }{4}+1,-2 s+\frac{\D }{2}-1\right)$ we close to the right, and all together we find
\begin{align}\label{13bulk1}
	\int_{\epsilon<r<\sqrt{1-a}}\sqrt{g}~&{V_{(1,3)}(\tilde\eta)}{}=R^2f_{\text{fin}}^{(1,3)}(m)-\frac{8 \pi  R^2 a^{\frac{\Delta }{2}-1}}{\Delta -2}(1+O(a))+\nonumber\\
	&\underset{\equiv R^2f^{(1,3)}_\epsilon (m)}{\underbrace{\frac{2^{-\Delta } R^2 \epsilon ^{2-2 \Delta } \sin \left(\frac{3 \pi  \Delta }{4}\right) \Gamma \left(\frac{3}{2}-\frac{\Delta }{2}\right) \Gamma \left(-\frac{3 \Delta}{4} \right) \Gamma \left(\frac{\Delta }{4}\right)^2 \Gamma \left(\frac{3 \Delta }{2}+1\right)}{\sqrt{\pi } (\Delta -1)^3 \Gamma (\Delta -1) \Gamma \left(-\frac{\Delta }{4}\right)}}}+O(\epsilon^\#)~,
\end{align}
where $O(\epsilon^\#)$ are positive powers of $\epsilon$, which will go to zero as we send $\epsilon\rightarrow 0$ while keeping $\D$ finite and
\begin{align}
f_{\text{fin}}^{(1,3)}(m)=-\pi  m -2\pi+O(m^{-1})~.
\end{align}
It is easy to see that the counterterm contribution of eq.~\eqref{V13ct} cancels exactly the IR divergence if we set
\begin{align}
c_{\text{c.t.}}= -\frac{R}{1-\frac{\Delta }{2}}~.
\end{align}
Putting all together we find
\begin{align}\label{resV13twopt}	
	I_{(1,3)}=R^2 f^{(1,3)}_\epsilon(m)+R^2 f_{\text{fin}}^{(1,3)}(m)~.
\end{align}

Finally, we have for the $V_{(1,5)}$ contribution
\begin{align}
	I_{(1,5)}\equiv \int_{\epsilon<r<\sqrt{1-a}}\sqrt{g}~{V_{(1,5)}(\tilde\eta)}{}+c_{c.t.} \int\sqrt{\indmetr}~{V_{(1,5)}(\tilde\eta)}{}\big\rvert_{r=\sqrt{1-a}}~.
\end{align}
As expected, both the bulk term and counterterm are regular as $a\rightarrow 0$. From the bulk term we find, up to terms that vanish in the $a,\epsilon \rightarrow 0$ limit
\begin{align}\label{15bulk1}
	&\int_{\epsilon<r<\sqrt{1-a}}\sqrt{g}~{V_{(1,5)}(\tilde\eta)}{}=\nonumber\\
	&\underset{\equiv R^2 f^{(1,5)}_\epsilon(m)}{\underbrace{\frac{8 R^2(\D +1) \epsilon ^{2-2 \D } \cos \left(\frac{\pi  \D }{4}\right) \Gamma \left(\frac{1}{2}-\frac{\D }{4}\right) \Gamma \left(\frac{3 \D }{4}+\frac{1}{2}\right) \Gamma \left(\frac{3}{4} (\D +2)\right)}{(\D -1) \D  \Gamma \left(\frac{5 \Delta }{4}+\frac{3}{2}\right)}}}+R^2 f_{\text{fin}}^{(1,5)}(m)~,
\end{align}
with
\begin{align}
	f_{\text{fin}}^{(1,5)}(m)=-4 \pi  m + \frac{8\pi}{3} +O(m^{-1})~.
\end{align}
Hence
\begin{align}\label{resV15twopt}	
	I_{(1,5)}=R^2 f^{(1,5)}_\epsilon(m)+R^2 f_{\text{fin}}^{(1,5)}(m)~.
\end{align}

\paragraph{Cancellation of UV divergences\\}
Putting together equations \eqref{resV11twopt}, \eqref{resV13twopt} and \eqref{resV15twopt}, one finds
\begin{align}
\begin{split}
    (2R)^{\Delta} \delta \langle \phi(0,0) \rangle_{{\bf a}} &=
    -g \mu^{2-\Delta} (2R)^{-\Delta} [(B_{(1,3)}^{{\bf a}\,(1,1)})^2 I_{(1,1)}+(B_{(1,3)}^{{\bf a}\,(1,3)})^2\ I_{(1,3)}+(B_{(1,3)}^{{\bf a}\,(1,5)})^2\ I_{(1,5)}]~\\
                                                             &\qquad + \text{UV counterterms}~.
\end{split}    
\end{align}
To cancel the UV divergences we define the renormalized operator
\begin{align}\label{mixings}
    \phiR = Z_\phi \phi + (2R)^{-\Delta} Z_{\phi \id}\id+\dots~,
\end{align}
and choose $Z_{\phi \id}$ to cancel the power-law divergences, i.e.
\begin{align}\label{cancellUVnew}
\begin{split}
    Z_{\phi \id} &= \frac{1}{4} g (2\mu R)^{2-\Delta} [(B_{(1,3)}^{{\bf a}\,(1,1)})^2 f^{(1,1)}_\epsilon (m)+ (B_{(1,3)}^{{\bf a}\,(1,3)})^2 f^{(1,3)}_\epsilon(m)+ (B_{(1,3)}^{{\bf a}\,(1,5)})^2 f^{(1,5)}_\epsilon (m)] \\
                 &= g (2 \mu R)^{2 - \Delta} \frac{ \pi  \epsilon ^{2-2 \Delta }}{\Delta -1}~.
\end{split}
\end{align}
Here the expression on the first line is what comes out of our computation, which used the boundary conformal block decomposition. The expression on the second line is what one would obtain from a short-distance expansion of the bulk two-point function and isolating the divergent power. The fact that the two expressions are identical is a consistency check for the bulk-boundary crossing discussed in appendix \ref{bulkbdcross}. 

The wave-function renormalization factor $Z_{\phi}$ is obtained from a standard bulk computation, and reads:
\begin{align}\label{phi13wfct}
Z_\phi = 1+ \delta Z_{\phi} &= 1 + \frac{1}{2} g\,C_{\phi\phi\phi} \frac{w_\Delta^2}{w_{2\Delta-2}}~,\quad  w_\Delta =  \frac{2^{2-\D} \pi \Gamma\left(\frac{2-\D}{2}\right)}{\Gamma(\D/2)}~,\nonumber\\
&=1+g\,\left(\frac{2 \pi  m}{\sqrt{3}}-\frac{4 \pi }{\sqrt{3}}-\frac{10 \pi }{\sqrt{3} m}+O(m^{-2})\right)~,
\end{align}
where in the last line we used $C_{\phi\phi\phi}$ of eq.~\eqref{bulkOPEcoeff}. As we discussed in subsection \ref{subsec:uvbehaviorgeneral}, in a general curved space one could have found additional divergences (terms linear in $m$) due to mixing with $(\text{Ric}) \id$ but their coefficient happens to vanish in AdS.

\paragraph{The final result\\}
Collecting all the terms we ultimately find the following one-loop correction to the one-point function of $\phi$:
\begin{equation}
    (2R)^\Delta \vev{\phi(0,0)}_{{\bf a}}  = B_{(1,3)}^{{\bf a}\,(1,1)} -  2\pi g \left( m  ( (2 \mu R)^{2-\Delta} - 1) + \frac{1}{2}  - \frac{8 \pi^2 (a_2^2 + \ldots)}{3 m} \right) + O(g^2)~.
\end{equation}
We recognize the first term as due to the anomalous dimension of $\phi(x)$. It can in principle be resummed to obtain an answer as a function of the running coupling, but at the fixed point we may just set $(2 \mu R) = 1$ which removes this term. Also at the fixed point we will have $g = O(m^{-1})$ so the $O(m^{-1})$ term we wrote may in principle get corrected at two loops. For this reason we did not write it out completely, although we did indicate the dependence on the boundary label $a_2$ that follows from our computation. Ultimately we obtain:
\begin{equation}\label{phi1ptcorrectedfull}
    (2R)^\Delta \vev{\phi(0,0)}_{{\bf a}}  = B_{(1,3)}^{{\bf a}\,(1,1)}  -  \frac{\sqrt{3}}{m}  +  \frac{16 \pi^2 a_2^2}{\sqrt{3} m^2} + O(m^{-2})~,
\end{equation}
at the fixed point $g_* = \sqrt{3}/\pi m$. We recall that
\begin{equation}
    \label{b13expansionmaintext}
    B_{(1,3)}^{{\bf a}\,(1,1)} = \sqrt{3} - \frac{2\pi^2(2 a_2^2 - 1)}{\sqrt{3} m^2} + O(m^{-3})~.
\end{equation}

\paragraph{Interpretation\\}
The interpretation of this result is perhaps a little disappointing. In flat space, the $\phi_{(1,3)}$ operator flows to $\phi_{(3,1)}$ at the IR fixed-point \cite{Zamolodchikov:1987ti}. For the latter we have:
\begin{equation}
    (2R)^\Delta \vev{\phi_{(3,1)}(0,0)}_{(b_1,b_2)_{m-1}} = \sqrt{3} - \frac{2 \pi^2 (2 b_1^2 -1)}{\sqrt{3} m^2} + O(m^{-3})~.
\end{equation}
If $(a_1,a_2)_m$ flows to $(a_2,a_1)_{m-1}$ then we see that this expression agrees with \eqref{b13expansionmaintext} to the given order, so we would naively expect the one-loop correction to be zero. Equation \eqref{phi1ptcorrectedfull} therefore presents us with two issues: the non-zero correction multiplying $m^{-1}$ and the non-trivial dependence on the boundary labels multiplying $m^{-2}$.

We believe that the first issue might be due to an incorrect identification of the infrared operator. We have chosen the wave function renormalization factor $Z_\phi$ such that the infrared operator is unit normalized in flat space, so the overall normalization should be fine. However, as we mentioned above, in curved space we can build new operators that involve the background curvature. In particular we cannot exclude the possibility that the infrared operator is any admixture of $\phi_{(3,1)}$ with the operator $\text{Ric} \id \sim R^{-2} \id$. If this is the case then the factor $\sqrt{3}/m$ in equation \eqref{phi1ptcorrectedfull} merely amounts to a determination of the relative coefficient between these two operators. We also note that such an admixture should be visible at higher orders in perturbation theory, because the latter operator has $\Delta = 2$ exactly, whereas $\Delta_{3,1} = 2 + 4/m$.

The final term in \eqref{phi1ptcorrectedfull} is subleading in $1/m$ and we could in principle have dismissed it as unreliable. It was however argued in \cite{Fredenhagen:2009tn} that the dependence on the boundary labels \emph{is} reliable at this order at least in flat space. If this is so then it poses a puzzle. This term cannot be explained by operator mixing, since the composition of the infrared operator should not depend on the boundary labels. We also verified that the one-loop correction to $\vev{\id}$ does not feature any dependence on the boundary labels at this order. However the analysis of \cite{Fredenhagen:2009tn} does not necessarily apply in AdS, since it was based purely on the large-$m$ behavior of the various OPE coefficients appearing in the computation. In our case we have seen that this can be offset by the integrals themselves, which can diverge polynomially in $m$. We therefore cannot conclude that this term is reliable without a more detailed analysis of the integrals involved in the higher-loop contributions.

\subsection{Irrelevant deformation}
In this section we study the $\phi_{(3,1)}$  deformation of unitary and diagonal minimal models $\mathcal{M}_{m+1,m}$ with conformal boundary conditions ${\bf a}= (a_1,a_2)_m$, in $\AdStwo$ and with $m>3$. We will compute some anomalous dimensions of boundary Virasoro primaries at finite $m$, at the first order in this deformation. In the large-$m$ limit, with $a_1,a_2\ll m$, the $\phi_{(3,1)}$ operator is the leading irrelevant deformation, with scaling dimension
\begin{align}\label{phi31dim}
	\D=2h_{3,1} = 2+\frac{4}{m}~.
\end{align}

\subsubsection{The perturbation}
We consider the following bulk perturbation by $\phi\equiv \phi_{(3,1)}$
\begin{align}\label{bulkpert31}
    \delta S = g R^{\D-2}\int \ddxsqrtg{2}{x}{g}\phi(x + i y, x - i y)+\text{counterterms}+O(g^2)~.
\end{align}
Again, for large but finite $m$ we can remove the near-boundary divergences without inducing a boundary RG flow, so that the only non-trivial $\beta$ function is for the bulk coupling and reads~\cite{Cardy:1996xt}
\begin{align}
	\dot g  = (2-\D)g - \pi\,C_{\phi \phi \phi}g^2+O(g^3)~.
\end{align}
At $m=\infty$ we have that $C_{\phi \phi \phi}={4}/{\sqrt{3}}$ (see eq.~\eqref{bulkOPEcoeffnew}), and so using eq.~\eqref{phi31dim} we find a UV fixed point at
\begin{align}\label{UVfixedpt}
	g_* = -\frac{\sqrt{3}}{\pi m}+O(m^{-2})~.
\end{align}
In the bulk, this is the inverted RG flow between consecutive minimal models, i.e. $\mathcal{M}_{m+1,m}\rightarrow \mathcal{M}_{m+2,m+1}$.

\subsubsection{Anomalous dimensions}

We present here some anomalous dimensions along the $\phi_{(3,1)}$ deformation of unitary and diagonal minimal models, in $\AdStwo$. The computations for this case are entirely similar to those of section~\ref{sec:anomaldimbd}, so we omit the details of the derivation.

The scaling dimension of $\psi_{(r,s)}$ after the perturbation of eq.~\eqref{bulkpert31} (which goes towards the UV) reads
\begin{align}\label{hrsfinitem31}
	\hD_{r,s}(g)=\hD_{r,s}+g\deltahD_{r,s}^{(1,1)}(m) +g\deltahD_{r,s}^{(3,1)}(m) +g\deltahD_{r,s}^{(5,1)}(m) +O(g^2)~,
\end{align}
where $\hD_{r,s} = h_{r,s}$ and
\begin{align}
	\deltahD_{r,s}^{(1,1)}(m)&=B_{(3,1)}^{{\bf a}\,(1,1)}\frac{\pi  (m+2) (m r-m s+r-1) (m r-m s+r+1)}{2^{\frac{m+4}{m}} (2-m) m (m+3)} \nonumber\\
		 &\times \sum_{k=0}^\infty\frac{\left(\frac{1}{2}\right)_k \left(-\frac{2}{m}\right)_k \left(\frac{r+m (r-s+1)-1}{m}\right)_k \left(\frac{-r+m (-r+s+1)-1}{m}\right)_k}{k! (2)_k \left(\frac{3}{2}-\frac{1}{m}\right)_k \left(-\frac{3}{m}\right)_k}~,\nonumber\\
	\deltahD_{r,s}^{(3,1)}(m)&=\alpha_{(3,1)}\frac{\sqrt{\pi }  \Gamma \left(\frac{1}{m}\right)}{2^{\frac{m+2}{m}} \Gamma \left(\frac{1}{2}+\frac{1}{m}\right)} \nonumber\\
		 &\times \sum_{k=0}^\infty \frac{\left(\frac{1}{m}\right)_k \left(-\frac{m+2}{2 m}\right)_k \left(\frac{1}{2}+\frac{r}{m}+r-s\right)_k \left(\frac{1}{2}-\frac{(m+1) r}{m}+s\right)_k}{k! \left(\frac{1}{2}+\frac{1}{m}\right)_k \left(\frac{3}{2}+\frac{1}{m}\right)_k \left(-\frac{m+4}{2 m}\right)_k}~,\nonumber\\
	\deltahD_{r,s}^{(5,1)}(m)&=\alpha_{(5,1)}\frac{\sqrt{\pi } 2^{\frac{2}{m}+2} \Gamma \left(\frac{3}{2}+\frac{3}{m}\right)}{\Gamma \left(2+\frac{3}{m}\right)}\nonumber\\
	&\times \sum_{k=0}^\infty \frac{\left(1+\frac{1}{m}\right)_k \left(\frac{3}{2}+\frac{3}{m}\right)_k \left(\frac{2+r+m (r-s+2)}{m}\right)_k \left(\frac{2-r+m (-r+s+2)}{m}\right)_k}{k! \left(\frac{5}{2}+\frac{2}{m}\right)_k \left(2+\frac{3}{m}\right)_k \left(3+\frac{3}{m}\right)_k}~.
\end{align}

\subsubsection{Large \texorpdfstring{$m$}{m}}
Finally we consider the large-$m$ limit with $a_1,a_2\ll m$. In this limit there are some extra $O(m)$ divergences, but they can be removed without affecting the anomalous dimensions as we explained below eq.~\eqref{largemectra}. The large-$m$ expansion for the coefficients is given in equations~\eqref{B3151} and \eqref{largemalpha31}, recalling that $\alpha_{(5,1)}=O(m^{-2})$. Plugging in the value of the UV fixed point in eq.~\eqref{UVfixedpt} we find
\begin{align}\label{hrsfinitemUV}
	\hD^{\text{UV}}_{r,s}=\hD_{r,s}+g\deltahD_{r,s}^{(1,1)}(\infty) +g\deltahD_{r,s}^{(1,3)}(\infty)+O(m^{-2})=\frac{1}{4} (r-s)^2-\frac{r^2-s^2}{4 m}+O(m^{-2})~.
\end{align}
By comparing this result with the boundary spectrum of the $\mathcal{M}_{m+2,m+1}$ minimal model we see that along the $\phi_{(3,1)}$ bulk RG flow (for $r\neq s$)
\begin{align}
	\psi_{(r,s)}\longrightarrow \psi_{(s,r)}~.
\end{align}
The spectral flow is again consistent with the following large-$m$ flow between conformal b.c.:
\begin{align}\label{RGflowbcinv}
	(a_1,a_2)_{m}\longrightarrow (a_2, a_1)_{m+1}~,\quad a_1,a_2\ll m~.
\end{align}
As expected, the $\phi_{(3,1)}$ deformation `inverts' the RG flow of eq.~\eqref{RGflowbc}.

\section{Outlook}
We have analyzed RG flows in AdS both in general terms and in the explicit example of the perturbed minimal models. To deal with the ultraviolet divergences one employs familiar quantum field theory in curved space techniques, and to deal with the infrared divergences one adds counterterms familiar from holographic renormalization. In the end one obtains finite renormalized correlation functions of both bulk and boundary operators.

In the future it would be interesting to resolve the somewhat puzzling status of the one-point functions in the perturbed minimal models. This will be essential for the identification of the correct infrared boundary conditions. This would require computing $\vev{\phi_{(1,3)}}$ to two-loop order. One might also compute $\vev{\phi_{(r,s)}}$ at one loop, for which the level of difficulty should be similar to what was presented in this work. To check our prescriptions at a structural level one might of course also try other controllable RG flows like the Wilson-Fisher fixed points in AdS. (See \cite{Carmi:2018qzm,Ankur:2023lum} for some other examples. More generally there is a vast literature on loop computations in AdS, but these mostly focus on Witten diagrams for boundary correlation functions and not on bulk one- or higher-point functions.)

This paper is not only an \'etude but also a pr\'elude to the numerical study of RG flows in AdS by applying conformal bootstrap methods to the boundary correlation functions. The viability of this method was demonstrated in \cite{Antunes:2021abs} where initial results for scalar field theories in AdS$_2$ were presented. In upcoming work \cite{Antunes:2024hrt} we will analyze the minimal model flows in AdS numerically and make essential use of the results presented above. In the future it will be interesting to include the sum rules presented in \cite{meineri2023renormalization} in the numerical analysis.

\section*{Acknowledgements}
We would like to thank A. Antunes for many discussions and collaboration on related topics and C. Behan, M. Bill\`o, L. Di Pietro, A. Hebbar, M. Hogervorst, N. Levine, M. Meineri, M. Paulos, M. Reehorst, A. Tilloy and E. Trevisani for discussions. EL and BvR are supported by the Simons Foundation grant $\#$488659 (Simons Collaboration on the non-perturbative bootstrap). This project is funded by the European Union: for EL by ERC ``QFT.zip'' with Project ID 101040260 (held by A. Tilloy) and for BvR by ERC ``QFTinAdS'' with Project ID 101087025. Views and opinions expressed are however those of the author(s) only and do not necessarily reflect those of the European Union or the European Research Council Executive Agency. Neither the European Union nor the granting authority can be held responsible for them.

\appendix

\section{Two-dimensional BCFTs}
\label{app:2dbcfts}

We will consider two-dimensional CFTs on the upper half-plane $$\uhp=\{z\in\mathbb{C}\;|\;\im~z\geq0\}~.$$ In Cartesian coordinates we write $z =  x + i y$ so $x$ is also a boundary coordinate.

We will consider various correlation functions between both bulk and boundary global primaries in a generic BCFT on the upper half-plane.\footnote{A Virasoro primary or descendant operator is called a `global primary' when it is a primary with respect to the global conformal group $SL(2,\mathbb{C})$.} In the present subsection, $\phi_i(z,\bar{z})$ is a scalar bulk global primary with dimension $\D_i$, and $\psi_i(x)$ is a scalar global boundary primary with dimension $\hD_i$. 

For a more detailed discussion on two-dimensional BCFTs and more references to the original literature we refer the reader to the books~\cite{Ginsparg:1988ui,DiFrancesco:1997nk,Mussardo:2010mgq,Recknagel:2013uja}.

\subsection{Global conformal invariance}
We begin with the general structure that follows from $SL(2,\mathbb{C})$ bulk invariance as well as the residual $SL(2,\mathbb{R})$ in the presence of a boundary. The OPEs in the various channels are
\begin{align}\label{OPEs}
	\phi_i(z_1,\bar{z}_1)\phi_j(z_2,\bar{z}_2)
	&=\sum_{k}C_{ij}{}^k\phi_k(z_2,\bar{z}_2)|z_1-z_2|%
	^{\Delta_k-\Delta_i-\Delta_j}+\cdots~,\nonumber\\
	\phi_i(x+iy,x-iy)&=\sum_k B_{i}{}^k\psi_k(x)(2y)^{\hD_k-\Delta_i}
	+\cdots~,\nonumber\\
	\psi_i(x_1)\psi_j(x_2)&=\sum_k\hC_{ij}{}^k\psi_k(x_2)
	(x_1-x_2)^{\hD_k-\hD_i-\hD_j} + \cdots~, \qquad(x_1>x_2)~.
\end{align}
The ellipses in the first (second and third) lines above denote $SL(2,\mathbb{C})$ ($SL(2,\mathbb{R})$) descendants. For both bulk and boundary operators, indices are raised and lowered using the two-point correlation function of the corresponding operator. Unless otherwise specified, we will always consider unit-normalized primary bulk (boundary) operators, and a unit-normalized identity operator: $\langle \id \rangle =\langle \hid \rangle =1$. The values of the OPE coefficients $B$ and $\hC$ depend on the specific conformal boundary condition.

We now provide the general form of some BCFT correlation functions as well as the structure of their conformal block decomposition.

\paragraph{Simple correlation functions} On the upper half-plane the simplest non-trivial correlation functions are ($x_{ij}\equiv x_i-x_j$ and $\hD_{ijk}\equiv \hD_i+\hD_j-\hD_k$)
\begin{align}\label{correlatorsuhpCFTconventions}
	\langle \psi(x_1)\psi(x_2)\rangle_{\uhp}&=\frac{1}{(x_{12}^2)^{\hD}}~,\qquad\qquad\qquad\qquad\quad(x_1>x_2)~\nonumber\\
	\langle\phi(x+iy,x-iy)\rangle_{\uhp}&=\frac{B_\phi}{(2y)^{\Delta}}~,\qquad\qquad\qquad\qquad\quad\quad(y>0)~\nonumber\\ 
	\langle\phi(x_1+iy_1,x_1-iy_1)\psi(x_2)\rangle_{\uhp} &=
	\frac{B_{\phi \psi}}{
		(2y_1)^{\Delta-\hD}(x_{12}^2+y_1^2)^{\hD}}~,~\qquad\quad(y_1>0)~\nonumber\\ 
	\langle\psi_i(x_1)\psi_j(x_2)\psi_k(x_3)\rangle_{\uhp} &=\frac{\hC_{ijk}}{
		(x_{12})^{\hD_{ijk}} (x_{23})^{\hD_{jki}}
		(x_{13})^{\hD_{ikj}}}~,\quad(x_{i}>x_{i+1})~.
\end{align}

\paragraph{Boundary four-point function} Consider the four-point correlation function between four global boundary conformal primaries $\psi_i$ with scaling dimensions $\hD_i$. By $SL(2,\mathbb{R})$ symmetry this correlation function takes the form (here and in the following $\hD_{ij}\equiv \hD_i-\hD_j$)
\begin{align}
	\langle \psi_1 (x_1)\psi_2 (x_2)\psi_3 (x_3)\psi_4 (x_4)\rangle_{\uhp} =\left(\frac{x_{14}}{x_{24}}\right)^{\hD_{21}} \left(\frac{x_{14}}{x_{13}}\right)^{\hD_{34}} \frac{\mathcal{G}^{1234}(\eta)}{(x_{12})^{\hD_{1}+\hD_{2}} (x_{34})^{\hD_{3}+\hD_{4}}}~,\quad x_{i}>x_{i+1}~,
\end{align}
with cross-ratio
\begin{align}\label{eta_def_line}
	\eta= \frac{x_{12}x_{34}}{x_{13}x_{24}}~,\quad 0<\eta<1~.
\end{align}
In the s-channel we have
\begin{align}\label{schanelblocksdec}
	\mathcal{G}^{1234}(\eta) = \sum_{k} \hC_{12}{}^{k} \hC_{34k}G(\hD_{21},\hD_{34},\hD_k,\eta)~,
\end{align}
with global blocks given by~\cite{Dolan:2011dv}
\begin{align}\label{schanelblocksgen}
	G(a,b,\D,\eta)= \eta^{\D } \, _2F_1(a+\D ,b+\D ;2 \D ;\eta)~.
\end{align}

\paragraph{Bulk two-point function} Consider the two-point correlation function between two bulk scalar global primaries $\phi_i$ with scaling dimensions $\D_i$.  By $SL(2,\mathbb{R})$ symmetry this two-point correlation function takes the following form~\cite{McAvity:1993ue,McAvity:1995zd,Liendo:2012hy}:
\begin{align}
	\langle \phi_1(x_1 + i y_1, x_1-i y_1)\phi_2(x_2+ i y_2, x_2-i y_2)\rangle_{\uhp} = \frac{f(\xi)}{(2y_1)^{\D_1}(2y_2)^{\D_2}}~.
\end{align}
The cross-ratio is
\begin{align}\label{eta_def_full}
	\xi\equiv \frac{(x_1-x_2)^2+(y_1-y_2)^2}{4 y_1 y_2}~.
\end{align}
The bulk and boundary channel expansions of $f(\xi)$ read
\begin{align}\label{twoptbulkexp}
	f(\xi)=\xi^{-\frac{1}{2}(\D_1+\D_2)}\sum_k C_{12}{}^k B_kf_{\text{bulk}}(\D_k;\xi)=\sum_k B_1{}^{ k}B_{2k}~f_{\text{bdy}}(\hD_k;\xi)~,
\end{align}
where~\cite{McAvity:1993ue,McAvity:1995zd,Liendo:2012hy}
\begin{align}\label{twoptblocks}
	f_{\text{bulk}}(\D_k;\xi)&= \xi^{\D_k/2}{}F_1 \left(\frac{\D_k}{2}, \frac{\D_k}{2},\D_k;-\xi\right)~,\nonumber\\
	f_{\text{bdy}}(\hD_k;\xi)&= \xi^{-\hD_k}{}F_1 \left(\hD_k, \hD_k,2\hD_k;-{1}/{\xi}\right)~.
\end{align}

\paragraph{Bulk-boundary-boundary function} Consider the correlation function between two boundary global primaries of scaling dimension $\hD_{1,2}$ and one (scalar) bulk global primary of scaling dimension $\D$.  Via the bulk-boundary OPE of $\phi$, this correlator can be written as
\begin{align}
	\langle \psi_1({x}_1)\psi_2({x}_2)&\phi(x+ i y,x - i y)\rangle_{\uhp}=\nonumber\\
	&\frac{\sum_{k} B_\phi{}^k {\hC}_{12 k} f(\hD_{12}, \hD_k, v)}{(y^2+(x_1-x)^2)^{{\hD_{12}/2}}(y^2+(x_2-x)^2)^{{\hD_{21}/2}}|{x}_{12}|^{\hD_1+\hD_2} (2y)^{\D}}~.
\end{align}
In our conventions the blocks computed in refs.~\cite{Lauria:2020emq,Nishioka:2022qmj} read
\begin{align}
\label{bulk_bd_bd_blocks_2d}
	f(\hD_{12}, \hD_k, v)=	(4v)^{\hD_k/2} \, _2F_1\left(\frac{\hD_k+\hD_{12}}{2},\frac{\hD_k-\hD_{12}}{2};\hD_k+\frac{1}{2};v\right)~,
\end{align}
with
\begin{align}\label{etatildedef}
v=\frac{y^2 ({x}_{12})^2}{ (y^2 + ({x}_1-{x})^2)(y^2 + ({x}_2-{x})^2)}\equiv \tieta/4~,\quad 0\leq v \leq 1~. 
\end{align}

\subsection{Local conformal invariance}
\label{subapp:localconformalinvariance}
Let $T_{\mu\nu}$ be the bulk stress tensor. Conservation and tracelessness imply
\begin{align}\label{holom_T}
	\bar{\partial}T=0~,\quad {\partial}\bar{T}=0~,
\end{align}
where $\partial \equiv \partial_z $, $\bar{\partial}\equiv \partial_{\bar z} $ and we have defined
\begin{align}
	T\equiv T(z)\equiv T_{zz}(z)~,\quad \bar{T}\equiv \bar{T}({\bar z})\equiv T_{{\bar z}\bar z}(\bar z)~.
\end{align}
Translation invariance along the boundary is equivalent to Cardy's condition~\cite{Cardy:1984bb,Cardy:1989ir,Cardy:2004hm}
\begin{align}\label{Cardyscond}
	T(z)=\bar{T}({\bar z})~,\quad \im~z=0~.
\end{align}
It follows from equations~\eqref{holom_T} and~\eqref{Cardyscond} that the boundary limit of $T$ is a global boundary primary operator of scaling dimension $\hD=2$, i.e.
\begin{align}\label{bOPET}
	\Disp (x) = T(x+i y)\rvert_{y=0}~.
\end{align}
We call this boundary operator  the `displacement'. The restriction of the anti-holomorphic component $\bar{T}$ of the stress tensor to the boundary defines the same displacement, thanks to Cardy's condition.

\subsubsection{Conformal Ward identities on the upper half-plane}
Let us consider an infinitesimal real-analytic transformation on the upper half-plane $\uhp$
\begin{align}\label{infrealan}
	z\rightarrow z+\epsilon(z),~\quad \epsilon(z)=\sum_n a_n z^{n+1}~,\quad a_n \in \mathbb{R}~.
\end{align}
The conserved charge on $\uhp$ is
\begin{align}\label{QHplus}
	Q_\epsilon=\frac{1}{2\pi i}\oint_{C^+} \epsilon(z)T(z)dz - \frac{1}{2\pi i}\oint_{C^+} \epsilon(\bar{z})\bar{T}(\bar{z})d\bar{z}~,
\end{align}
where $C^+$ is a contour in $\uhp$, we regarded $\epsilon(\bar{z})$ as $\epsilon^*(z)$ and  $\bar{T}(\bar{z})$ is the analytic continuation of $T(z)$ to the lower half-plane ($\lhp$). Because of Cardy's condition~\eqref{Cardyscond}, the operator $Q_\epsilon$ is well defined even when correlation functions are evaluated on the real axis. We can then define the Virasoro generators in analogy with those in the full complex plane as
\begin{align}\label{VirGenBD}
	L_n&=\frac{1}{2\pi i}\oint_{C^+} z^{n+1}T(z)dz - \frac{1}{2\pi i}\oint_{C^+} {\bar z}^{n+1}\bar{T}(\bar{z})d\bar{z}~\nonumber\\
	&=\frac{1}{2\pi i}\oint_{C \in \mathbb{C}} z^{n+1}T(z)dz~,
\end{align}
where in the last line we used Cardy's condition~\eqref{Cardyscond} to combine the two addends into a single contour integral on the full complex plane. The generators so defined then obey the commutation relations of a holomorphic Virasoro algebra, with central charge $c$
\begin{align}\label{Vir}
	[L_n\,,L_m]=(n-m)L_{n+m}+\frac{c}{12}n(n^2-1)\delta_{n+m,0}~.
\end{align}
The elements $\{L_{\pm 1},L_0\}$ of the Virasoro algebra generate the conformal group $SL(2,\mathbb{R})$.  
On the real axis, we can define Virasoro primary operators $\psi_i$ with respect to this holomorphic Virasoro algebra as
\begin{equation}\label{boundaryvirasoroprimary}
	[L_{0},\psi_i(0)]=\hD_i\, \psi_i(0) ~,\quad	[L_{n},\psi_i(0)]=0~,\quad n\geq 1~.
\end{equation}
Using the definition above it follows that, under the (finite) real-analytic conformal transformations
\begin{align}\label{real_transf_prim}
	z\rightarrow  f(z)~,\quad	\psi_i'(f(z))=\left(\frac{\partial f}{\partial z}\right)^{-\hD_i}\psi_i(z)~.
\end{align}
Namely $\psi_i$ transforms as a `holomorphic Virasoro primary', by which we mean a Virasoro primary with weights $h_i=\hD_i,~\bar{h}_i=0$. Equivalently eq.~\eqref{VirGenBD} implies that
\begin{align}
	T(z)\psi_i(w)=\frac{\hD_i}{(z-w)^2}{\psi_i}(w)+ \frac{1}{z-w}\partial{\psi_i}(w)+O((z-w)^0)~.
\end{align}
Boundary Virasoro descendants are obtained by repeatedly applying the $L_{-k}$'s with $k>0$ to the primary ${\psi_i}$. Using eq.~\eqref{VirGenBD} we have that
\begin{align}\label{desc}
	[L_{-k},{\psi_i}(w)]\equiv {\psi_i}^{(-k)}(w)=\frac{1}{2\pi i}\oint_{C_w}\,dz\, (z-w)^{1-k}T(z){\psi_i}(w)~,
\end{align}
where the contour $C_w$ lies on the full complex plane and encircles $w$. We can use eq.~\eqref{desc} to compute the correlation function of ${\psi_i}^{(-k)}(w)$ with any number of either bulk or boundary Virasoro primaries from the correlation function of ${\psi_i}(w)$ with the same number of either bulk or boundary Virasoro primaries. In terms of the string of Virasoro primaries (in defining ${\mathcal{X}}$ we disregard its anti-holomorphic dependence as it plays no role in the following) 
\begin{align}\label{chi_string_def}
	{\mathcal{X}}^{(n,m)}(\{z_i\})\equiv \psi_1(z_1)\dots \psi_n (z_n)\phi_{n+1} (z_{n+1},\bar{z}_{n+1})\dots\phi_{n+m} (z_{n+m},\bar{z}_{n+m})~,
\end{align}
we have that
\begin{align}\label{hphichi_corr}
	\langle\psi^{(-k)}(w)&{\mathcal{X}}^{(n,m)}(\{z_i\})\rangle \nonumber\\
	&=
	\frac{1}{2\pi i}\oint_{C_w}\frac{dz}{(z-w)^{k-1}}
	\langle T(z)\psi(w){\mathcal{X}}^{(n,m)}(\{z_i\})\rangle~,\nonumber\\
	&=-\frac{1}{2\pi i}\oint_{C_{\{z_i\}}}\frac{dz}{(z-w)^{k-1}}
	\langle T(z)\psi(w){\mathcal{X}}^{(n,m)}(\{z_i\})\rangle~,\nonumber\\
	&=-\frac{1}{2\pi i}\oint_{C_{\{z_i\}}}\frac{dz}{(z-w)^{k-1}}
	\sum_{i=1}^{n+m} \left(\frac{1}{z-z_i}\partial_i+\frac{h_i}{(z-z_i)^2}\right)\langle\psi(w){\mathcal{X}}^{(n,m)}(\{z_i\})\rangle~,\nonumber\\
	&= \mathcal{L}_{-k}^{(w)}~\langle\psi(w){\mathcal{X}}^{(n,m)}(\{z_i\})\rangle~,
\end{align}
and we denoted the weight of each boundary Virasoro primary $\psi_i$ as $\hD_i=h_i$.
In the third line we modified the integration contour in order to pick up the singularities at $z=z_i$. These singularities are simple and double poles, as dictated by the OPE of $T$ with either $\phi_i$ or $\psi_i$
\begin{align}\label{Tphi_OPE}
	T(z){\mathcal{X}}^{(n,m)}(\{z_i\})\underset{z\rightarrow z_k}{\sim} \frac{h_i}{(z-z_k)^2}{\mathcal{X}}^{(n,m)}(\{z_i\})+\frac{1}{z-z_k}\partial_k{\mathcal{X}}^{(n,m)}(\{z_i\})+\text{regular}~.
\end{align}
In the last step, after computing the integral using the residue theorem, we introduced the important differential operator
\begin{align}\label{calLdiffopLn}
	\mathcal{L}_{-k}^{(w)}\equiv\sum_{i=1}^{n+m}\left(\frac{ (k-1)h_i}{(z_i-w)^k}-\frac{1}{(z_i-w)^{k-1}}\partial_i\right)~.
\end{align}
The result in eq.~\eqref{hphichi_corr} implies that if $\psi^{(-k)}$ is a null descendant of $\psi$ at level $k$, correlation functions with $\psi$ insertions satisfy a linear differential equation of order $k$. By construction, this will be the same differential equation satisfied by a holomorphic Virasoro primary with weight $(h=\hD,~\bar{h}=0)$.

\subsubsection{The method of images}
Bulk correlation functions on the upper half-plane $\uhp$ can be computed using the method of images~\cite{Cardy:1984bb} (see also Chapter 11.2 of~\cite{DiFrancesco:1997nk}) and bulk-boundary crossing symmetry. To see how it works, let us consider the effect of the transformation \eqref{infrealan} on a correlator between a string of bulk Virasoro primaries
\begin{align}\label{stringchi}
	{\mathcal{X}}(z_1,\bar{z}_1;\dots;z_n,\bar{z}_n)\equiv \phi_1(z_1,\bar{z}_1)\dots \phi_n(z_n,\bar{z}_n)~,
\end{align}
with $\Delta_i = h_i +\bar{h}_i$. Note that here $h_i \neq \bar{h}_i$, generically.  This transformation is (recall eq.~\eqref{QHplus})
\begin{align}
	\frac{1}{2\pi i}\oint_{C^+} \epsilon(z)dz\langle T(z){\mathcal{X}}(z_1,\bar{z}_1;\dots;z_n,\bar{z}_n)\rangle_{\uhp} - \frac{1}{2\pi i}\oint_{C^+} \epsilon(\bar{z})d\bar{z}\langle \bar{T}(\bar{z}){\mathcal{X}}(z_1,\bar{z}_1;\dots;z_n,\bar{z}_n)\rangle_{\uhp}~.
\end{align}
We can regard the dependence of the correlators on antiholomorphic coordinates $\bar{z}_i$ on the upper half-plane $\uhp$ as a dependence on the holomorphic coordinates $z_i^* = \bar{z}_i$ on the lower half-plane $\lhp$. This `mirroring' is nothing but a parity transformation with respect to the real axis, and as such holomorphic indices change into antiholomorphic ones e.g. $T(z^*)=\bar{T}(z)$.\footnote{If a given bulk operator is purely holomorphic, so $\bar{h}=0$, then its antiholomorphic part is the identity operator. As a consequence there is no need to double holomorphic operators, such as $T(z)$.} Since $T=\bar{T}$ on the real axis by Cardy's condition, when we combine the two addends above to write a contour integral over the full complex plane, the total contribution from the real axis vanishes, and we are left with a contour integral of a correlator with $2n$ holomorphic Virasoro primaries on the full complex plane:
\begin{align}
	\frac{1}{2\pi i}\oint_{C} \epsilon(z)dz\langle T(z){\mathcal{X}}(z_1,z'_1;\dots; z_n,z'_n)\rangle~,\quad z_i'\equiv z^*_i~.
\end{align}
Hence, the correlation function $\langle\phi_1(z_1,\bar{z}_1)\dots \phi_n(z_n,\bar{z}_n)\rangle_{\uhp}$ regarded as a function of $(z_1,\bar{z}_1;\dots; z_n,\bar{z}_n)$ satisfies the same differential equation as the correlation function on homogeneous space, $\langle \phi_1(z_1,\bar{z}_1)\dots \phi_n(z_{2n},\bar{z}_{2n})\rangle $, as a function of $(z_1,\dots,z_{2n})$. 

As an application, let us compute the one-point correlation function on $\uhp$ of a Virasoro primary $\phi$ of weight $\D=h+\bar{h}$ and spin $\ell=h-\bar{h}$, denoted as $\phi_{h,\bar h} (z,\bar z)$.\footnote{The attentive reader will have noticed a small change of notation: here and in the next section the Virasoro primaries are labelled by their dimensions $(h,\bar{h})$, and not by a pair of integers $(r,s)$ as everywhere else in this work. This choice is convenient in order to explain the application of the method of images.}
By the method of images, we are instructed to consider
\begin{align}
	\langle \phi_{h,\bar{h}}(z,\bar{z})\rangle_{\uhp}\propto \langle \phi_h(z)\phi_{\bar h}(z^*)\rangle~,
\end{align}
where we have regarded $z^*$ as a holomorphic coordinate, and $\phi_{\bar h}$ as a holomorphic field with holomorphic dimension $\bar{h}$ . Note that the $\propto$ symbol is appropriate, since the method of images cannot fix overall constants in correlation functions. Now, the two-point correlation function on the right-hand side above is fixed by conformal symmetry as in eq.~\eqref{phiphi}. In particular we immediately learn that if $\ell\neq 0$, then $\phi$ has a vanishing one-point correlation function on $\uhp$. Alternatively, if $\ell=0$, writing $z=x+ i y$ with $y> 0$, we get
\begin{align}\label{phi1pt}
	\langle  \phi_{h,{h}}(x + i y, x- i y)\rangle_{\uhp} = \frac{B_{\phi}}{(2y)^{\D}}~,\quad \D = 2h~.
\end{align}
The constant $B_{\phi}\equiv B_{\phi\id}$, which is not fixed by the method of images, is a physical datum of the BCFT.
Finally let us consider the bulk-boundary two-point correlation function between the scalar bulk Virasoro primary $\phi$  and a boundary Virasoro primary $\psi$ with scaling dimension $\hD$. Applying again the method of images, we compute
\begin{align}
	\langle \phi_{h,{h}}(z,\bar{z})\psi(0)\rangle_{\uhp}\propto \langle \phi_h(z)\phi_{ h}(z^*)\psi(0)\rangle~,
\end{align}
where we have regarded $\psi$ as a holomorphic operator with holomorphic dimension $\hD$. Choosing again $z=x+ i y$ with $y> 0$, we then find
\begin{align}
	\langle \phi_{h,{h}}(x + i y, x- i y)\psi(0)\rangle_{\uhp}= \frac{B_{\phi \psi}}{(2y)^{\D-\hD}(y^2+x^2)^{\hD}}~.
\end{align}
The constant $B_{\phi \psi}$ is another physical datum of the BCFT.

\subsection{Some universal correlation functions}
\label{subapp:universalcorrelators}
Correlation functions involving the stress tensor $T(z)$, $\bar T(\bar z)$ and the displacement operator $\Disp(x)$ are to a large extent fixed by the conformal Ward identities. In the instances below we will find several such correlation functions. They are universal in the sense that they do not depend on the specific BCFT and instead only depend on $c$ and the scaling dimension of the operators involved. By taking coincident points we also obtain correlation functions of $\Disp^2(x)$ and $\TTb(z,\bar z)$.

\subsubsection{Interlude: the stress tensor on the plane}
\label{app:Tplanecorr}
In a generic 2d CFT, correlation functions with stress tensor insertions are determined by the conformal Ward identities~\cite{Belavin:1984vu}. Consider first the simple case with just one $T$ insertion in between a string of $n$ Virasoro primaries $\phi_i(z_i,\bar{z}_i)$, with weights $(h_i,\bar{h}_i)$. In the notation introduced in eq.~\eqref{chi_string_def}, this correlator will be denoted as
\begin{align}
	\langle T(w){\mathcal{X}}^{(0,n)}(\{z_i\})\rangle ~.
\end{align}
Because the $T$ is a level-two descendant of the identity operator $\id$, from eq.~\eqref{hphichi_corr} we immediately find
\begin{align}\label{Tchi_disp}
	\langle T(w){\mathcal{X}}^{(0,n)}(\{z_i\})\rangle = \mathcal{L}_{-2}^{(w)}\langle{\mathcal{X}}^{(0,n)}(\{z_i\})\rangle~.
\end{align}

Let us now consider two insertions of the stress tensor. 
\begin{align}
	\langle T(w) T(w') {\mathcal{X}}^{(0,n)}(\{z_i\})\rangle~.
\end{align}
We can apply the same procedure that led to the formula in eq.~\eqref{hphichi_corr} to reduce the number of $T$ insertions, but this time we get additional contributions from the additional poles in the $T\times T$ OPE when deforming the integration contour. We recall that
\begin{align}\label{TTOPE}
	T(z)T(0)=\frac{c/2}{z^4}\id+\frac{2}{z^2}T(0)+\frac{1}{z} T'(0)+\frac{3}{10}T''(0)+T^2(0)+O(z^2)~,
\end{align}
where primed operators are derivatives with respect to $z$, $O(z^2)$ denotes higher order contributions and $T^2$ is the unique global primary at level four in the $T\times T$ OPE.\footnote{This operator is obtained (up to normalization) by acting with $L_{-4}-\frac{5}{3}L_{-2}^2$ on the Virasoro identity module.}
Using the $T\times T$ OPE we then get
\begin{align}\label{TTany}
	\langle T(w) T(w') &{\mathcal{X}}^{(0,n)}(\{z_i\})\rangle=\mathcal{L}_{-2}^{(w)}\langle T(w') {\mathcal{X}}^{(0,n)}(\{z_i\})\rangle+\frac{c/2}{(w-w')^4}\langle {\mathcal{X}}^{(0,n)}(\{z_i\})\rangle~,
\end{align}
and $\mathcal{L}_{-2}^{(w)}$ acts on $T(w')$, as well as on the string ${\mathcal{X}}$. 
As a particular case of the equations above, let us take ${\mathcal{X}}$ to be a string of two identical holomorphic Virasoro primaries with two-point function
\begin{align}\label{phiphi}
	\langle \phi_i(z_1) {\phi_i}(z_2)\rangle = \frac{1}{(z_{12})^{2 h_i}}~,\quad z_{ij}\equiv z_i-z_j~.
\end{align}
From eq.~\eqref{Tchi_disp} we obtain
\begin{align}\label{phiphiandTphiphi}
	\langle T(z_1) \phi_i(z_2) {\phi_i}(z_3)\rangle = \frac{C_{T\phi\phi}}{(z_{12})^2 (z_{23})^{2 h_i-2} (z_{31})^2}~, \quad C_{T\phi\phi}=h_i~.
\end{align}
From eq.~\eqref{TTany} we obtain
\begin{align}\label{TTphiphi}
	\langle &T(z_0) T(z_1) {\phi_i}(z_2) {{\phi_i}}(z_3)\rangle = \frac{\mathcal{G}^{TT\phi\phi}(\eta)}{(z_{01})^4(z_{23})^{2 h_i}}~,
	&&\mathcal{G}^{TT\phi\phi}(\eta)=\left(\frac{c}{2}+\frac{h_i \eta ^2 (\eta  (h_i \eta -2)+2)}{(\eta -1)^2} \right)~.
\end{align}
The cross-ratio $\eta$ is
\begin{align}\label{etadefcomplex}
	\eta\equiv \frac{ z_{01}z_{23}}{z_{02}z_{13}}~.
\end{align}
We can recursively apply the algorithm used above to add more insertions of $T$. For example:
\begin{align}\label{TTTTdiff}
	&\langle T(z_0) T(z_1)T(z_2) {\mathcal{X}}^{(0,n)}(\{z_i\})\rangle= \nonumber\\
	&\mathcal{L}_{-2}^{(z_0)}\langle T(z_1) T(z_2) {\mathcal{X}}^{(0,n)}(\{z_i\})\rangle+\frac{c/2}{(z_{01})^4}\langle T(z_2) {\mathcal{X}}^{(0,n)}(\{z_i\})\rangle+\frac{c/2}{(z_{02})^4}\langle T(z_1) {\mathcal{X}}^{(0,n)}(z_i)\rangle~.
\end{align}

As a particular case of eq.~\eqref{TTTTdiff} , when ${\mathcal{X}}^{(0,n)}$ is the identity operator and using
\begin{align}\label{T2pt3pt}
	\langle T(z_2) T(z_3)\rangle = \frac{c/2}{(z_{23})^{4}}~, \quad \langle T(z_1) T(z_2) T(z_3)\rangle = \frac{c}{(z_{12})^2 (z_{23})^{2} (z_{31})^2}~,
\end{align}
we find the four-point function of $T(z)$
\begin{align}\label{T4ptfunction}
	&\langle T(z_0) T(z_1) T(z_2) T(z_3)\rangle = \frac{\mathcal{G}^{TTTT}(\eta)}{(z_{01})^4 (z_{23})^4}~,\nonumber\\
	\mathcal{G}^{TTTT}(\eta)&=\frac{2c~\eta ^2 ((\eta -1) \eta +1)}{(\eta -1)^2}+\frac{c^2}{4}\left(1+\eta ^4+\left(\frac{\eta }{\eta -1}\right)^4\right)~.
\end{align}

The four-point correlation function of $T(z)$ is a physical correlator, and as such it admits a decomposition into the $SL(2,\mathbb{R})$ conformal blocks of eq.~\eqref{schanelblocksgen} (with positive coefficients in unitary theories). 
In the $s$-channel we find
\begin{align}\label{schanelblocksdecTTTT}
	\mathcal{G}^{TTTT}(\eta) = \frac{c^2}{4}G(0,0,0,\eta)+\sum_{n = 2,4,\dots}C_{TT}{}^n C_{TTn}\,G(0,0,n,\eta)~,
\end{align}
with squared OPE coefficients
\begin{align}\label{T4ptdecompositionOPE}
	C_{TT}{}^n C_{TTn}=\frac{\sqrt{\pi } c  (c (n-3) (n-2) (n-1) \Gamma (n+3)+288 ((n-1) n-1) \Gamma (n))}{9\times2^{2 n+1} \Gamma \left(n-\frac{1}{2}\right)}~.
\end{align}
In particular, the squared OPE coefficients with $T$ and $T^2$ in the $T\times T$ OPE read
\begin{align}\label{T4ptdecomposition}
	C_{TT}{}^T C_{TTT}=2c~,\quad C_{TT}{}^{T^2} C_{TT T^2}=\frac{c}{10}(22+5 c)~.
\end{align}
Note that eq.~\eqref{TTOPE} implies that $C_{TT}{}^{T^2}=1$, and so from the rightmost equation above we get
\begin{align}\label{T2twopt}
	\langle T^2(0)T^2(\infty)\rangle \equiv C_{T^2}=C_{TTT^2}= \frac{c}{10}(22+5 c)~.
\end{align}

\subsubsection{Correlation functions of the stress tensor}
Let us now return to the BCFT setup. We recall that $n$-point functions of the stress tensor $T(z)$ on the upper half-plane are the same as in flat space. The reason for this is holomorphy, which states that
\begin{align}
	\bar{\partial}_i\langle T(z_1)T(z_2)\dots T(z_n)\rangle_{\uhp} =0~.
\end{align}
and so the correlation functions are invariant under a translation with respect to the transverse coordinate orthogonal to the boundary. Therefore, they must take the same form as in the full complex plane~\cite{Quella:2006de,Billo:2016cpy, Meineri:2019ycm}. For two- and three-point correlation functions of $T$, the bulk self-OPE of $T$ together with $SL(2,\mathbb{R})$ symmetry dictate that
\begin{align}\label{TTandTTTuhp}
	\langle T(z_1){T}(z_2)\rangle_{\uhp} =\frac{c/2}{(z_{12})^4}~,\quad \langle T(z_1){T}(z_2)T(z_3)\rangle_{\uhp} =\frac{c}{(z_{12})^{2}(z_{23})^{2}(z_{31})^{2}}~.
\end{align}
Analogous results hold for higher-point correlation functions of $T$ or of $\bar{T}$. Note that eq.~\eqref{Cardyscond} implies that correlation functions of $\bar{T}$ are those of $T$ on the upper half-plane, analytically continued to the lower half-plane, so for example
	\begin{align}
		\langle T(z)\bar{T}(\bar{w})\rangle_{\uhp} =\langle T(z){T}(w^*)\rangle_{\uhp}~.
\end{align}

\subsubsection{Correlation functions of the displacement operator}
The $n$-point functions of the displacement operator are directly obtained from eq.~\eqref{TTandTTTuhp}, by simply restricting all $T$-insertions to the real line. We  choose an operator ordering along the line so that $x_i > x_{i+1}$, and so we find
\begin{align}\label{Displ2pt3pt}
	\langle \Disp(x_1) \Disp(x_2)\rangle_{\uhp} = \frac{\Cd}{x_{12}^4}~, &\quad \langle \Disp(x_1) \Disp(x_2) \Disp(x_3)\rangle_{\uhp} = \frac{\hC_{\Disp\Disp\Disp}}{(x_{12})^{2}(x_{23})^{2}(x_{31})^{2}}~,\nonumber\\
	&\Cd=c/2~,\quad \hC_{\Disp\Disp\Disp}=c~.
\end{align}
Similarly, the four-point correlation function can be obtained from eq.~\eqref{T4ptfunction}, e.g.
\begin{align}\label{D4ptfunction}
	&\langle \Disp(x_1) \Disp(x_2) \Disp(x_3) \Disp(x_4)\rangle_{\uhp}= \frac{\mathcal{G}^{\Disp\Disp\Disp\Disp}(\eta)}{(x_{12})^4 (x_{34})^4}~,\quad x_i > x_{i+1}~,\nonumber\\
	&\mathcal{G}^{\Disp\Disp\Disp\Disp}(\eta)=\frac{2c~\eta ^2 ((\eta -1) \eta +1)}{(\eta -1)^2}+\frac{c^2}{4}\left(1+\eta ^4+\left(\frac{\eta }{\eta -1}\right)^4\right)~.
\end{align}
The cross-ratio $\eta$, which can be found in eq.~\eqref{eta_def_line}, is the restriction of the familiar holomorphic cross-ratio in eq.~\eqref{etadefcomplex} to the real line. We can decompose this correlator into the $s$-channel $SL(2,\mathbb{R})$ blocks defined in eq.~\eqref{D4ptfunction}. The resulting squared OPE coefficients equal the ones of eq.~\eqref{schanelblocksdecTTTT}, so in particular from eq.~\eqref{T4ptdecomposition} we find
\begin{align}\label{CDDDandCDDD2}
	\hC_{\Disp\Disp}{}^{\Disp}\hC_{\Disp\Disp\Disp}=2c~,\quad \hC_{\Disp\Disp}{}^{\Disp^2}\hC_{\Disp\Disp\Disp^2}=\frac{c}{10}(22+5 c)~.
\end{align}
In the equation above we identified the global primary operator $\Disp^2$ as the boundary limit of the operator $T^2$ in the bulk, which is defined in eq.~\eqref{TTOPE}. From eq.~\eqref{T2twopt} we find
\begin{align}\label{D2twopt}
	\langle \Disp^2(0) \Disp^2(\infty)\rangle_{\uhp} \equiv  \Cdd=\frac{c}{10}(22+5 c)~,
\end{align}
and, combining with eq.~\eqref{CDDDandCDDD2}
\begin{align}\label{CD2andCDDD2}
	\hC_{\Disp\Disp}{}^{\Disp^2}=1~,\quad \hC_{\Disp\Disp\Disp^2} = \Cdd~.
\end{align}

\subsubsection{Correlation functions of the displacement with boundary operators}
Let us now consider a conformal boundary condition that allows for a boundary Virasoro primary $\psi$ with two-point correlation function
\begin{align}\label{hphitwopt}
	\langle \psi(x_1)\psi(x_2)\rangle_{\uhp} = \frac{1}{x_{12}^{2\hD}}~, \quad x_1 > x_{2}~.
\end{align}
As we discussed in the previous subsection, as far as Ward identities are concerned, boundary Virasoro primaries behave as holomorphic Virasoro primaries. Correlation functions between $\Disp$ and $\psi$ can be obtained from correlation functions on the upper half-plane between $T$ and a holomorphic Virasoro primary $\phi$ with weight $h=\hD$.
From eq.~\eqref{phiphiandTphiphi}, upon using the bulk-boundary OPE of $T$ in eq.~\eqref{bOPET} and choosing the operator ordering $x_i > x_{i+1}$ on the line we immediately get
\begin{align}\label{Dispphiphi}
	\langle \Disp(x_1)\psi(x_2)\psi(x_3)\rangle_{\uhp} &=\frac{\hC_{\psi\psi \Disp}}{(x_{12})^{2}(x_{23})^{2\hD-2}(x_{31})^{2}}~,\quad \hC_{\psi\psi \Disp}=\hC_{\Disp\psi\psi}=\hD~.
\end{align}
This result, when combined with eq.~\eqref{Displ2pt3pt}, tells us that the displacement operator will enter the $SL(2,\mathbb{R})$ block decomposition of the four-point correlation function of $\psi$ as
\begin{align}\label{phiphiDcoeff}
	\hC_{\psi\psi}{}^{\Disp}\hC_{\psi\psi\Disp}=\frac{\hD^2}{\Cd}=\frac{2\hD^2}{c}~.
\end{align}
The four-point correlation function $\langle \Disp \Disp \psi\psi\rangle $ in this ordering can be obtained from the result of eq.~\eqref{TTphiphi} and reads ($\eta$ is defined in eq.~\eqref{eta_def_line})
\begin{align}\label{DDphiphiuhp}
	\langle &\Disp(x_1) \Disp(x_2) \psi(x_3) {\psi}(x_4)\rangle_{\uhp} = \frac{\mathcal{G}^{\Disp\Disp\psi\psi}(\eta)}{(x_{12})^4(x_{34})^{2 \hD}}~,
	&&\mathcal{G}^{\Disp\Disp\psi\psi}(\eta)=\left(\frac{c}{2}+\frac{\hD \eta ^2 (\eta  (\hD \eta -2)+2)}{(\eta -1)^2} \right)~.
\end{align}
The $SL(2,\mathbb{R})$ s-channel block decomposition reads
\begin{align}\label{schanelblocksdecDDhphihphi}
	\mathcal{G}^{\Disp\Disp\psi\psi}(\eta) =\frac{c}{2}G(0,0,0,\eta)+ \sum_{n=2,4,\dots} \hC_{\Disp\Disp}{}^n \hC_{\psi\psi n}G(0,0,n,\eta)~,
\end{align}
with blocks given in eq.~\eqref{D4ptfunction} and coefficients
\begin{align}
	\hC_{\Disp\Disp}{}^n \hC_{\psi\psi n}= \frac{\sqrt{\pi }\hD}{2^{2n-3}\Gamma \left(n-\frac{1}{2}\right)}(\hD (n-2) (n+1)+2) \Gamma (n)~.
\end{align}
In particular for the first few coefficients we have
\begin{align}
	\hC_{\Disp\Disp}{}^{\Disp} \hC_{\psi\psi \Disp} = 2 \hD~,&\quad \hC_{\Disp\Disp}{}^{\Disp^2} \hC_{\psi\psi \Disp^2}=\frac{\hD}{5} (5 \hD+1)~,
\end{align}
and so, once we combine with eq.~\eqref{D2twopt} and eq.~\eqref{CD2andCDDD2}, we find
\begin{align}\label{hphihphiD2}
	\hC_{\psi\psi \Disp^2}=\frac{\hD}{5} (5 \hD+1)~.
\end{align}
We can also consider a different ordering, e.g.
\begin{align}\label{DphiDphiuhp}
	\langle \Disp(x_1) \psi(x_2)& \Disp(x_3) {\psi}(x_4)\rangle_{\uhp} =\left(\frac{x_{14}}{x_{24}}\right)^{\hD-2} \left(\frac{x_{14}}{x_{13}}\right)^{2-\hD} \frac{\mathcal{G}^{\Disp\psi\Disp\psi}(\eta)}{(x_{12})^{2+\hD} (x_{34})^{2+\hD}} ~,\quad x_i > x_{i+1}~,\nonumber\\
	&\mathcal{G}^{\Disp\psi\Disp\psi}(\eta)=\eta ^{\hD+2} \left(\frac{c}{2}+ \frac{\hD (2 (\eta -1) \eta +\hD)}{\eta^2(\eta-1)^2}\right)~,
\end{align}
with $\eta$ defined as in eq.~\eqref{eta_def_line}. The $s$-channel $SL(2,\mathbb{R})$ block decomposition reads
\begin{align}\label{schanelblocksdecDhphiDhphi}
	\mathcal{G}^{\Disp\psi\Disp\psi}(\eta) = \hD^2G(\hD-2,2-\hD,\hD,\eta)+\sum_{n=2,3,4,\dots} \hC_{\Disp\psi}{}^n {\hC}_{\Disp\psi n}G(\hD-2,2-\hD,\hD+n,\eta)~,
\end{align}
and the first non-zero coefficients are
\begin{align}
	\hC_{\Disp\psi}{}^2 \hC_{\Disp\psi 2} &= \frac{c}{2}+\hD \left(4-\frac{9}{2 \hD+1}\right)~,\nonumber\\
	\hC_{\Disp\psi}{}^3 \hC_{\Disp\psi 3} &=-\frac{2 \hD}{(\hD+1) (\hD+2)}\left((c-7) \hD+c+3 \hD^2+2\right) ~,\nonumber\\
	\hC_{\Disp\psi}{}^4 \hC_{\Disp\psi 4} &= \frac{\hD (5 c (4 \hD (\hD+2)+3)+4 \hD (\hD (8 \hD-19)+26)-15)}{(\hD+3) (2 \hD+3) (2 \hD+5)}~.
\end{align}
Note that the negative term on the second line is due to a parity-odd operator at level 3, see for example appendix K of \cite{Homrich:2019cbt}.

\subsubsection{Correlation functions of the displacement with a bulk operator}
\label{subsubapp:ddphi}
Consider a scalar bulk Virasoro primary $\phi_{h,h}(z,\bar{z})$ with scaling dimension $\D=2h$. We want to compute its two-point correlation function with the displacement
\begin{align}
	\langle \Disp(0) \phi_{h,h}(x+ i y,x - iy)\rangle_{\uhp}~.
\end{align}
As we have discussed, $\Disp$ can be traded for an insertion of $T$ on $\uhp$. Since $T$ is holomorphic,  it does not need to be doubled, and by the method of images
\begin{align}
	\langle \Disp(0) \phi_{h,h}(z,\bar{z})\rangle_{\uhp} \propto \langle T(0) \phi_h(z)\phi_h(z^*)\rangle~.
\end{align}
Upon using eq.~\eqref{phiphiandTphiphi} we learn that
\begin{align}\label{Dphiphi}
	\langle \Disp(0)\phi_{h,h}(x + i y, x- i y)\rangle_{\uhp} = \frac{B_{\phi \Disp}}{(2y)^{\D-2}(x^2+y^2)^2}~.
\end{align}
The bulk-to-boundary coefficient $B_{\phi \Disp}$ is further related to the one-point function in eq.~\eqref{phi1pt} by means of a special Ward identity~\cite{McAvity:1993ue,McAvity:1995zd}
\begin{align}\label{bphiD}
	B_{\phi \Disp}=-B_{\phi}\D/2~.
\end{align}
The same logic applies to higher-point correlators, e.g.
\begin{align}
	\langle \Disp(x_1)  \Disp(x_2) \phi_{h,h}(z,\bar{z})\rangle_{\uhp} &\propto 	\langle T(x_1)  T(x_2) \phi_h(z)\phi_h(z^*)\rangle~,
\end{align}
and from eq.~\eqref{TTphiphi} we find
\begin{align}\label{DDphiphi}
	\langle \Disp(x_1)  \Disp(x_2) \phi_{h,h}(x+ i y,x-i y)\rangle_{\uhp} =\frac{B_\phi}{(2y)^{\D}(x_{12})^4}\,\left(\frac{c}{2}-\D  \tieta \left(1-\frac{\D  \tieta}{4}\right) \right)~.
\end{align}
The cross-ratio $\tieta$ is defined in eq.~\eqref{tietadef}.
The normalization of eq.~\eqref{DDphiphi} is chosen such that the $y\rightarrow 0$ expansion of the correlator above gives the identity contribution in the $\Disp\times\Disp$ OPE, with the correct normalization. 
As a check of this result, we can decompose eq.~\eqref{DDphiphi} into the bulk-boundary-boundary conformal blocks of eq.~\eqref{bulk_bd_bd_blocks_2d}. The second addend in eq.~\eqref{DDphiphi} can be decomposed as follows
\begin{align}\label{DDphiphi_dec_bd_3pt}
B_{\phi} \sum_{n=2,4,6\dots} \frac{\sqrt{\pi } \D   e^{\frac{i \pi  n}{2}} (\D  (n-2) (n+1)+4) \Gamma (n)}{2^{2 n-1}\Gamma \left(n-\frac{1}{2}\right)}\,f(0,n,\tieta/4)~.
\end{align}
In particular, we learn that
\begin{align}
	\hC_{\Disp\Disp}{}^{\Disp}B_{\phi\Disp} = -B_{\phi}~\D, \quad 		\hC_{\Disp\Disp}{}^{\Disp^2} B_{\phi\Disp^2} = \frac{1}{20} B_{\phi} \D  (5 \D +2)~.
\end{align}
As a check of this result, we note that the leftmost piece of the equation above reproduces eq.~\eqref{bphiD} after using eq.~\eqref{Displ2pt3pt}. From the rightmost piece, upon combining with eq.~\eqref{CD2andCDDD2} we find
\begin{align}\label{bphiD2}
	B_{\phi\Disp^2} = \frac{1}{20} B_{\phi} \D  (5 \D +2)~.
\end{align}

\subsubsection{Correlation functions with \texorpdfstring{$\Disp^2$}{D2}}
\label{subsubapp:dsquaredcorrs}
The operator $\Disp^2$ is the unique global primary at level four in the boundary identity module. It is defined upon restricting the $T\times T$ OPE in eq.~\eqref{TTOPE} to the real line
\begin{align}\label{DDOPE}
	\Disp(x)\Disp(0)=\frac{c/2}{x^4}\hid+\frac{2}{x^2}\Disp(0)+\frac{1}{x}\Disp'(0)+\frac{3}{10} \Disp''(0)+\Disp^2(0)+O(x^2)~,
\end{align}
where primed operators are derivatives with respect to $x$, $O(x^2)$ denotes higher order contributions and $\Disp^2$ is the restriction of $T^2$ to the boundary.

To compute correlation functions with $\Disp^2$ on the upper half-plane we apply the following strategy. Starting from correlation functions with two insertions of $T$ on $\uhp$, each $T^2$ insertion is obtained as
\begin{align}\label{T2def}
	\langle T^2(z)\dots \rangle \equiv  \lim_{z\rightarrow 0} \langle \left(T(z)T(0)-\frac{c/2}{z^4}\id-\frac{2}{z^2}T(0)-\frac{1}{z} T'(0)-\frac{3}{10}T''(0) \right)\dots\rangle ~.
\end{align}
Finally, each $T^2(z)$ insertion defines a $\Disp^2(x)$ insertion upon taking the boundary limit on $z$, the latter being regular, as dictated by holomorphy.

Our first target is
\begin{align}
	\langle \Disp^2 (x_1) \Disp^2 (x_2) \phi (x+ i y, x -i y)\rangle_{\uhp}~,
\end{align}
with $\phi$ a bulk Virasoro primary of dimension $\D$. For this we should consider
\begin{align}
	\langle T(z_0) T(z_1)T(z_2) T(z_3)\phi(x+ i y, x -i y)\rangle_{\uhp}~,
\end{align}
which is determined by the conformal Ward identity in terms of the correlator of $\phi$ with zero, one, two, and three $T$ insertions, as explained above. Note that in order to take the limit \eqref{T2def}, we need to combine this five-point correlator with the correlator of $\phi$ with one, two, and three $T$ insertions. Taking the limit \eqref{T2def} of coincident points (say, between the first two and the second two $T$ insertions), and finally the boundary limit of each $T^2$ insertion, we find
\begin{align}\label{D2D2phiall}
	\langle \Disp^2 (x_1) \Disp^2 (x_2) \phi (x +i y, x- iy)\rangle_{\uhp} = \frac{\mathcal{G}^{\Disp^2\Disp^2\phi}(\tieta)}{(2y)^{\D}(x_{12})^8}~,
\end{align}
where the cross-ratio is that of eq.~\eqref{etatildedef} and
\begin{align}
	\mathcal{G}^{\Disp^2\Disp^2\phi}(\tieta)={B_{\phi}}&\left(\frac{c}{10}  (5 c+22)-\frac{2\D}{5}  (5 c+22)\tieta+\frac{\D}{50}  (5 (5 c+64) \D+84)\tieta^2\right.\nonumber\\
	&~~~~\left.-\frac{\D}{25}  (5 \D+2)^2\tieta^3+\frac{\D^2}{400} (5 \D+2)^2\tieta^4\right)~.
\end{align}
The $B_\phi$ is that of eq.~\eqref{phi1pt}.
This result can also be expanded into the bulk-boundary-boundary blocks of eq.~\eqref{bulk_bd_bd_blocks_2d} to find
\begin{align}\label{D2D2phiphi_dec_bd_3pt}
\frac{B_{\phi} }{10} c (5 c+22)	+ B_{\phi} \sum_{n=2,4,6\dots} \left(c_n^{(1)} \D+c_n^{(2)}\D^2+c_n^{(3)}\D^3+c_n^{(4)}\D^4\right)\,f(0,n,\tieta/4)~,
\end{align}
where
\begin{align}\label{cndefs}
c_n^{(1)}&=	e^{\frac{i \pi  n}{2}}\frac{\sqrt{\pi } 2^{3-2 n} (50 c+((n-1) n+8) ((n-1) n+20)) \Gamma (n)}{25 \Gamma \left(n-\frac{1}{2}\right)}~,\nonumber\\
c_n^{(2)}&=	e^{\frac{i \pi  n}{2}}\frac{\sqrt{\pi } 2^{-2 (n+1)} (n-2) (2 n-1) (1800 c+(n-1) n ((n-1) n+678)+14760)  \Gamma (n+2)}{225 n \Gamma \left(n+\frac{1}{2}\right)}~,\nonumber\\
c_n^{(3)}&=	e^{\frac{i \pi  n}{2}}\frac{\sqrt{\pi } 2^{-2 n-1} (n-4) (n-2) (n+3) ((n-1) n+150)  \Gamma (n+2)}{45 n \Gamma \left(n-\frac{1}{2}\right)}~,\nonumber\\
c_n^{(4)}&=	e^{\frac{i \pi  n}{2}}\frac{\sqrt{\pi } 2^{-2 n-3} (n-6) (n-4) (n-2) (n+3) (n+5) \Gamma (n+2)}{9 n \Gamma \left(n-\frac{1}{2}\right)}~.
\end{align}

\subsection{Correlation functions with \texorpdfstring{$\TTb$}{TTb}}
\label{app:TTbcorrelators}
Let us now discuss how we compute correlation functions with insertions of $\TTb$ on $\uhp$. The first observation is that the $T \times \bar{T}$ bulk OPE is regular, as dictated by holomorphy, and the limit of coincident points defines a unique scalar bulk global primary of scaling dimension equal to four. Therefore
\begin{align}\label{prescr_TTb}
	\langle \dots T\bar{T} (z) \rangle_{\uhp} & \equiv \lim_{z'\rightarrow z}  \langle \dots T(z')\bar{T} (z) \rangle_{\uhp}~\nonumber\\
	&=  \lim_{z'\rightarrow z} \langle \dots T(z') T (z^*) \rangle_{\uhp} =  \langle \dots T(z) T (z^*) \rangle_{\uhp}~.
\end{align}
As an immediate application of the prescription above, we learn that
\begin{align}\label{TTb1pt}
	\langle \TTb( x + i y,x- i y) \rangle_{\uhp}  \equiv \frac{B_{\TTb}}{(2y)^4}= \langle  T(x+ i y) T (x - iy) \rangle_{\uhp} = \frac{c/2}{(2 y)^4}~,
\end{align}
and so $ B_{\TTb} = c/2$.

Let us now compute the three-point correlation function between $\TTb$ and a boundary Virasoro primary $\psi$ of scaling dimension $\hD$. Starting from eq.~\eqref{TTphiphi} (with $\phi\rightarrow \psi$), we apply the prescription in eq.~\eqref{prescr_TTb} to obtain
\begin{align}\label{phiphiTTbfull}
	&\langle \psi(x_1)\psi(x_2)\TTb(x+i y,x-i y)\rangle_{\uhp} = \frac{\mathcal{G}^{\psi\psi \TTb}(\tieta)}{(2y)^4(x_{12})^{2\hD}}~, \quad x_1>x_2~.
\end{align}
where
\begin{align}
\mathcal{G}^{\psi\psi \TTb}(\tieta)= \frac{c}{2}-2\hD \tieta+\hD^2 \tieta^2~.
\end{align}
The cross-ratio $\tieta$ is that of eq.~\eqref{etatildedef}.
The normalization is fixed by the $\psi$ self-OPE limit, and by eq.~\eqref{TTb1pt}. 

Expanding this result into the bulk-boundary-boundary blocks of eq.~\eqref{bulk_bd_bd_blocks_2d} we find
\begin{align}
	\frac{c}{2} +\sum_{n=2,4,6\dots}( \alpha^{(1)}_n \hD +\alpha^{(2)}_n \hD^2)~\,f(0,n,\tieta/4)~,
\end{align}
where
\begin{align}
	\alpha^{(0)}_n=\frac{\sqrt{\pi } e^{\frac{i \pi  n}{2}} \Gamma (n)}{4^{n-2}\Gamma \left(n-\frac{1}{2}\right)}~,\quad 
	\alpha^{(1)}_n=\frac{\sqrt{\pi }  e^{\frac{i \pi  n}{2}} (n-2) (n+1) \Gamma (n)}{2^{2 n-3}\Gamma \left(n-\frac{1}{2}\right)}~.
\end{align}

Another interesting correlator is the three-point correlation function between two insertions of the displacement and $\TTb$ on $\uhp$. This can be obtained from the four-point function of $T$ in  eq.~\eqref{T4ptfunction}, upon applying the prescription in eq.~\eqref{prescr_TTb} (e.g. for the last two insertions of $T$) and taking the boundary limit on the first two $T$ insertions. The normalization is again fixed by the bulk-boundary limit, so we find
\begin{align}\label{DDTTb3ptfunctionfull}
	\langle \Disp(x_1) \Disp(x_2)\TTb(x+ i y, x - i y)\rangle_{\uhp}   = \frac{\mathcal{G}^{\Disp\Disp \TTb}(\tieta)}{(2y)^4(x_{12})^4}~,
\end{align}
where again $\tieta$ is that of eq.~\eqref{etatildedef} and
\begin{align}
\mathcal{G}^{\Disp\Disp\TTb}(\tieta) &= \frac{c^2}{4}\left(1-\frac{8 \tieta}{c}+\left(\frac{8}{c}+2\right) \tieta^2-4 \tieta^3+\tieta^4\right)~.
\end{align}
Expanding this result into the bulk-boundary-boundary blocks of eq.~\eqref{bulk_bd_bd_blocks_2d} we find
\begin{align}
	\frac{c^2}{4} +2c\sum_{n=2,4,6\dots}( \alpha^{(0)}_n+\alpha^{(1)}_n c)~\,f(0,n,\tieta/4)~,
\end{align}
where
\begin{align}
	\alpha^{(0)}_n=\frac{\sqrt{\pi }  e^{\frac{i \pi  n}{2}} ((n-1) n-1) \Gamma (n)}{2^{2 n-3}\Gamma \left(n-\frac{1}{2}\right)}~,\quad
	\alpha^{(1)}_n=	\frac{\sqrt{\pi }  e^{\frac{i \pi  n}{2}} (n-3) (n-2) (n-1) \Gamma (n+3)}{9\times4^{1+n}\Gamma \left(n-\frac{1}{2}\right)}~.
\end{align}

\subsection{A correlation function with \texorpdfstring{$\Disp^2$}{D2} and \texorpdfstring{$\TTb$}{TTb}}
We conclude with the three-point correlation function between two insertions of $\Disp^2$ and $\TTb$. The starting point is the six-point correlation function of $T$ on $\uhp$, which however is completely fixed by the conformal Ward identities.
Taking the various limits, so eq.~\eqref{T2def} in the first and second $T\times T$ insertions, and eq.~\eqref{prescr_TTb} in the last $T\times T$ insertion we find ($\tieta$ in eq.~\eqref{etatildedef}.)
\begin{align}\label{D2D2TTball}
	\langle \Disp^2 (x_1) \Disp^2 (x_2) \TTb (x+i y, x - iy)\rangle_{\uhp} =\frac{\mathcal{G}^{\Disp^2\Disp^2\TTb}(\tieta)}{(2y)^{4}(x_{12})^8}~,
\end{align}

\begin{align}
\mathcal{G}^{\Disp^2\Disp^2\TTb}(\tieta)&= \frac{c^2}{20} (5 c+22)\left(1-\frac{16 \tieta}{c} +\left(\frac{248}{5 c}+4\right) \tieta^2-\left(\frac{176}{5 c}+8\right) \tieta^3+\left(\frac{44}{5 c}+2\right) \tieta^4\right)~.
\end{align}
We can expand this result into the bulk-boundary-boundary blocks of eq.~\eqref{bulk_bd_bd_blocks_2d} to find
\begin{align}
\frac{c^2}{20}(22+5c) -\frac{4}{5} c (5 c+22)\sum_{n=2,4,6\dots}( \alpha^{(0)}_n+\alpha^{(1)}_n c)~\,f(0,n,\tieta/4)~,
\end{align}
where
\begin{align}
	 \alpha^{(0)}_n&=-\frac{\sqrt{\pi } 2^{-2 n-1} e^{\frac{i \pi  n}{2}} ((n-1) n (11 (n-1) n ((n-1) n-8)+1572)-2160) \Gamma (n)}{45 \Gamma \left(n-\frac{1}{2}\right)}~,\nonumber\\
 \alpha^{(1)}_n&=	 -\frac{\sqrt{\pi } 4^{-n-1} e^{\frac{i \pi  n}{2}} (n-3) (n-2) (n-1) \Gamma (n+3)}{9 \Gamma \left(n-\frac{1}{2}\right)}~.
\end{align}

\section{Review of minimal model BCFTs}
\label{app:minmodelsbcft}
We will consider unitary and diagonal minimal models $\mathcal{M}_{m+1,m}$ with central charge
\begin{equation}\label{central}
	c=1-\frac{6}{m(m+1)}~,\qquad {m=3,\,4,\,5,\ldots\,}
\end{equation}
The holomorphic Virasoro primaries are labelled by a pair of positive integers $(r,s)$ as
\begin{align}
	\phi_{(r,s)}(z)~,\quad 1\leq r\leq m-1~,\quad 1\leq s\leq m~,\quad (r,s) \cong (m-r,m+1-s)~,
\end{align}
and have the following scaling dimensions
\begin{equation}\label{VirasoroPrimaries}
	h_{r,s}= \frac{\bigl( (m+1)r-ms \bigr)^2-1}{4m (m+1)}~.
\end{equation}
The (holomorphic) fusion rules read
\begin{align}\label{fusionrulesholo}
	\phi_{(r,s)} \times \phi_{(r',s')} = 
	\sum_{\substack{r''=|r-r'|+1 \\ r+r'+r'' {\rm odd}}}^{r_{\rm max}} \,\,
	\sum_{\substack{s''=|s-s'|+1 \\ s+s'+s'' {\rm odd}}}^{s_{\rm max}} \
	\phi_{(r'',s'')}~,
\end{align}
where 	$r_{\rm max}={\rm min}(r+r'-1,\;2m-r-r'-1)$ and $s_{\rm max}={\rm min}(s+s'-1,\;2m-s-s'+1)$.

The diagonal minimal models are obtained by gluing a copy of the holomorphic sector with its anti-holomorphic counterpart, so that the physical spectrum of the theory contains only scalar Virasoro primaries. The so-obtained diagonal minimal models enjoy a $\mathbb{Z}_2$ symmetry. The charge of a Virasoro primary with labels $(r,s)$ in $\mathcal{M}_{m+1,m}$ under such symmetry can be chosen to be~\cite{Cardy:1986gw,Cappelli:1986hf} (see also \cite{Ruelle:1998zu} for a generic analysis on unitary minimal models)
\begin{align}\label{rsparity}
	\epsilon^{(m)}_{(r,s)}=(-1)^{(m+1) r+m s+1}~.
\end{align}

\subsection{Conformal boundary conditions}\label{subapp:confbcminmods}
Let us consider placing a unitary and diagonal minimal model $\mathcal{M}_{m+1,m}$ on the upper half-plane. The `elementary' conformal boundary conditions for $\mathcal{M}_{m+1,m}$ are parameterised by the highest weight representations of the Virasoro algebra~\cite{Cardy:1984bb,Cardy:2004hm,Cardy:1989ir}
\begin{align}\label{bclabel}
	{\bf a}= (a_1,a_2)_m~,\quad 1\leq a_1\leq m-1~,\quad 1\leq a_2\leq m~,\quad (a_1,a_2) \cong (m-a_1,m+1-a_2)~.
\end{align}
In radial quantization, the elementary conformal boundary condition~\eqref{bclabel} defines a Cardy state~\cite{Cardy:1989ir}
\begin{equation}\label{Cardystate}
	|{{{\bf a}}}\rangle=\sum_{(r,s)}
	\frac{S_{(a_1,a_2)}^{(r,s)}}%
	{\sqrt{S_{(1,1)}^{\;(r,s)}}}\,\ish{r,s}\,~.
\end{equation}
In the equation above, $\ish{r,s}$ is the Ishibashi state in the 
$(r,s)$ sector~\cite{Ishibashi:1988kg} and
\begin{equation}\label{S-matrix}
	S_{(a_1,a_2)}^{(r,s)}=\sqrt{\frac{8}{m (m+1)}}(-1)^{1+a_1s+a_2r}\,
	\sin\left(\tfrac{m+1}{m}\pi a_1r\right)
	\sin\left(\tfrac{m}{m+1}\pi a_2s\right)~,
\end{equation}
are the entries of the modular $S$-matrix. The Ishibashi state $\ish{r,s}$ carries the same parity as $(r,s)$ under the bulk $\mathbb{Z}_2$ symmetry, see eq.~\eqref{rsparity}, so an elementary conformal boundary condition preserves this $\mathbb{Z}_2$ if the corresponding Cardy state contains only $\mathbb{Z}_2$-even Ishibashi states.

Along with the Virasoro primaries that characterize the spectrum of bulk local operators, in the presence of a boundary we can have local excitations located at the boundary. These excitations are captured by the boundary Virasoro primaries, which are highest-weight representations of a single copy of the Virasoro algebra preserved by the boundary. Boundary Virasoro primaries $\psi_{(r,s)}$ are labelled by a pair of integers
\begin{align}\label{bdVirasoroPrimariesrs}
	\psi_{(r,s)}(x)~,\quad 1\leq r\leq m-1~,\quad 1\leq s\leq m~,\quad (r,s) \cong (m-r,m+1-s)~,
\end{align}
and they have scaling dimensions
\begin{align}
	\psi_{(r,s)}(x):~\quad \hD_{r,s}=h_{r,s}~.
\end{align}
Their fusion rules are the same as the holomorphic fusion rules of eq.~\eqref{fusionrulesholo}.

\begin{table}
	\begin{center}
		$\begin{array}{|c|c|c|l|}
			\hline
			\hline
			m &(a_1,a_2)_m&  \mathbb{Z}_2 \text{ - preserving}& \text{Boundary spectrum} \\
			\hline
			\hline
			3 &(1,2)_3 &\checkmark& \hid, \psi_{(1,3)} \\
			\hline
			4 &(1,2)_4 && \hid, \psi_{(1,3)}   \\
			&(1,3)_4 && \hid, \psi_{(1,3)}   \\
			&(2,1)_4 &\checkmark& \hid, \psi_{(3,1)}   \\
			&(2,2)_4 &\checkmark& \hid, \psi_{(1,3)}, \psi_{(3,1)}, \psi_{(3,3)}   \\
			\hline
			5 &(1,2)_5 && \hid, \psi_{(1,3)}   \\
			&(1,3)_5 &\checkmark& \hid, \psi_{(1,3)}, \psi_{(1,5)}  \\
			&(1,4)_5 && \hid, \psi_{(1,3)}   \\
			&(2,1)_5 && \hid, \psi_{(3,1)}   \\
			&(2,2)_5 && \hid, \psi_{(1,3)}, \psi_{(3,1)}, \psi_{(3,3)}   \\
			&(2,3)_5 &\checkmark& \hid, \psi_{(1,3)}, \psi_{(1,5)}, \psi_{(3,1)}, \psi_{(3,3)}, \psi_{(3,5)}  \\
			&(2,4)_5 && \hid, \psi_{(1,3)}, \psi_{(3,1)}, \psi_{(3,3)}   \\
			&(2,5)_5 && \hid, \psi_{(3,1)}   \\
			\hline
			6 &(1,2)_6 && \hid, \psi_{(1,3)}   \\
			&(1,3)_6 && \hid, \psi_{(1,3)}, \psi_{(1,5)}  \\
			&(1,4)_6 && \hid, \psi_{(1,3)}, \psi_{(1,5)}  \\
			&(1,5)_6 && \hid, \psi_{(1,3)}   \\
			&(2,1)_6 && \hid, \psi_{(3,1)}  \\
			&(2,2)_6 && \hid, \psi_{(1,3)}, \psi_{(3,1)}, \psi_{(3,3)}   \\
			&(2,3)_6 && \hid, \psi_{(1,3)}, \psi_{(1,5)}, \psi_{(3,1)}, \psi_{(3,3)}, \psi_{(3,5)}  \\
			&(2,4)_6 && \hid, \psi_{(1,3)}, \psi_{(1,5)}, \psi_{(3,1)}, \psi_{(3,3)}, \psi_{(3,5)}  \\
			&(2,5)_6 && \hid, \psi_{(1,3)}, \psi_{(3,1)}, \psi_{(3,3)}  \\
			&(2,6)_6 && \hid, \psi_{(3,1)}  \\
			&(3,1)_6 &\checkmark& \hid, \psi_{(3,1)}, \psi_{(5,1)}  \\
			&(3,2)_6 &\checkmark& \hid, \psi_{(1,3)}, \psi_{(3,1)}, \psi_{(3,3)}, \psi_{(5,1)}, \psi_{(5,3)}  \\
			&(3,3)_6 &\checkmark& \hid, \psi_{(1,3)}, \psi_{(1,5)}, \psi_{(3,1)}, \psi_{(3,3)}, \psi_{(3,5)}, \psi_{(5,1)},  \psi_{(5,3)},  \psi_{(5,5)}  \\
			\hline
			\hline
		\end{array}$
	\end{center}
\caption{Elementary conformal boundary conditions for diagonal and unitary minimal models with $m \leq 6$. $\mathbb{Z}_2$-preserving boundary conditions are those corresponding to $\mathbb{Z}_2$-even Cardy states.	
	We have not included the conformal b.c. labeled by the identity, nor its $\mathbb{Z}_2$-conjugate $(1, m)_m$: they are always possible and allow only for $\hid$ at the boundary.}
\label{tbl:smallconfbc}
\end{table}

The spectrum of allowed boundary Virasoro primaries in a given conformal b.c. is constrained by modular invariance. With the elementary conformal boundary condition ${\bf a} =(a_1,a_2)$, the allowed boundary Virasoro primaries are those that can appear in the self-OPE of a bulk Virasoro primary labelled by ${\bf a}$. This is expressed in terms of the annulus partition function with boundary condition $\bf a$ as follows:
\begin{align}
	Z_{\bf a a}(\delta)=\sum_{\phi_{(r,s)} \in \phi_{\bf a}\times \phi_{\bf a}}  N^{(r,s)}_{\bc a\bc a}\chi_{(r,s)}(q)~,\quad q\equiv e^{-\pi \delta}~,
\end{align}
where $N^{(r,s)}_{\bc a\bc a}$ is the fusion coefficient of $\phi_{\bf a}\times \phi_{\bf a}$ into $\phi_{(r,s)}$, $\chi_{(r,s)}(q)$ is the (holomorphic) Virasoro character of the $(r,s)$ module and $\delta$ is the width of the annulus. A table with a survey of elementary conformal b.c. for diagonal minimal models with $m\leq 6$ can be found in Table~\ref{tbl:smallconfbc}.

Let us now briefly discuss some aspects of the OPE data. The coefficients $B$, $\hC$ and $C$ are determined via the `sewing' constraints of Lewellen~\cite{Lewellen:1991tb}. For unitary and diagonal minimal models with elementary conformal b.c., these were solved by Runkel in ref.~\cite{Runkel:1998he} (see~\cite{Runkel:1999dz} for the extension to the D-series of minimal models), who has found explicit expressions for such $B$, $\hC$ and $C$ in terms of the so-called `fusion matrices', or F-matrices. Of course the $C$ coefficients are the same as those of the homogeneous minimal model, and these can be efficiently computed via the Coulomb gas formalism of refs.~\cite{Dotsenko:1984nm,Dotsenko:1984ad,Dotsenko:1985hi}.\footnote{In their supplementary material, the authors of ref.~\cite{Esterlis:2016psv} provided a very useful Mathematica notebook that implements these formulae.}

Note that the bulk-boundary coefficient of $\phi_{(r,s)}$ with the boundary identity in the second line of eq.~\eqref{correlatorsuhpCFTconventions} is determined by the Cardy state to be~\cite{Cardy:1991tv,Lewellen:1991tb}
\begin{align}\label{bulkboundaryidentity}
	B_{(r,s)}^{{\bf a} \,(1,1)} = \frac{S_{(a_1,a_2)}^{(r,s)} \sqrt{S_{(1,1)}^{(1,1)}}}{S_{(a_1,a_2)}^{(1,1)} \sqrt{S_{(1,1)}^{(r,s)}}}~.
\end{align}

\subsection{\texorpdfstring{The $\phi_{(1,3)}$ and $\psi_{(1,3)}$ operators}{The phi13 and psi13 operators}}
In the main text we study the deformation of the minimal models by the $\phi_{(1,3)}$ operator. In the remainder of this review we will therefore study its properties and explicitly compute several correlation functions involving this operator. We are also interested in the near-boundary operator $\psi_{(1,3)}$. We note first of all that
\begin{equation}
	\Delta_{1,3} = 2h_{1,3} = 2\frac{m-1}{m+1}~.
\end{equation}

\paragraph{Bulk OPE}
The bulk self-OPE of $\phi_{(1,3)}$ reads (see eq.~\eqref{fusionrulesholo})
\begin{align}
	\phi_{(1,3)}(z,\bar{z})  \phi_{(1,3)}(0,0) &= \frac{\hid}{|z|^{2\D_{1,3}}} +\text{desc.}&& m=3\nonumber\\
	\phi_{(1,3)}(z,\bar{z})  \phi_{(1,3)}(0,0) &= \frac{\id}{|z|^{2\D_{1,3}}} + \frac{C_{(1,3)(1,3)}{}^{(1,3)}}{|z|^{\D_{1,3}}}\phi_{(1,3)}(0,0)+\text{desc.}&& m=4\nonumber\\
	\phi_{(1,3)}(z,\bar{z})  \phi_{(1,3)}(0,0) &= \frac{\id}{|z|^{2\D_{1,3}}} + \frac{C_{(1,3)(1,3)}{}^{(1,3)}}{|z|^{\D_{1,3}}}\phi_{(1,3)}(0,0)+\frac{C_{(1,3)(1,3)}{}^{(1,5)}}{|z|^{2\D_{1,3}-\D_{1,5}}}\phi_{(1,5)}(0,0)+\text{desc.}&& m\geq 5~,
\end{align}
up to Virasoro descendants.  The OPE coefficients are found to be
\begin{align}\label{bulkOPEcoeff}
	C_{(1,3)(1,3)(1,3)} &= -\frac{\Gamma \left(\frac{2}{m+1}\right) \Gamma \left(\frac{1-m}{m+1}\right) \Gamma \left(\frac{3 m-1}{m+1}\right)}{\Gamma \left(\frac{2-2 m}{m+1}\right) \Gamma \left(\frac{2 m}{m+1}\right)^2}\sqrt{-\frac{\Gamma \left(\frac{2-m}{m+1}\right) \Gamma \left(\frac{m}{m+1}\right)^3}{\Gamma \left(\frac{1}{m+1}\right)^3 \Gamma \left(\frac{2 m-1}{m+1}\right)}}~,&&\text{for } m>3~\nonumber\\
	&=\frac{4}{\sqrt{3}}-\frac{4 \sqrt{3}}{m}-\frac{8}{\sqrt{3} m^2}+\frac{32 \sqrt{3} \zeta (3)}{m^3}+O(m^{-4})~,\nonumber\\
	C_{(1,3)(1,3)(1,5)}&=\frac{m-1}{3 m-1}\sqrt{\frac{\Gamma \left(\frac{1-2 m}{m+1}\right) \Gamma \left(\frac{2-m}{m+1}\right) \Gamma \left(\frac{m}{m+1}\right) \Gamma \left(\frac{4 m-1}{m+1}\right)}{\Gamma \left(\frac{1}{m+1}\right) \Gamma \left(\frac{2-3 m}{m+1}\right) \Gamma \left(\frac{3 m}{m+1}\right) \Gamma \left(\frac{2 m-1}{m+1}\right)}}~,&& \text{for } m>4~\nonumber\\
	&=\frac{\sqrt{5}}{3}-\frac{7 \sqrt{5}}{9 m}-\frac{53 \sqrt{5}}{54 m^2}+\frac{\sqrt{5} (2592 \zeta (3)-691)}{324 m^3}+O(m^{-4})~.&&
\end{align}

\paragraph{Bulk-boundary OPE}

The bulk-boundary OPE of $\phi_{(1,3)}$ reads
\begin{align}\label{bOPEphi13}
	\phi_{(1,3)}(x+i y,x-i y) &=\frac{B_{(1,3)}^{{\bf a}\,(1,1)}}{(2y)^{\D_{1,3}}}\hid+\text{desc.}&& m=3\nonumber\\
	\phi_{(1,3)}(x+i y,x-i y) &=\frac{B_{(1,3)}^{{\bf a}\,(1,1)}}{(2y)^{\D_{1,3}}}\hid + \frac{B_{(1,3)}^{{\bf a}\,(1,3)}}{(2y)^{\D_{1,3}-h_{1,3}}}\psi_{(1,3)}(x)+\text{desc.}&& m=4\nonumber\\
	\phi_{(1,3)}(x+i y,x-i y) &=\frac{B_{(1,3)}^{{\bf a}\,(1,1)}}{(2y)^{\D_{1,3}}}\hid + \frac{B_{(1,3)}^{{\bf a}\,(1,3)}}{(2y)^{\D_{1,3}-h_{1,3}}}\psi_{(1,3)}(x)+\frac{B_{(1,3)}^{{\bf a}\,(1,5)}}{(2y)^{\D_{1,3}-h_{1,5}}}\psi_{(1,5)}(x)+\text{desc.}&& m\geq5
\end{align}
up to boundary Virasoro descendants. The bulk-boundary OPE coefficient $B_{(1,3)}^{{\bf a}\,(1,1)}$ is given in eq.~\eqref{B1315}. The squared coefficients $(B_{(1,3)}^{{\bf a}\,(1,3)})^2$ and $(B_{(1,3)}^{{\bf a}\,(1,5)})^2$ are determined by the bulk-boundary crossing symmetry and are reported in section~\ref{bulkbdcross}.

\subsection{\texorpdfstring{The bulk two-point function of $\phi_{(1,3)}$}{The bulk two-point function of phi13}}\label{sec:phi13twoptsol}
We start from the bulk two-point function of $\phi_{(1,3)}$ on the upper half-plane:
\begin{align}
	\langle \phi_{(1,3)}(x_1+ i y_1,x_1- i y_1)\phi_{(1,3)}(x_2+ i y_2, x_2- i y_2)\rangle_{\uhp}~.
\end{align}
To compute such a correlator, we employ the method of images and solve
\begin{align}\label{thirdgen}
	\left({\mathcal L}_{-3}^{(z_4)}-\frac{2}{h_{1,3}+2}{\mathcal L}_{-1}{\mathcal L}_{-2}^{(z_4)}+\frac{1}{(h_{1,3}+1)(h_{1,3}+2)}{\mathcal L}_{-1}^3\right)\langle \phi_{(1,3)}(z_1)\phi_{(1,3)}(z_2)\phi_{(1,3)}(z_3)\phi_{(1,3)}(z_4)\rangle=0~,
\end{align}
where ${\cal L}_{-n}^{(\cdot)}$ is the differential operator defined in eq.~\eqref{calLdiffopLn} and ${\cal L}_{-1}=\partial_{z_4}$. This holomorphic correlator takes the following form
\begin{align}
	\langle \phi_{(1,3)}(z_1)\phi_{(1,3)}(z_2)\phi_{(1,3)}(z_3)\phi_{(1,3)}(z_4)\rangle=\frac{\mathcal{G}(\eta)}{(z_{12}z_{34})^{2h_{1,3}}}~.
\end{align}
The cross-ratio $\eta$ is defined as ($z_{ij}\equiv z_i-z_j$)
\begin{align}\label{eta_def}
	\eta&= \frac{z_{12}z_{34}}{z_{13}z_{24}}=-\frac{4 y_1 y_2}{(y_1-y_2)^2+(x_1-x_2)^2}~.
\end{align}
The differential equation for $\mathcal{G}(\eta)$ reads
\begin{equation}
	\begin{split}
		0&=\eta ^2 (\eta -1)^3 (m+1)^3 \mathcal{G}'''(\eta )+2 \eta  (\eta -1)^2 (m+1)^2 (\eta  (m+3)+m-3) \mathcal{G}''(\eta )\\
		&+2 (\eta -1) (m+1) \left(\eta  \left(m^2+m-6\right)+\eta ^2 (m (3-2 m)+3)-3 m+3\right) \mathcal{G}'(\eta )\\
		&+4 (\eta -2) \eta (m-1)^2  m\,\mathcal{G}(\eta)~.
	\end{split}
\end{equation}
In order to solve this equation, it is convenient to define another function
\begin{align}\label{GtoGtilde}
	\mathcal{G}(\eta) =\tilde{ \mathcal{G}}\left(\tieta\equiv\frac{\eta^2}{1-\eta}\right)~.
\end{align}
The cross-ratio $\tieta$ is real and positive in the physical region. In fact, in terms of the cross-ratio $\xi$ defined in eq.~\eqref{eta_def_full} we have that
\begin{align}\label{tietadef}
	\tieta= \xi^{-1}(1+\xi)^{-1}~=\frac{16 y_1^2 y_2^2}{\left((y_1-y_2)^2+(x_1-x_2)^2\right) \left((y_1+y_2)^2+(x_1-x_2)^2\right)}~.
\end{align}
There are three independent solutions to this equation, corresponding to the boundary operators $\hid, \psi_{(1,3)}$ and $\psi_{(1,5)}$, see e.g. eq.~\eqref{bOPEphi13}. The three Virasoro blocks at finite $m$ read 
\begin{align}\label{phi13twoptblocks}
	V_{(1,1)}(\tilde\eta)&=\, _3F_2\left(\frac{1-m}{m+1},\frac{2m}{m+1},-\frac{2m-2}{m+1};\frac{3+m}{2m+2},\frac{2-m}{m+1};-\frac{\tieta}{4}\right)~,\nonumber\\
	V_{(1,3)}(\tilde\eta)&=\tieta^{{h_{1,3}}/{2}} \, _3F_2\left(\frac{5 m-1}{2m+2},\frac{1-m}{2m+2},-\frac{3m-3}{2m+2};\frac{3-m}{2m+2},\frac{3 m+1}{2m+2};-\frac{\tieta}{4}\right),\nonumber\\
	V_{(1,5)}(\tilde\eta)&=\tieta^{{h_{1,5}}/{2}}\, _3F_2\left(\frac{1}{m+1},\frac{m}{m+1},\frac{4 m-1}{m+1};\frac{3 m}{m+1},\frac{5 m+1}{2m+2};-\frac{\tieta}{4}\right)~.
\end{align}
The final solution is
\begin{align}\label{phi13twopt}
	\langle &\phi_{(1,3)}(x_1 + i y_1,x_1 - i y_1)\phi_{(1,3)}(x_2+ i y_2,x_2 - i y_2)\rangle_{\uhp}=\frac{\tilde{ \mathcal{G}}(\tilde\eta)}{(4y_1 y_2)^{2h_{1,3}}}~,\nonumber\\
	&\tilde{ \mathcal{G}}(\tilde\eta)=(B_{(1,3)}^{{\bf a}\,(1,1)})^2V_{(1,1)}(\tilde\eta)+	(B_{(1,3)}^{{\bf a}\,(1,3)})^2V_{(1,3)}(\tilde\eta)+	(B_{(1,3)}^{{\bf a}\,(1,5)})^2V_{(1,5)}(\tilde\eta)~.
\end{align}

We will fix the coefficients using bulk-boundary crossing symmetry.

\subsubsection{Bulk-boundary crossing}\label{bulkbdcross}
The correlator in eq.~\eqref{phi13twopt} is manifestly written in terms of the boundary data. By expanding it around $\tieta=\infty$ we can read off the bulk-channel data. For a systematic expansion around $\tieta=\infty$ it is convenient to define the following basis of bulk-channel blocks\footnote{These can be obtained upon `inverting' the blocks of eq.~\eqref{phi13twoptblocks} using standard hypergeometric transformations and picking up the correct linear combination of `inverted blocks' consistently with the bulk OPE limit.}
\begin{align}\label{phi13twoptblocksbulk}
	V^{\text{bulk}}_{(1,1)}(\tieta)&=\tieta^{2h_{1,3}}\, _3F_2\left(\frac{1}{m+1},\frac{2-2m}{m+1},\frac{3-3m}{2 m+2};\frac{2}{m+1},\frac{3-3m}{m+1};-\frac{4}{\tieta}\right)~,\nonumber\\
	V^{\text{bulk}}_{(1,3)}(\tieta)&=\tieta^{h_{1,3}}\, _3F_2\left(\frac{m}{m+1},\frac{1-m}{m+1},\frac{1-m}{2 m+2};\frac{2 m}{m+1},\frac{2-2m}{m+1};-\frac{4}{\tieta}\right)~,\nonumber\\
	V^{\text{bulk}}_{(1,5)}(\tieta)&=\tieta^{2h_{1,3}-h_{1,5}} \, _3F_2\left(\frac{2 m}{m+1},\frac{4 m-1}{m+1},\frac{5 m-1}{2 m+2};\frac{4 m}{m+1},\frac{5 m-1}{m+1};-\frac{4}{\tieta}\right)~.
\end{align}
The bulk-boundary crossing is then simply the statement that 
\begin{align}
	&(B_{(1,3)}^{{\bf a}\,(1,1)})^2V_{(1,1)}(\tilde\eta)+	(B_{(1,3)}^{{\bf a}\,(1,3)})^2V_{(1,3)}(\tilde\eta)+	(B_{(1,3)}^{{\bf a}\,(1,5)})^2V_{(1,5)}(\tilde\eta) = \nonumber\\
	& V^{\text{bulk}}_{(1,1)}(\tieta)+C_{(1,3)(1,3)(1,3)}B_{(1,3)}^{{\bf a}\,(1,1)}V^{\text{bulk}}_{(1,3)}(\tieta)+C_{(1,3)(1,3)(1,5)}B_{(1,5)}^{{\bf a}\,(1,1)}V^{\text{bulk}}_{(1,5)}(\tieta)~.
\end{align}
The bulk coefficients are given in eq.~\eqref{bulkOPEcoeff}, while from eq.~\eqref{bulkboundaryidentity} we have that
\begin{align}\label{B1315}
	B_{(1,3)}^{{\bf a}\,(1,1)} &=\left(1+2 \cos \left(\frac{2 \pi  a_2 m}{m+1}\right)\right)\sqrt{\frac{{\sin \left(\frac{\pi }{m}\right) \sin \left(\frac{\pi  m}{m+1}\right)} }{{\sin \left(\frac{\pi }{m}\right) \sin \left(\frac{3 \pi  m}{m+1}\right)}}}\nonumber\\
	&=\sqrt{3}-\frac{2 \pi ^2 \left(2 a_2^2-1\right)}{\sqrt{3} m^2}+\frac{4 \pi ^2 \left(2a_2^2-1\right)}{\sqrt{3} m^3}+O(m^{-4})~,\nonumber\\
	B_{(1,5)}^{{\bf a}\,(1,1)} &= \sin \left(\frac{5 \pi  a_2 m}{m+1}\right) \csc \left(\frac{\pi  a_2 m}{m+1}\right)\sqrt{\frac{{\sin \left(\frac{\pi }{m}\right) \sin \left(\frac{\pi  m}{m+1}\right)}}{{\sin \left(\frac{\pi }{m}\right) \sin \left(\frac{5 \pi  m}{m+1}\right)}}}\nonumber\\
	&=\sqrt{5}+\frac{2 \pi ^2 \sqrt{5} \left(1-2 a_2^2\right)}{m^2}+\frac{4 \pi ^2 \sqrt{5} \left(2 a_2^2-1\right)}{m^3}+O(m^{-4})~,
\end{align}
and the large-$m$ limit is taken assuming a generic conformal b.c.~${\bf a}$, with $a_1,a_2\ll m$.
The coefficients $(B_{(1,3)}^{{\bf a}\,(1,3)})^2$ and $(B_{(1,3)}^{{\bf a}\,(1,5)})^2$ are determined by the bulk-boundary crossing symmetry for any finite $m$.

\paragraph{Solution to bulk-boundary crossing for $m=3$ and $m=4$\\} Let us start from the simplest case, $m=3$, where only the identity is present in both the bulk and the boundary channels. Crossing imposes that $(B_{(1,3)}^{{\bf a}\,(1,1)})^2 =1$ which, upon comparing to the first of eq.~\eqref{B1315}, is true if $a_2=1,2,3$, as it should.

For the case with $m=4$ we have only two operators in both channels: $\hid$ and $\psi_{(1,3)}$ in the boundary channel, $\id$ and $\phi_{(1,3)}$ in the bulk channel. The only remaining coefficient is $(B_{(1,3)}^{{\bf a}\,(1,3)})^2$, which can be non-zero only if $a_2=2,3$ and is fixed by crossing as follows
\begin{align}\label{bulkbd4m}
	(B_{(1,3)}^{{\bf a}\,(1,3)})^2=-\frac{24 \sqrt{2-\frac{2}{\sqrt{5}}} \left(25\ 2^{2/5} \sqrt{\pi\left(9-4 \sqrt{5}\right)  } \Gamma \left(\frac{3}{5}\right) \Gamma \left(\frac{7}{10}\right)-12 \pi  \Gamma \left(-\frac{6}{5}\right)\right)}{3125 \Gamma \left(\frac{3}{5}\right) \Gamma \left(\frac{8}{5}\right) \Gamma \left(\frac{13}{5}\right)}\simeq0.663053~.
\end{align}

\paragraph{Solution to bulk-boundary crossing for general $m>4$\\}
We find
\begin{align}\label{bulkbdgenm}
(B_{(1,3)}^{{\bf a}\,(1,3)})^2&=\frac{\sqrt{\pi } \sec \left(\frac{2 \pi }{m+1}\right) \Gamma \left(\frac{m}{m+1}\right) \Gamma \left(\frac{1-m}{2 m+2}\right) \Gamma \left(\frac{3 m-1}{m+1}\right)}{2 \Gamma \left(\frac{m-1}{m+1}\right) \Gamma \left(\frac{2 m-1}{m+1}\right) \Gamma \left(\frac{3-m}{2 m+2}\right) \Gamma \left(\frac{3 m+1}{2 m+2}\right)}\nonumber\\
&-\frac{1}{\pi}{\sin \left(\frac{2 \pi }{m+1}\right) \Gamma \left(\frac{2 m}{m+1}\right) \Gamma \left(\frac{2}{m+1}-1\right)} C_{(1,3)(1,3)(1,3)}B_{(1,3)}^{{\bf a}\,(1,1)}\nonumber\\
&+\frac{2^{\frac{6}{m+1}-4} \Gamma \left(\frac{2-m}{m+1}\right) \Gamma \left(\frac{3-3 m}{2 m+2}\right) \Gamma \left(\frac{1-m}{2 m+2}\right) \Gamma \left(\frac{m+3}{2 m+2}\right) \Gamma \left(\frac{5 m-1}{2 (m+1)}\right)}{\pi ^{3/2} \left(\sec \left(\frac{\pi }{m+1}\right)+\sec \left(\frac{3 \pi }{m+1}\right)\right) \Gamma \left(\frac{2-2 m}{m+1}\right)^2 \Gamma \left(\frac{2 m}{m+1}\right) \Gamma \left(\frac{3 m+1}{2 m+2}\right)}~\nonumber\\
&~~~~\times \left(\Gamma \left(\frac{1-m}{m+1}\right) \Gamma \left(\frac{2 m}{m+1}\right)-\pi  \csc \left(\frac{4 \pi }{m+1}\right)\right)(B_{(1,3)}^{{\bf a}\,(1,1)})^2~,\nonumber\\
&\nonumber\\
(B_{(1,3)}^{{\bf a}\,(1,5)})^2&=\frac{(m-1) \left(\sec \left(\frac{2 \pi }{m+1}\right)-2\right) \Gamma \left(-\frac{1}{m+1}\right) \Gamma \left(4-\frac{5}{m+1}\right)}{2 (m-2) (3 m-1) \Gamma \left(\frac{m-2}{m+1}\right) \Gamma \left(\frac{3 m}{m+1}\right)}~\nonumber\\
& +\frac{\sqrt{\pi } \sec \left(\frac{2 \pi }{m+1}\right) \Gamma \left(\frac{2 m}{m+1}\right) \Gamma \left(4-\frac{5}{m+1}\right) \Gamma \left(\frac{3}{2}-\frac{1}{m+1}\right)}{2 \Gamma \left(\frac{3 m}{m+1}\right) \Gamma \left(3-\frac{4}{m+1}\right) \Gamma \left(\frac{5}{2}-\frac{2}{m+1}\right) \Gamma \left(\frac{1}{m+1}-\frac{1}{2}\right)}C_{(1,3)(1,3)(1,3)}B_{(1,3)}^{{\bf a}\,(1,1)}\nonumber\\
&-\frac{2^{\frac{6}{m+1}-4} \Gamma \left(\frac{2-m}{m+1}\right) \Gamma \left(\frac{m+3}{2 m+2}\right) \Gamma \left(\frac{4 m-1}{m+1}\right) }{\Gamma \left(\frac{2-2 m}{m+1}\right) \Gamma \left(\frac{1-m}{m+1}\right) \Gamma \left(\frac{2 m}{m+1}\right) \Gamma \left(\frac{3 m}{m+1}\right) \Gamma \left(\frac{5 m+1}{2 m+2}\right)}~\nonumber\\
&~~~~\times \frac{\left(\pi  \csc \left(\frac{4 \pi }{m+1}\right) \sec \left(\frac{\pi }{m+1}\right)+\sec \left(\frac{3 \pi }{m+1}\right) \Gamma \left(\frac{1-m}{m+1}\right) \Gamma \left(\frac{2 m}{m+1}\right)\right)}{\left(\sec \left(\frac{\pi }{m+1}\right)+\sec \left(\frac{3 \pi }{m+1}\right)\right)}(B_{(1,3)}^{{\bf a}\,(1,1)})^2~,
\end{align}
with $C_{(1,3)(1,3)(1,3)}$ and $B_{(1,3)}^{{\bf a}\,(1,1)}$ given in equations~\eqref{bulkOPEcoeff} and~\eqref{B1315} (respectively). These results are consistent with those obtained using F-matrices in ref.~\cite{Runkel:1998he}.

\paragraph{Large-$m$ expansion\\} 
In the large-$m$ expansion (we have in mind a sufficiently generic conformal b.c. ${\bf a}$, with $a_1,a_2\ll m$), the solutions of eq.~\eqref{bulkbdgenm} read
\begin{align}\label{B13B15sqrt}
	(B_{(1,3)}^{{\bf a}\,(1,3)})^2&=\frac{32 \pi ^2 \left(a_2^2-1\right)}{3 m^2}-\frac{160 \pi ^2 \left(a_2^2-1\right)}{3 m^3}-\frac{32 \pi ^2 \left(a_2^2-1\right) \left(\pi ^2 \left(4 a_2^2-7\right)-21\right)}{9 m^4}+O(m^{-5})~,\nonumber\\
	(B_{(1,3)}^{{\bf a}\,(1,5)})^2&=\frac{8\pi^4 \left(a_2^2-1\right) \left(a_2^2-4\right)}{9 m^4}+O(m^{-5})~.
\end{align}

\subsection{Correlator between four \texorpdfstring{$\psi_{(1,3)}$}{psi13}}\label{13threept}
Consider the four-point correlation function of $\psi_{(1,3)}$. The associated differential equation and its solutions where given in section~\ref{sec:phi13twoptsol}. The fusion rules read
\begin{align}\label{psi13fusion}
	\psi_{(1,3)}\times \psi_{(1,3)}&=\hid&& m=3\nonumber\\
	\psi_{(1,3)}\times \psi_{(1,3)}&=\hid+\psi_{(1,3)}&& m=4\nonumber\\
	\psi_{(1,3)}\times \psi_{(1,3)}&=\hid+\psi_{(1,3)}+\psi_{(1,5)}&& m\geq 5~.
\end{align}
The solution is 
\begin{align}\label{psi13fourpt}
	&\langle \psi_{(1,3)}(x_1)\psi_{(1,3)}(x_2)\psi_{(1,3)}(x_3)\psi_{(1,3)}(x_4)\rangle_{\uhp}=\frac{\tilde{ \mathcal{G}}(\tilde\eta)}{(x_{12}x_{34})^{2h_{1,3}}}~,\nonumber\\
	\tilde{ \mathcal{G}}&(\tilde\eta)=V_{(1,1)}(\tilde\eta)+({\hC}_{(1,3)(1,3)}^{{\bf a}(1,3)})^2V_{(1,3)}(\tilde\eta)+	({\hC}_{(1,3)(1,3)}^{{\bf a}(1,5)})^2V_{(1,5)}(\tilde\eta)~.
\end{align}
The form of the $V_{(r,s)}(\tieta)$ is given in eq.~\eqref{phi13twoptblocks}, while $\tieta$ is defined in terms of the standard cross ratio $\eta$ of eq.~\eqref{eta_def_line} as
\begin{align}
	\tieta=\frac{\eta ^2}{1-\eta}~=\frac{x_{12}^2 x_{34}^2}{x_{12}x_{14}  x_{23}x_{24}}~,
\end{align}
which is positive if $x_i > x_{i+1}$. 

\subsubsection{Boundary crossing}
The squared boundary OPE coefficients in eq.~\eqref{psi13fourpt} are fixed by crossing symmetry, which is the statement that (recall the definition in eq.~\eqref{GtoGtilde})
\begin{align}
		\mathcal{G}(\eta)=\left(\frac{1-\eta }{\eta }\right)^{-2 h_{1,3}}	\mathcal{G}(1-\eta)~.
\end{align}
For $m=3$ we have that ${\hC}_{(1,3)(1,3)(1,3)}^{{\bf a}}={\hC}_{(1,3)(1,3)(1,5)}^{{\bf a}}=0$ and boundary crossing symmetry is ensured by the identity block alone. For $m=4$, or for those conformal b.c. that do not allow $\psi_{(1,5)}$ we have that ${\hC}_{(1,3)(1,3)(1,5)}^{{\bf a}}=0$ and crossing is satisfied if
\begin{align}\label{C13131315hm4}
	({\hC}_{(1,3)(1,3)(1,3)}^{{\bf a}})^2=	-\frac{\sqrt{\pi}\,\Gamma \left(3-\frac{4}{m+1}\right) \Gamma \left(\frac{1}{m+1}-\frac{1}{2}\right) \left(\pi  \csc \left(\frac{3 \pi }{m+1}\right)-\Gamma \left(\frac{1}{m+1}\right) \Gamma \left(\frac{m}{m+1}\right)\right)}{\Gamma \left(\frac{1}{m+1}\right) \Gamma \left(\frac{m-1}{m+1}\right) \Gamma \left(2-\frac{3}{m+1}\right) \Gamma \left(\frac{3}{2}-\frac{1}{m+1}\right) \Gamma \left(\frac{2}{m+1}-\frac{1}{2}\right)}~,
\end{align}
which vanishes identically for $m=3$, but is positive otherwise. This result is consistent with the known results from the literature, see e.g. ref.~\cite{Runkel:1998he}.
For $m>4$ we find generically a one-parameter family of solutions to the crossing equation, so that additional correlators (and crossing constraints) are needed in order to compute ${\hC}_{(1,3)(1,3)(1,3)}^{{\bf a}}$ and ${\hC}_{(1,3)(1,3)(1,5)}^{{\bf a}}$.

\subsection{Correlator between two \texorpdfstring{$\psi_{(r,s)}$ and one $\phi_{(1,3)}$}{psirs and one phi13}}\label{app:bdbdbulk3pt}

Next, we compute the boundary-boundary-bulk correlation function between a boundary Virasoro primary $\psi_{(r,s)}$ (assuming it exists) and $\phi_{(1,3)}$,
\begin{align}\label{bdbdphi13gen}
	\langle \psi_{(r,s)}(x_1)\psi_{(r,s)}(x_2)\phi_{(1,3)}(x+i y,x-iy)\rangle_{\uhp}~,\quad x_1>x_2~.
\end{align}
We recall that the scaling dimension of $\psi_{(r,s)}$ is $\D_{r,s} = h_{r,s}$.
The correlator of eq.~\eqref{bdbdphi13gen} satisfies a third order differential equation. To compute such a correlator, we employ the method of images and solve
\begin{align}\label{thirdgengen}
	\left({\mathcal L}_{-3}^{(z_4)}-\frac{2}{h_{1,3}+2}{\mathcal L}_{-1}{\mathcal L}_{-2}^{(z_4)}+\frac{1}{(h_{1,3}+1)(h_{1,3}+2)}{\mathcal L}_{-1}^3\right)\langle \psi_{(r,s)}(z_1)\psi_{(r,s)}(z_2)\phi_{(1,3)}(z_3)\phi_{(1,3)}(z_4)\rangle=0~,
\end{align}
where ${\cal L}_{-n}^{(\cdot)}$ is the differential operator defined in eq.~\eqref{calLdiffopLn} and ${\cal L}_{-1}=\partial_{z_4}$. 
This holomorphic correlator takes the following form:
\begin{align}
	\langle \psi_{(r,s)}(z_1)\psi_{(r,s)}(z_2)\phi_{(1,3)}(z_3)\phi_{(1,3)}(z_4)\rangle=\frac{\mathcal{G}(\eta)}{(z_{12})^{2h_{r,s}} (z_{34})^{2h_{1,3}}}~.
\end{align}
The cross-ratio $\eta$ is defined as
\begin{align}\label{eta_def2}
	\eta&= \frac{z_{12}z_{34}}{z_{13}z_{24}}=\frac{2 i y x_{12}}{(x_1-z) (x_2-z^*)}~.
\end{align}
The differential equation for $\mathcal{G}(\eta)$ reads
\begin{equation}
	\begin{split}
		0&=(\eta -1)^3 \eta ^2 (m+1)^3 \mathcal{G}'''(\eta )+2 (\eta -1)^2 \eta  (m+1)^2 (\eta  (m+3)+m-3)\mathcal{G}''(\eta )\\
		&-(\eta -1) (m+1) \left(12 \eta -2 \eta  m (m+1)+6 m-6\right)\mathcal{G}'(\eta )\\
			&-(\eta -1)\eta ^2 (m+1) \left(m ((r-s) ((m+2) r-m s)-2)+r^2-7\right)\mathcal{G}'(\eta )\\
		&+(\eta -2) \eta  (m-1) (m r-m s+r-1) (m r-m s+r+1)  \mathcal{G}(\eta)~.
	\end{split}
\end{equation}
In order to solve this equation, it is convenient to define another function
\begin{align}
	\mathcal{G}(\eta) =\tilde{ \mathcal{G}}\left(\tieta\equiv\frac{\eta^2}{\eta-1}\right)~.
\end{align}
The cross-ratio $\tieta$ so defined is real and $\tieta \in [0,4]$ (its explicit expression in terms of the coordinates is given in eq.~\eqref{etatildedef}).
There are three independent solutions to this equation corresponding to the boundary operators $\hid, \psi_{(1,3)}$ and $\psi_{(1,5)}$ (see the fusion rules in eq.~\eqref{psi13fusion}). These Virasoro blocks at finite $m$ are found to be
\begin{align}\label{allblocks}
	V_{(1,1)}(\tilde\eta)&=\, _3F_2\left(\frac{1-m}{m+1},\frac{1+m q-r}{m+1},\frac{1-m q+r}{m+1};\frac{2-m}{m+1},\frac{m+3}{2 m+2};\frac{\tieta}{4}\right)~,\nonumber\\
	V_{(1,3)}(\tilde\eta)&=\tieta^{h_{1,3}/2} \, _3F_2\left(\frac{1-m}{2m+2},\frac{m+2 m q-2 r+1}{2m+2},\frac{m-2 m q+2 r+1}{2m+2};\frac{3-m}{2m+2},\frac{3 m+1}{2m+2};\frac{\tieta}{4}\right),\nonumber\\
	V_{(1,5)}(\tilde\eta)&=\tieta^{h_{1,5}/2}\, _3F_2\left(\frac{m}{m+1},\frac{2 m+m q-r}{m+1},\frac{2 m-mq+r}{m+1};\frac{3 m}{m+1},\frac{5 m+1}{2m+2};\frac{\tieta}{4}\right)~,
\end{align}
in terms of the parameter $q$ defined as
\begin{align}
	q\equiv s-r \in \mathbb{Z}~.
\end{align}
The final solution is then
\begin{align}\label{finalcorrectgen}
	\langle \psi_{(r,s)}&(x_1)\psi_{(r,s)}(x_2)\phi_{(1,3)}(x+iy,x-iy)\rangle_{\uhp}=\frac{\tilde{ \mathcal{G}}(\tilde\eta)}{(x_{12})^{2h_{r,s}} (2y)^{2h_{1,3}}}~,\nonumber\\
	&\tilde{ \mathcal{G}}(\tilde\eta)=B_{(1,3)}^{{\bf a}\,(1,1)}V_{(1,1)}(\tilde\eta)+	\alpha_{(1,3)}V_{(1,3)}(\tilde\eta)+	\alpha_{(1,5)}V_{(1,5)}(\tilde\eta)~,
\end{align}
where $B_{(1,3)}^{{\bf a}\,(1,1)}$ is the one-point function coefficient of $\phi_{(1,3)}$ computed in eq.~\eqref{B1315} and we denoted
\begin{align}
	\alpha_{(1,3)}\equiv {\hC}_{(r,s)(r,s)(1,3)}^{{\bf a}}B_{(1,3)}^{{\bf a}\,(1,3)}~,\quad \alpha_{(1,5)}\equiv {\hC}_{(r,s)(r,s)(1,5)}^{{\bf a}}B_{(1,3)}^{{\bf a}\,(1,5)}~.
\end{align}
Of course some of the three-point function coefficients ${\hC}_{(r,s)(r,s)(1,3)}^{{\bf a}},~{\hC}_{(r,s)(r,s)(1,5)}^{{\bf a}}$ can be zero in a given conformal b.c. Instead, $B_{(1,3)}^{{\bf a}\,(1,3)},~B_{(1,3)}^{{\bf a}\,(1,5)}$ are generically non-zero (see e.g. eq.~\eqref{B13B15sqrt}).

\subsubsection{Removing unphysical singularities}\label{app:removingunphdiv}
The r.h.s. of eq.~\eqref{finalcorrectgen} features unphysical singularities when the configuration is such that $\tieta\in [4,\infty]$ ends up on the branch cut of the Virasoro blocks of eq.~\eqref{allblocks}. In order to remove these singularities we must require that $\text{Disc}~\tilde{ \mathcal{G}}=0$ across that cut. This is one of the `sewing constraints' of Lewellen~\cite{Lewellen:1991tb} for the open-open-closed string amplitude on the upper half-plane.  The condition that unphysical singularities in defect-defect-bulk blocks must disappear from the physical correlation functions must hold in higher dimensions as well, and it has been exploited in refs.~\cite{Lauria:2020emq,Herzog:2022jqv} to prove `triviality' of certain free theory conformal defects and in refs.~\cite{Behan:2020nsf,Behan:2021tcn} to constrain the space of conformal boundary conditions for a theory of a free massless scalar field. See also refs.~\cite{Nishioka:2022odm,Nishioka:2022qmj} for applications of this idea for $O(N)$ models with boundaries of defects and \cite{Levine:2023ywq} for applications in QFTs in AdS.

\paragraph{Case with $m=3$\\}
The only non-identity boundary primary allowed to appear in any conformal b.c is $\psi_{(1,3)}$  with self-OPE
\begin{align}
	\psi_{(1,3)}\times \psi_{(1,3)}&= \hid.
\end{align}
For this special case, eq.~\eqref{finalcorrectgen} contains only the identity block $V_{(1,1)}$, which becomes a polynomial in $\tieta$. Hence $\text{Disc}~\tilde{ \mathcal{G}}=0$ and all is well.

\paragraph{Case with $m=4$\\}
This is the first non-trivial case. The non-identity boundary primaries allowed to appear in any conformal b.c are $\psi_{(1,3)}, \psi_{(3,1)}, \psi_{(3,3)}$ (see Table~\ref{tbl:smallconfbc})
with self-OPEs
\begin{align}\label{tricOPE2}
	\psi_{(1,3)}\times \psi_{(1,3)}&=\hid+\psi_{(1,3)}~,\nonumber\\
	\psi_{(3,1)}\times \psi_{(3,1)}&=\hid~,\nonumber\\
	\psi_{(3,3)}\times \psi_{(3,3)}&=\hid+\psi_{(1,3)}~.
\end{align}
We have to inspect the singularity structure of the correlator~\eqref{finalcorrectgen} for each of the possible allowed boundary operators. When $\psi_{(r,s)}=\psi_{(3,1)}$ there is only $V_{(1,1)}$, and it is easy to check that $\text{Disc}~\tilde{ \mathcal{G}}=0$ automatically. For $\psi_{(1,3)}$ and $\psi_{(3,3)}$ this condition implies that
\begin{align}\label{OPEcoeffnosing}
\alpha_{(1,3)}= B_{(1,3)}^{{\bf a}\,(1,1)}\frac{ \Gamma \left(-\frac{3}{10}\right) \Gamma \left(\frac{7}{10}\right) \sec \left(\pi  \left(r-\frac{4 s}{5}\right)\right)}{4 \sqrt[5]{2} \Gamma \left(r-\frac{4 s}{5}+\frac{1}{5}\right) \Gamma \left(-r+\frac{4 s}{5}+\frac{1}{5}\right)}~.
\end{align}
As a check, the case of $r=1$, $s=3$ in the equation above is consistent with the (squared) OPE coefficient obtained from bulk-boundary crossing in eq.~\eqref{bulkbdgenm} times the (squared) OPE coefficient obtained from boundary crossing in eq.~\eqref{C13131315hm4}, as it should. One can also check that eq.~\eqref{OPEcoeffnosing} is consistent with the known results obtained from F-matrices in ref.~\cite{Runkel:1998he}.

\paragraph{General $m$, only $V_{(1,1)}$ and $V_{(1,3)}$\\}
 Consider a conformal b.c.~${\bf a}=(a_1,a_2)_m$ for which the following two properties are satisfied:
\begin{enumerate}
	\item The boundary primary $\psi_{(r,s)}$ is allowed to exist;
	\item $\alpha_{(1,3)}\equiv {\hC}_{(r,s)(r,s)(1,3)}^{{\bf a}}B_{(1,3)}^{{\bf a}\,(1,3)}\neq0$, but $\alpha_{(1,5)}\equiv {\hC}_{(r,s)(r,s)(1,5)}^{{\bf a}}B_{(1,3)}^{{\bf a}\,(1,5)}= 0$ (either because $\psi_{(1,5)}$ does not exist, or ${\hC}_{(r,s)(r,s)(1,5)}^{{\bf a}}=0$) or it is subleading with respect to $\alpha_{(1,3)}$ in the large-$m$ limit,
\end{enumerate}
then the condition $\text{Disc}~\tilde{ \mathcal{G}}=0$ requires that
\begin{align}\label{largemalpha13exact}
\alpha_{(1,3)}=-B_{(1,3)}^{{\bf a}\,(1,1)}\frac{ \Gamma \left(\frac{2-m}{m+1}\right) \Gamma \left(\frac{1-m}{2 m+2}\right) \Gamma \left(\frac{m+3}{2 m+2}\right) \Gamma \left(\frac{2 r m-2 s m+m+2 r+1}{2 m+2}\right) \Gamma \left(\frac{-2 r m+2 s m+m-2 r+1}{2 m+2}\right)}{2^{\frac{m-1}{m+1}}\Gamma \left(\frac{1-m}{m+1}\right) \Gamma \left(\frac{3-m}{2 m+2}\right) \Gamma \left(\frac{3 m+1}{2 m+2}\right) \Gamma \left(\frac{m r+r-m s+1}{m+1}\right) \Gamma \left(\frac{m s+1-m r-r}{m+1}\right)}~.
\end{align}
This is consistent with earlier results in the literature computed from F-matrices, see~\cite{Runkel:1998he}.
For finite $(r,s)$, the large-$m$ expansion of the former result gives  (in taking this limit we take $a_1,a_2\ll m$)
\begin{align}\label{largemalpha13}
	\alpha_{(1,3)}=\frac{2 \pi  (r-s) (s-\text{sgn}(r-s))}{\sqrt{3} m}+\frac{2 \pi  (s+1) \left(s+2 (s-r) H_{s-r}-1\right)}{\sqrt{3} m^2}+O(m^{-3})~,
\end{align}
with $H_{n}$ the $n^{\text{th}}$ harmonic number. 

\paragraph{General $m$, all blocks\\}
For a generic choice of conformal b.c. and of the external field $\psi_{(r,s)}$ we find a one-parameter family of solutions to the condition that $\text{Disc}~\tilde{ \mathcal{G}}=0$, so that only a linear combination of $\alpha_{(1,3)}$ and $\alpha_{(1,5)}$ gets fixed. In order to determine $\alpha_{(1,3)}$ and $\alpha_{(1,5)}$ completely, one has to then consider the full set of `sewing' constraints.

\subsection{The \texorpdfstring{$\phi_{(3,1)}$}{phi31} and \texorpdfstring{$\psi_{(3,1)}$}{psi31} operators}

We are also interested in the irrelevant deformation of the minimal models generated by the $\phi_{(3,1)}$ operator with dimension
\begin{equation}
	\Delta_{3,1}= 2 h_{3,1}= 2+4/m~.
\end{equation}

\paragraph{Bulk OPE} The bulk self-OPE of $\phi_{(3,1)}$ reads (see eq.~\eqref{fusionrulesholo})
\begin{align}
	\phi_{(3,1)}(z,\bar{z})  \phi_{(3,1)}(0,0) &= \frac{\id}{|z|^{2\D_{3,1}}} +\text{desc.}&& m=4\nonumber\\
	\phi_{(3,1)}(z,\bar{z})  \phi_{(3,1)}(0,0) &= \frac{\id}{|z|^{2\D_{3,1}}} + \frac{C_{(3,1)(3,1)}{}^{(3,1)}}{|z|^{\D_{3,1}}}\phi_{(3,1)}(0,0)+\text{desc.}&& m=5\nonumber\\
	\phi_{(3,1)}(z,\bar{z})  \phi_{(3,1)}(0,0) &= \frac{\id}{|z|^{2\D_{3,1}}} + \frac{C_{(3,1)(3,1)}{}^{(3,1)}}{|z|^{\D_{3,1}}}\phi_{(3,1)}(0,0)+\frac{C_{(3,1)(3,1)}{}^{(5,1)}}{|z|^{2\D_{3,1}-\D_{5,1}}}\phi_{(5,1)}(0,0)+\text{desc.}&& m\geq 6.
\end{align}
The bulk OPE coefficients are found to be~\cite{Dotsenko:1984nm,Dotsenko:1984ad,Dotsenko:1985hi}
\begin{align}\label{bulkOPEcoeffnew}
	C_{(3,1)(3,1)(3,1)} &=\frac{2^{\frac{4}{m}+2} (m+2) \Gamma \left(\frac{3}{2}+\frac{2}{m}\right) \Gamma \left(-\frac{m+2}{2 m}\right) }{m^2 \Gamma \left(\frac{3}{2}+\frac{1}{m}\right) \Gamma \left(-\frac{m+4}{2 m}\right)}\sqrt{\frac{\Gamma \left(\frac{1}{m}\right) \Gamma \left(-\frac{m+3}{m}\right)}{\Gamma \left(2+\frac{3}{m}\right) \Gamma \left(\frac{m-1}{m}\right)}}~,&&\text{for } m>4~\nonumber\\
	&=\frac{4}{\sqrt{3}}+\frac{4 \sqrt{3}}{m}-\frac{20}{\sqrt{3} m^2}+\frac{28-96 \zeta (3)}{m^3}+O(m^{-4})~,\nonumber\\
	C_{(3,1)(3,1)(5,1)}&=-\frac{m (m+2) \Gamma \left(\frac{1}{m}\right) \Gamma \left(-\frac{m+3}{m}\right)}{(2 m+3) (3 m+4) \Gamma \left(2+\frac{3}{m}\right) \Gamma \left(-\frac{1}{m}\right)}\sqrt{\frac{\Gamma \left(4+\frac{5}{m}\right) \Gamma \left(\frac{m-1}{m}\right)}{\Gamma \left(-3-\frac{5}{m}\right) \Gamma \left(\frac{1}{m}\right)}}~,&& \text{for } m>5~\nonumber\\
	&=\frac{\sqrt{5}}{3}+\frac{7 \sqrt{5}}{9 m}-\frac{95 \sqrt{5}}{54 m^2}+\frac{\sqrt{5} (1579-2592 \zeta (3))}{324 m^3}+O(m^{-4})~.
\end{align}

\paragraph{Bulk-boundary OPE} The bulk-boundary OPE of $\phi_{(3,1)}$ reads
\begin{align}\label{bOPEphi31}
	\phi_{(3,1)}(x+i y,x-i y) &=\frac{B_{(3,1)}^{{\bf a}\,(1,1)}}{(2y)^{\D_{3,1}}}\hid+\text{desc.}&& m=4\nonumber\\
	\phi_{(3,1)}(x+i y,x-i y) &=\frac{B_{(3,1)}^{{\bf a}\,(1,1)}}{(2y)^{\D_{3,1}}}\hid + \frac{B_{(3,1)}^{{\bf a}\,(3,1)}}{(2y)^{\D_{3,1}-h_{3,1}}}\psi_{(3,1)}(x)+\text{desc.}&& m=5\nonumber\\
	\phi_{(3,1)}(x+i y,x-i y) &=\frac{B_{(3,1)}^{{\bf a}\,(1,1)}}{(2y)^{\D_{3,1}}}\hid + \frac{B_{(3,1)}^{{\bf a}\,(3,1)}}{(2y)^{\D_{3,1}-h_{3,1}}}\psi_{(3,1)}(x)+\frac{B_{(3,1)}^{{\bf a}\,(5,1)}}{(2y)^{\D_{3,1}-h_{5,1}}}\psi_{(5,1)}(x)+\text{desc.}&& m\geq6.
\end{align}
The bulk-boundary OPE coefficient $B_{(3,1)}^{{\bf a}\,(1,1)}$ is given in eq.~\eqref{B3151}. The squared coefficients $(B_{(3,1)}^{{\bf a}\,(3,1)})^2$ and $(B_{(3,1)}^{{\bf a}\,(5,1)})^2$ are determined by the bulk-boundary crossing symmetry, as we show in the next section.

\subsection{The bulk two-point function of \texorpdfstring{$\phi_{(3,1)}$}{phi31}}\label{sec:phi31twoptsol}
We start from the bulk two-point function of $\phi_{(3,1)}$ on the upper half-plane.
To compute such a correlator, we employ the method of images and solve
\begin{align}
	\left({\mathcal L}_{-3}^{(z_4)}-\frac{2}{h_{3,1}+2}{\mathcal L}_{-1}{\mathcal L}_{-2}^{(z_4)}+\frac{1}{(h_{3,1}+1)(h_{3,1}+2)}{\mathcal L}_{-1}^3\right)\langle \phi_{(3,1)}(z_1)\phi_{(3,1)}(z_2)\phi_{(3,1)}(z_3)\phi_{(3,1)}(z_4)\rangle=0~,
\end{align}
where ${\cal L}_{-n}^{(\cdot)}$ is the differential operator defined in eq.~\eqref{calLdiffopLn} and ${\cal L}_{-1}=\partial_{z_4}$. This holomorphic correlator takes the following form
\begin{align}
	\langle \phi_{(3,1)}(z_1)\phi_{(3,1)}(z_2)\phi_{(3,1)}(z_3)\phi_{(3,1)}(z_4)\rangle=\frac{\mathcal{G}(\eta)}{(z_{12}z_{34})^{2h_{3,1}}}~.
\end{align}
The cross-ratio $\eta$ is defined as in eq.~\eqref{eta_def}. We have three independent solutions to this equation, corresponding to the boundary operators $\hid, \psi_{(3,1)}$ and $\psi_{(5,1)}$. In terms of the function $\tilde{\mathcal{G}}$ defined as in eq.~\eqref{GtoGtilde} with cross-ratio \eqref{tietadef}, the three Virasoro blocks at finite $m$ read 
\begin{align}\label{phi31twoptblocks}
	V_{(1,1)}(\tieta)&=\, _3F_2\left(-\frac{4}{m}-2,-\frac{2}{m}-1,\frac{2}{m}+2;-\frac{3}{m}-1,\frac{1}{2}-\frac{1}{m};-\frac{\tieta}{4}\right)~,\nonumber\\
	V_{(3,1)}(\tieta)&=\tieta^{{ h_{3,1}}/{2}}\, _3F_2\left(-\frac{3}{m}-\frac{3}{2},-\frac{1}{m}-\frac{1}{2},\frac{3}{m}+\frac{5}{2};-\frac{2}{m}-\frac{1}{2},\frac{1}{m}+\frac{3}{2};-\frac{\tieta}{4}\right)~,\nonumber\\
	V_{(5,1)}(\tieta)&=\tieta^{{ h_{5,1}}/{2}} \, _3F_2\left(\frac{1}{m}+1,\frac{5}{m}+4,-\frac{1}{m};\frac{2}{m}+\frac{5}{2},\frac{3}{m}+3;-\frac{\tieta}{4}\right)~.
\end{align}
The final solution is
\begin{align}\label{phi31twopt}
	\langle &\phi_{(3,1)}(x_1 + i y_1,x_1 - i y_1)\phi_{(3,1)}(x_2+ i y_2,x_2 - i y_2)\rangle_{\uhp}=\frac{\tilde{ \mathcal{G}}(\tilde\eta)}{(4y_1 y_2)^{2h_{3,1}}}~,\nonumber\\
	&\tilde{ \mathcal{G}}(\tilde\eta)=(B_{(3,1)}^{{\bf a}\,(1,1)})^2V_{(1,1)}(\tilde\eta)+	(B_{(3,1)}^{{\bf a}\,(3,1)})^2V_{(3,1)}(\tilde\eta)+	(B_{(3,1)}^{{\bf a}\,(5,1)})^2V_{(5,1)}(\tilde\eta)~.
\end{align}
In the bulk channel we find
\begin{align}\label{phi31twoptblocksbulk}
	V^{\text{bulk}}_{(1,1)}(\tieta)&=\tieta^{2h_{3,1}}\, _3F_2\left(-\frac{4}{m}-2,-\frac{3}{m}-\frac{3}{2},-\frac{1}{m};-\frac{6}{m}-3,-\frac{2}{m};-\frac{4}{\tieta}\right)~,\nonumber\\
	V^{\text{bulk}}_{(3,1)}(\tieta)&=\tieta^{h_{3,1}}\, _3F_2\left(\-\frac{2}{m}-1,-\frac{1}{m}-\frac{1}{2},\frac{1}{m}+1;-\frac{4}{m}-2,\frac{2}{m}+2;-\frac{4}{\tieta}\right)~,\nonumber\\
	V^{\text{bulk}}_{(5,1)}(\tieta)&=\tieta^{2h_{3,1}-h_{5,1}} \, _3F_2\left(\frac{2}{m}+2,\frac{3}{m}+\frac{5}{2},\frac{5}{m}+4;\frac{4}{m}+4,\frac{6}{m}+5;-\frac{4}{\tieta}\right)~.
\end{align}
Bulk-boundary crossing symmetry requires that
\begin{align}
	&(B_{(3,1)}^{{\bf a}\,(1,1)})^2V_{(1,1)}(\tilde\eta)+	(B_{(3,1)}^{{\bf a}\,(3,1)})^2V_{(3,1)}(\tilde\eta)+	(B_{(3,1)}^{{\bf a}\,(5,1)})^2V_{(5,1)}(\tilde\eta) = \nonumber\\
	& V^{\text{bulk}}_{(1,1)}(\tieta)+C_{(3,1)(3,1)(3,1)}B_{(3,1)}^{{\bf a}\,(1,1)}V^{\text{bulk}}_{(3,1)}(\tieta)+C_{(3,1)(3,1)(5,1)}B_{(5,1)}^{{\bf a}\,(1,1)}V^{\text{bulk}}_{(5,1)}(\tieta)~.
\end{align}
From eq.~\eqref{bulkboundaryidentity} we have that
\begin{align}\label{B3151}
	B_{(3,1)}^{{\bf a}\,(1,1)} &=\left(1+2 \cos \left(\frac{2 \pi  a_1 (m+1)}{m}\right)\right)\sqrt{\frac{\sin \left(\frac{\pi }{m}\right) \sin \left(\frac{\pi  m}{m+1}\right)}{\sin \left(\frac{3 \pi }{m}\right) \sin \left(\frac{\pi  m}{m+1}\right)}}\nonumber\\
	&=\sqrt{3}+\frac{2 \pi ^2 \left(1-2 a_1^2\right)}{\sqrt{3} m^2}+O(m^{-4})~,\nonumber\\
	B_{(5,1)}^{{\bf a}\,(1,1)} &=\sin \left(\frac{5 \pi  a_1 (m+1)}{m}\right) \csc \left(\frac{\pi  a_1 (m+1)}{m}\right)\sqrt{\frac{\sin \left(\frac{\pi }{m}\right) \sin \left(\frac{\pi  m}{m+1}\right)}{\sin \left(\frac{5 \pi }{m}\right) \sin \left(\frac{\pi  m}{m+1}\right)}} \nonumber\\
	&=\sqrt{5}+\frac{2 \sqrt{5} \pi ^2 \left(1-2 a_1^2\right)}{m^2}+O(m^{-4}).
\end{align}
The coefficients $(B_{(3,1)}^{{\bf a}\,(3,1)})^2$ and $(B_{(3,1)}^{{\bf a}\,(5,1)})^2$ are determined by the bulk-boundary crossing symmetry for any finite $m$.

\subsubsection{Solution to bulk-boundary crossing}
 In the simplest cases of $m=4$, only the identity is present in both the bulk and the boundary channels. Crossing imposes that $(B_{(3,1)}^{{\bf a}\,(1,1)})^2 =1$, which upon comparing to eq.~\eqref{B3151}, happens if $a_1=1,2$, as it should.  For the case with $m=5$ we have only two operators in both channels: $\hid$ and $\psi_{(3,1)}$ in the boundary channel, $\id$ and $\phi_{(3,1)}$ in the bulk channel. The only remaining coefficient is $(B_{(3,1)}^{{\bf a}\,(3,1)})^2$, which can be non-zero only if $a_1=2$, and it is fixed by crossing symmetry as follows:
\begin{align}\label{bulkbd4mmm}
	(B_{(3,1)}^{{\bf a}\,(3,1)})^2&=\frac{3 \Gamma \left(-\frac{14}{5}\right) \Gamma \left(\frac{19}{10}\right)}{3125\ 2^{7/10} \pi ^{3/2} \Gamma \left(-\frac{8}{5}\right)}\nonumber\\
	\times &\left(10584 \sqrt{5-\sqrt{5}} \Gamma \left(-\frac{14}{5}\right) \Gamma \left(-\frac{7}{5}\right)-15625 \sqrt[10]{2} \left(3 \sqrt{5}-7\right) \sqrt{\pi}\,\Gamma \left(\frac{13}{10}\right)\right) \simeq2.28878~.
\end{align}
For $m>5$ we have 
\begin{align}\label{bulkbdgenm31}
	(B_{(3,1)}^{{\bf a}\,(3,1)})^2&=C_{(3,1)(3,1)(3,1)}B_{(3,1)}^{{\bf a}\,(1,1)}\nonumber\\
	&+(B_{(3,1)}^{{\bf a}\,(1,1)})^2\frac{\left(\csc \left(\frac{2 \pi }{m}\right)+\csc \left(\frac{4 \pi }{m}\right)\right) \Gamma \left(\frac{5}{2}+\frac{3}{m}\right) \Gamma \left(-\frac{m+2}{2 m}\right)^2 \Gamma \left(-\frac{2 (m+3)}{m}\right)}{6 \left(\sec \left(\frac{\pi }{m}\right)+\sec \left(\frac{3 \pi }{m}\right)\right) \Gamma \left(\frac{3}{2}+\frac{1}{m}\right) \Gamma \left(2+\frac{2}{m}\right) \Gamma \left(-\frac{2 (m+2)}{m}\right)^2}~,\nonumber\\
	&\nonumber\\
	(B_{(3,1)}^{{\bf a}\,(5,1)})^2&=C_{(3,1)(3,1)(3,1)}B_{(3,1)}^{{\bf a}\,(1,1)}\frac{\pi  2^{-\frac{4}{m}-3} \sec \left(\frac{2 \pi }{m}\right) \Gamma \left(\frac{3}{2}+\frac{1}{m}\right) \Gamma \left(4+\frac{5}{m}\right)}{\Gamma \left(\frac{3}{2}+\frac{2}{m}\right) \Gamma \left(\frac{5}{2}+\frac{2}{m}\right) \Gamma \left(3+\frac{3}{m}\right) \Gamma \left(-\frac{m+2}{2 m}\right)}\nonumber\\
	&+(B_{(3,1)}^{{\bf a}\,(1,1)})^2\frac{2^{-\frac{2 (m+1)}{m}} m \sec \left(\frac{2 \pi }{m}\right) \Gamma \left(\frac{1}{2}-\frac{1}{m}\right) \Gamma \left(4+\frac{5}{m}\right) \Gamma \left(-\frac{m+3}{m}\right)}{\sqrt{\pi } (3 m+4) \Gamma \left(3+\frac{3}{m}\right) \Gamma \left(-\frac{m+2}{m}\right)}~,
\end{align}
with $C_{(3,1)(3,1)(3,1)}$ and $B_{(3,1)}^{{\bf a}\,(1,1)}$ given (respectively) in equations~\eqref{bulkOPEcoeffnew} and~\eqref{B3151}.
In the large-$m$ expansion (with $a_1,a_2\ll m$)
\begin{align}\label{B31B51sqrt}
	(B_{(3,1)}^{{\bf a}\,(3,1)})^2&=\frac{32 \pi ^2 \left(a_1^2-1\right)}{3 m^2}+\frac{32 \pi ^2 \left(a_1^2-1\right)}{m^3}-\frac{32 \pi ^2 \left(\pi ^2 \left(4 a_1^2-7\right)+15\right) \left(a_1^2-1\right)}{9 m^4}+O(m^{-5})~,\nonumber\\
	(B_{(3,1)}^{{\bf a}\,(5,1)})^2&=\frac{8 \pi ^4 (a_1^2-4) (a_1^2-1)}{9 m^4}+O(m^{-5})~.
\end{align}

\subsection{Correlator between four \texorpdfstring{$\psi_{(3,1)}$}{psi31}}\label{31threept}
Consider the four-point correlation function of $\psi_{(3,1)}$. The associated differential equation and its solutions where given in section~\ref{sec:phi31twoptsol}. The fusion rules read
\begin{align}\label{psi31fusion}
	\psi_{(3,1)}\times \psi_{(3,1)}&=\hid&& m=4\nonumber\\
	\psi_{(3,1)}\times \psi_{(3,1)}&=\hid+\psi_{(3,1)}&& m=5\nonumber\\
	\psi_{(3,1)}\times \psi_{(3,1)}&=\hid+\psi_{(3,1)}+\psi_{(5,1)}&& m\geq 6~.
\end{align}
The solution is
\begin{align}\label{psi31fourpt}
	&\langle \psi_{(3,1)}(x_1)\psi_{(3,1)}(x_2)\psi_{(3,1)}(x_3)\psi_{(3,1)}(x_4)\rangle_{\uhp}=\frac{\tilde{ \mathcal{G}}(\tilde\eta)}{(x_{12}x_{34})^{2h_{3,1}}}~,\nonumber\\
	\tilde{ \mathcal{G}}&(\tilde\eta)=V_{(1,1)}(\tilde\eta)+({\hC}_{(3,1)(3,1)}^{{\bf a}(3,1)})^2V_{(3,1)}(\tilde\eta)+	({\hC}_{(3,1)(3,1)}^{{\bf a}(5,1)})^2V_{(5,1)}(\tilde\eta)~.
\end{align}
The form of the $V_{(r,s)}(\tieta)$ is given in eq.~\eqref{phi31twoptblocks}, while $\tieta$ is defined in terms of the standard cross ratio $\eta$ as
\begin{align}
	\tieta=\frac{\eta ^2}{1-\eta}~=\frac{x_{12}^2 x_{34}^2}{x_{12}x_{14}  x_{23}x_{24}}~,
\end{align}
which is positive if $x_i > x_{i+1}$. 

\subsubsection{Boundary crossing} The squared boundary OPE coefficients in eq.~\eqref{psi31fourpt} are fixed by crossing symmetry, which in terms of the  $	\mathcal{G}$ previously defined, requires that:
\begin{align}
		\mathcal{G}(\eta)=\left(\frac{1-\eta }{\eta }\right)^{-2 h_{3,1}}	\mathcal{G}(1-\eta)~.
\end{align}
For $m=4$ we have that ${\hC}_{(3,1)(3,1)(3,1)}^{{\bf a}}={\hC}_{(3,1)(3,1)(5,1)}^{{\bf a}}=0$ and boundary crossing symmetry is ensured by the identity block alone. For $m=5$, or for those conformal b.c. that do not allow $\psi_{(5,1)}$ we have that ${\hC}_{(3,1)(3,1)(5,1)}^{{\bf a}}=0$ and crossing is satisfied if
\begin{align}
	({\hC}_{(3,1)(3,1)(3,1)}^{{\bf a}})^2=\frac{ 2^{\frac{m-2}{m}}\pi \cos \left(\frac{2 \pi }{m}\right) \Gamma \left(3+\frac{4}{m}\right) \Gamma \left(-\frac{m+2}{2 m}\right)}{\left(2 \cos \left(\frac{2 \pi }{m}\right)+1\right) \Gamma \left(\frac{1}{2}+\frac{1}{m}\right) \Gamma \left(\frac{3}{2}+\frac{1}{m}\right) \Gamma \left(2+\frac{3}{m}\right) \Gamma \left(-\frac{m+4}{2 m}\right)}~,
\end{align}
which vanishes identically for $m=4$, but is positive otherwise.
For $m>5$ we find generically a one-parameter family of solutions to the crossing equation, so that additional correlators (and crossing constraints) are needed in order to compute ${\hC}_{(3,1)(3,1)(3,1)}^{{\bf a}}$ and ${\hC}_{(3,1)(3,1)(5,1)}^{{\bf a}}$, see e.g. ref.~\cite{Runkel:1998he}.

\subsection{Correlator between two \texorpdfstring{$\psi_{(r,s)}$}{psirs} and one \texorpdfstring{$\phi_{(3,1)}$}{phi31}}\label{app:bdbdbulk31pt}
Next, we consider the boundary-boundary-bulk correlation function
\begin{align}\label{bdbdphi31gen}
	\langle \psi_{(r,s)}(x_1)\psi_{(r,s)}(x_2)\phi_{(3,1)}(x+i y,x-iy)\rangle_{\uhp}~,\quad x_1>x_2~.
\end{align}
This correlator satisfies the third order differential equation of~\eqref{thirdgengen}, now with $h_{1,3}\rightarrow h_{3,1}$. There are three independent solutions (corresponding to the boundary operators $\hid, \psi_{(3,1)}$ and $\psi_{(5,1)}$):
\begin{align}\label{allblocks31}
	V_{(1,1)}(\tilde\eta)&=\, _3F_2\left(-\frac{2}{m}-1,\frac{r}{m}-\frac{1}{m}+r-s,-\frac{r}{m}-\frac{1}{m}-r+s;-\frac{3}{m}-1,\frac{1}{2}-\frac{1}{m};\frac{\tieta}{4}\right)~,\nonumber\\
	V_{(3,1)}(\tilde\eta)&=\tieta^{h_{3,1}/2} \, _3F_2\left(-\frac{1}{m}-\frac{1}{2},\frac{r}{m}+r-s+\frac{1}{2},-\frac{r}{m}-r+s+\frac{1}{2};-\frac{2}{m}-\frac{1}{2},\frac{1}{m}+\frac{3}{2};\frac{\tieta}{4}\right),\nonumber\\
	V_{(5,1)}(\tilde\eta)&=\tieta^{h_{5,1}/2}\, _3F_2\left(\frac{1}{m}+1,\frac{r}{m}+\frac{2}{m}+r-s+2,-\frac{r}{m}+\frac{2}{m}-r+s+2;\frac{2}{m}+\frac{5}{2},\frac{3}{m}+3;\frac{\tieta}{4}\right)~.
\end{align}
The cross-ratio $\tieta$ is defined as in eq.~\eqref{etatildedef}.
The final solution is then
\begin{align}\label{treptphi31}
	\langle \psi_{(r,s)}(x_1)&\psi_{(r,s)}(x_2)\phi_{(3,1)}(x+iy,x-iy)\rangle_{\uhp}=\frac{\tilde{ \mathcal{G}}(\tilde\eta)}{(x_{12})^{2h_{r,s}} (2y)^{2h_{3,1}}}~,\nonumber\\
	&\tilde{ \mathcal{G}}(\tilde\eta)=B_{(3,1)}^{{\bf a}\,(1,1)}V_{(1,1)}(\tilde\eta)+	\alpha_{(3,1)}V_{(3,1)}(\tilde\eta)+	\alpha_{(5,1)}V_{(5,1)}(\tilde\eta)~,
\end{align}
where $B_{(3,1)}^{{\bf a}\,(1,1)}$ is the one-point function coefficient of $\phi_{(3,1)}$ computed in eq.~\eqref{B3151} and we denoted
\begin{align}
	\alpha_{(3,1)}\equiv {\hC}_{(r,s)(r,s)(3,1)}^{{\bf a}}B_{(3,1)}^{{\bf a}\,(3,1)}~,\quad \alpha_{(5,1)}\equiv {\hC}_{(r,s)(r,s)(5,1)}^{{\bf a}}B_{(3,1)}^{{\bf a}\,(5,1)}~.
\end{align}

\subsubsection{Removing unphysical singularities}
Let us now look for possible unphysical singularities in the r.h.s. of eq.~\eqref{treptphi31}. There is a branch cut at $\tieta\in [4,\infty]$, so we must require that $\text{Disc}~\tilde{ \mathcal{G}}=0$ across that cut. For the b.c.~where $\psi_{(3,1)}$ and $\psi_{(5,1)}$ are not allowed in the boundary spectrum (see Table~\ref{tbl:smallconfbc}), $V_{(1,1)}$ alone ensures that $\text{Disc}~\tilde{ \mathcal{G}}=0$. 

For generic $m>5$, and any conformal b.c. ${\bf a} =(a_1,a_2)_m$ such that
\begin{enumerate}
	\item The boundary primary $\psi_{(r,s)}$ is allowed to exist;
	\item $\alpha_{(3,1)}\neq0$, but $\alpha_{(5,1)}\equiv {\hC}_{(r,s)(r,s)(5,1)}^{{\bf a}}B_{(3,1)}^{{\bf a}\,(5,1)}= 0$ (either because $\psi_{(5,1)}$ does not exist, or ${\hC}_{(r,s)(r,s)(5,1)}^{{\bf a}}=0$) or it is subleading with respect to $\alpha_{(3,1)}$ in the large-$m$ limit,
\end{enumerate}
then the condition $\text{Disc}~\tilde{ \mathcal{G}}=0$ requires that
\begin{align}
	\alpha_{(3,1)}=-B_{(3,1)}^{{\bf a}\,(1,1)}\frac{ 2^{-\frac{6}{m}-4} \sqrt{\pi }\, \Gamma \left(\frac{1}{2}-\frac{1}{m}\right) \Gamma \left(-\frac{m+2}{2 m}\right) \Gamma \left(-\frac{m+3}{m}\right) \sec \left(\pi  \left(\frac{r}{m}+r-s\right)\right)}{\Gamma \left(\frac{3}{2}+\frac{1}{m}\right) \Gamma \left(-\frac{2 (m+2)}{m}\right) \Gamma \left(s-\frac{m r+r+1}{m}\right) \Gamma \left(\frac{m r+r-m s-1}{m}\right)}~.
\end{align}
For finite $(r,s)$, the leading large-$m$ expansion (with $a_1,a_2\ll m$) of the former result gives
\begin{align}\label{largemalpha31}
	\alpha_{(3,1)}=\frac{2 \pi  (r-s) (\text{sgn}(r-s)+r)}{\sqrt{3} m}+O(m^{-2})~.
\end{align}

\section{Some integrals on \texorpdfstring{AdS${}_{p+1}$}{AdSp1}}\label{app:integrals}
We would like to compute the following integrals:
\begin{align}\label{integrruleapppdim}
	I_p(\alpha)\equiv \int d^p \vec{x}\int_{a}^\infty \frac{d z} {z^{p+1}}~v^\alpha~,\quad 
	\hat{I}_p(\alpha)\equiv\int\frac{d^p \vec{x}}{a^p}~v^\alpha\bigg\rvert_{z=a}~,
\end{align}
with $a,\alpha>0$ and
\begin{align}
	v\equiv \frac{z^2 (x_1-x_2)^2}{ (z^2 + (x_1-x)^2)(z^2 + (x_2-x)^2)}~,
\end{align}
being $\vec{x}_1, \vec{x}_2, \vec{x}$ parallel coordinates along the AdS${}_{p+1}$ boundary (in Poincaré coordinates) whereas $z$ is transverse, and $x_{12}\equiv |\vec{x}_1-\vec{x_2}|$. Let us start with the first integral. After going to polar coordinates, for the angular integral we find
\begin{align}
I_p(\alpha) &= S_{p-1} \int_{a}^\infty \frac{d z} {z^{p+1}}\int_{0}^\infty r^{r-1}dr \int_0^{2\pi} (\sin \theta )^{p-3} ~v^\alpha\nonumber\\
&=S_{p}\int_{0}^\infty dr \int_{a}^\infty dz~{ \frac{r^{p-1} x_{12}^{2 \alpha } z^{2 \alpha -p-1}}{[\left(r^2+z^2\right) \left(r^2+x_{12}^2+z^2\right)]^{\alpha } } \, _2F_1\left(\frac{\alpha }{2},\frac{\alpha +1}{2};\frac{p}{2};\frac{4 r^2 x_{12}^2}{\left(r^2+x_{12}^2+z^2\right)^2}\right)}{}~.
\end{align}
Next we Taylor expand $\, _2F_1$, so that the above equation becomes an infinite sum of elementary functions. Each term in this expansion can be further mellinized, so that we find
\begin{align}
	I_p(\alpha) &= \sum_{n=0}^\infty  \frac{2^{2 n+1} \pi ^{p/2} \left(\frac{\alpha }{2}\right)_n \left(\frac{\alpha +1}{2}\right)_n}{n! \left(\frac{p}{2}\right)_n\Gamma\left(\frac{p}{2}\right)}\int_{-i\infty}^{i\infty} \frac{d t}{2\pi i} ~c_n (\alpha,t)\,u^{4t-2 \alpha -4 n-p}~,
\end{align}
where $u\equiv a/x_{12}$ and
\begin{align}
	c_n (\alpha,t) \equiv\frac{\Gamma \left(n+\frac{p}{2}\right) \Gamma \left(2t-n-\frac{p}{2}\right) \Gamma (2 n-2t+2\alpha ) \Gamma \left(3 n+\frac{p}{2}-2 t+\alpha \right)}{  (2 \alpha +4 n+p-4 t) \Gamma (2 n+\alpha )\Gamma \left(3 n+\frac{p}{2}-2 t+2 \alpha \right)}~.
\end{align}
We can close the contour to the left, and pick up poles at $-n - p/2 + 2 t = -k$, $k=0,1,2,3,\dots$. After summing over all these residues we find
\begin{align}
R_n(\alpha)&\equiv 	\int_{-i\infty}^{i\infty} \frac{d t}{2\pi i} ~c_n (\alpha,t)\,u^{4t-2 \alpha -4 n-p} \nonumber\\
	& =  \sum_{k=0}^\infty \frac{(-1)^k \Gamma \left(n+\frac{p}{2}\right) \Gamma (k+2 n+\alpha )  \Gamma \left(k+n+2 \alpha -\frac{p}{2}\right)}{4 k! (\alpha +k+n) \Gamma (2 n+\alpha ) \Gamma (k+2 n+2\alpha )} \left(\frac{1}{u^2}\right)^{\alpha + k+n}.
\end{align}
Plugging this result into the expression for $I_p(\alpha) $ we get
\begin{align}\label{Ipalpha1}
	I_p(\alpha) &= \sum_{n=0}^\infty  \frac{2^{2 n+1} \pi ^{p/2} \left(\frac{\alpha }{2}\right)_n \left(\frac{\alpha +1}{2}\right)_n}{n! \Gamma \left(\frac{p}{2}\right) \left(\frac{p}{2}\right)_n} R_n(\alpha)~\nonumber\\
	&= \frac{2^{-2 \alpha } \pi ^{\frac{p+1}{2}}  \Gamma \left( 2\alpha -p/2\right) }{  \Gamma (1+\alpha ) \Gamma \left(\alpha+1/2\right)}\left(\frac{1}{u^2}\right)^{\alpha } \, _3F_2\left(\alpha ,\alpha ,2 \alpha -\frac{p}{2};\alpha +\frac{1}{2},\alpha +1;-\frac{1}{4 u^2}\right)~.
\end{align}
For small $a$ and finite $x_{12}$ (namely small $u$) we find
\begin{align}\label{Ismalla}
	I_p(\alpha)&=\frac{\pi ^{p/2} \Gamma \left(\alpha -\frac{p}{2}\right)}{\Gamma (\alpha )}\left( \log (1/u^2)-\psi(\alpha )+\psi \left(\alpha -\frac{p}{2}\right)\right)\nonumber\\
	&+u^{2 \alpha -p} \left(\frac{\pi ^{\frac{p}{2}+2} \csc ^2\left(\frac{1}{2} \pi  (p-2 \alpha )\right) \Gamma \left(2 \alpha -\frac{p}{2}\right)}{\Gamma (\alpha )^2 \Gamma (p-2 \alpha +1) \Gamma \left(-\frac{p}{2}+\alpha +1\right)^2}+O(u^2)\right)~,
\end{align}
where $\psi(x)$ is the digamma function.
Similar tricks can be used in order to compute the second integral in \eqref{integrruleapppdim}, for which we find
\begin{align}\label{Ipalpha2}
	\hat{I}_p(\alpha) &= \sum_{n,k=0}^\infty \frac{(-1)^k \pi ^{p/2} \Gamma (k+2 n+\alpha ) \Gamma \left(k+n+2 \alpha -\frac{p}{2}\right)}{\Gamma (\alpha ) \Gamma (k+1) \Gamma (n+1) \Gamma (k+2 (n+\alpha ))}\left(\frac{1}{u^2}\right)^{\alpha + k+n}~.
\end{align}
Therefore after performing the infinite sum we obtain
\begin{align}\label{Ipsmalla}
\hat{I}_p(\alpha) =	\frac{2^{1-2 \alpha } \pi ^{\frac{p+1}{2}} \Gamma \left(2\alpha-p/2\right)}{\Gamma (\alpha ) \Gamma \left(\alpha+1/2\right)}\left(\frac{1}{u^2}\right)^{\alpha}  \, _2F_1\left(\alpha ,2 \alpha -\frac{p}{2};\alpha +\frac{1}{2};-\frac{1}{4 u^2}\right)~.
\end{align}
For small $u$ we have that
\begin{align}
	\hat{I}_p(\alpha)&=\frac{2\pi ^{p/2} \Gamma \left(\alpha -\frac{p}{2}\right)}{\Gamma (\alpha )}+u^{2 \alpha -p} \left(\frac{\pi ^{\frac{p+1}{2}} 2^{2 \alpha -p+1} \Gamma \left(\frac{p}{2}-\alpha \right) \Gamma \left(2 \alpha -\frac{p}{2}\right)}{\Gamma (\alpha )^2 \Gamma \left(\frac{p}{2}-\alpha +\frac{1}{2}\right)}+O(u^2)\right)~.
\end{align}

\bibliography{bib}
\bibliographystyle{JHEP}

\end{document}